\documentclass[aps,prl,10pt,twocolumn,a4paper,superscriptaddress, longbibliography, notitlepage]{revtex4-1}

\usepackage{amsmath}
\usepackage{amssymb}
\usepackage{textcomp}
\usepackage{upgreek}
\usepackage{units}
\usepackage[pdftex]{graphicx}
\usepackage{microtype}
\usepackage{enumitem}
\usepackage{xcolor}
\usepackage{gnuplottex}
\usepackage{multirow}
\usepackage{sansmath}
\usepackage{soul} %to yellow highlight  text
\usepackage{hyperref}
\hypersetup{colorlinks=true, citecolor=blue, urlcolor=blue, linkcolor=blue}

\usepackage{txfonts}

\renewcommand{\vec}[1]{\ensuremath{\boldsymbol{#1}}}

\usepackage{pifont}
\usepackage{braket}
\usepackage{comment}

\begin{document}

\title{
$\mathbb{Z}_2$ Vortex Crystals in Tetrahedral Antiferromagnets: Fractional Charges and Topological Magnons
%$\mathbb{Z}_2$ Vortex Crystals and Topological Magnons in a Tetrahedral Antiferromagnet
}

\author{Tomoki Hirosawa}
\affiliation{Department of Physical Sciences, Aoyama Gakuin University, Sagamihara, Kanagawa 252-5258, Japan}
\author{Alexander Mook}
\affiliation{University of Münster, Institute of Solid State Theory, 48149 Münster, Germany}
\affiliation{Institute of Physics, Johannes Gutenberg University Mainz, Mainz D-55128, Germany}
\author{Maria Azhar}
\affiliation{Faculty of Physics and Center for Nanointegration Duisburg-Essen (CENIDE), University of Duisburg-Essen, 47057 Duisburg, Germany}

\begin{abstract}
    We report the formation of a $\mathbb{Z}_2$ vortex crystal in the tetrahedral antiferromagnetic order on a triangular lattice.
    The noncoplanar tetrahedral state consists of four sublattices with spins oriented along the faces of a tetrahedron in spin space.
    The long-range order characterized by a $\mathbb{Z}_2$ topology arises due to the Dzyaloshinskii–Moriya interaction and appears at zero temperature and without external fields. 
    Each vortex carries a half-integer electric charge relative to the uniform background in itinerant magnets, enabling the emergence of anyonic excitations.
    Its magnetic excitations include magnetically active gyrotropic and breathing modes, which---under an external magnetic field---carry nontrivial Chern numbers that stabilize chiral magnon edge states.
\end{abstract}
\maketitle

Magnetic systems provide a rich platform for exploring topological defects~\cite{zang2018topology}, which are robust against perturbations, making them promising candidates for information applications.
For example, magnetic skyrmions are topological defects characterized by an integer quantized skyrmion number (also known as topological charge), defined as $N_\textrm{sk}=
\frac{1}{4\pi} \int dr^2 \,\vec{n}\cdot[
\partial_x \vec{n}\times \partial_y\vec{n}]$ where $\vec{n}(\vec{r})$ is a unit vector field~\cite{nagaosaTopologicalPropertiesDynamics2013, everschorPerspectiveMagneticSkyrmions2018}. They are stabilized through various mechanisms, including Dzyaloshinskii–Moriya~(DM) interactions~\cite{bogdanovThermodynamicallyStableMagnetic1994, rosslerSpontaneousSkyrmionGround2006,muhlbauerSkyrmionLatticeChiral2009, yuRealspaceObservationTwodimensional2010}, frustrated interactions~\cite{Okubo2012,Leonov2015,hayamiEffectiveBilinearbiquadraticModel2017a, kurumajiSkyrmionLatticeGiant2019}, dipolar interactions~\cite{ezawaGiantSkyrmionsStabilized2010,yuMagneticStripesSkyrmions2012}, and four-spin interactions~\cite{Heinze2011}.
Noncoplanar magnetic textures induce an effective magnetic field for electrons, giving rise to topological phenomena such as the quantum anomalous Hall effect~\cite{ohgushiSpinAnisotropyQuantum2000, nagaosaAnomalousHallEffect2006,Martin2008} and the topological Hall effect~\cite{neubauerTopologicalHallEffect2009, jonietzSpinTransferTorques2010, schulzEmergentElectrodynamicsSkyrmions2012}. 
Furthermore, spin-wave quanta, or magnons, exhibit topological phases in skyrmion crystals~(SkXs)~\cite{muhlbauerSkyrmionLatticeChiral2009}, including Chern insulators~\cite{roldan-molinaTopologicalSpinWaves2016, garstCollectiveSpinExcitations2017, diazChiralMagnonicEdge2020, mookQuantumDampingSkyrmion2020, hirosawaLaserControlledRealReciprocalSpace2022} and second-order topological insulators~\cite{hirosawaMagnonicQuadrupoleTopological2020}.

Another example is the $\mathbb{Z}_2$ vortex, a point-like topological defect of an $\mathrm{SO}(3)$ order parameter~\cite{Kawamura1984}. These vortices emerge as thermal excitations in a Heisenberg antiferromagnetic~(AFM) triangular lattice with nearest-neighbor interactions, where the classical spin ground state adopts a noncollinear 120$^\circ$ order due to frustration. 
While the experimental observation of $\mathbb{Z}_2$ vortices had been elusive, the signature of $\mathbb{Z}_2$ vortices was recently reported using quasielastic neutron scattering measurements~\cite{tomiyasuObservationTopological2022}.
A $\mathbb{Z}_2$ vortex corresponds to a $2\pi$ winding of the 120$^\circ$ order around a line defect. Since the 120$^\circ$ order is invariant under $\mathrm{SO}(3)$ rotations, $\mathbb{Z}_2$ vortices with opposite chiralities can be continuously transformed into each other. Consequently, pairs of $\mathbb{Z}_2$ vortices can annihilate one another.

%=========================================
\begin{figure}[t!]
	\centering
	\includegraphics[width=0.5\textwidth]{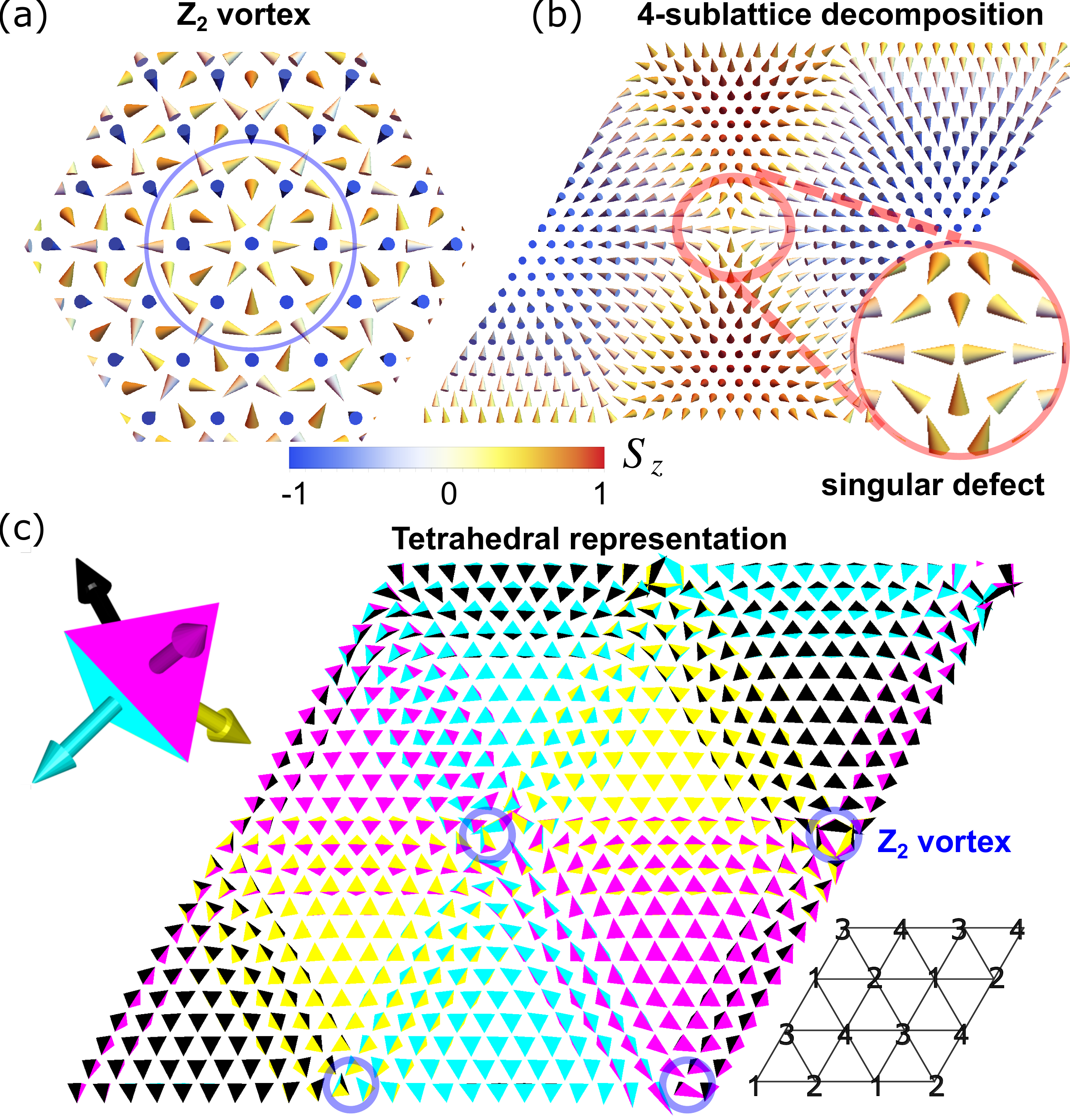}
	\caption{
    The $\mathbb{Z}_2$ vortex crystal in the noncoplanar four-sublattice tetrahedral AFM order consists of interpenetrating triple-$q$ orders. (a)~The full spin configuration near the $\mathbb{Z}_2$ vortex, (b) its decomposition into one of the four sublattices, and (c) the tetrahedral representation of the spin structure, where each tetrahedron corresponds to four lattice sites. 
    A magnetic unit cell of the $\mathbb{Z}_2$ vortex crystal is plotted in (b) and (c).
    Faces of tetrahedra are orthogonal to spin vectors of each sublattice in (c), which are colored magenta, cyan, black, and yellow. 
    The rotation centers of the tetrahedra (encircled in blue) are characterized by a nontrivial $\mathbb{Z}_2$ vorticity.
    The sublattice spin texture exhibits singular defects at $\mathbb{Z}_2$ vortex cores as highlighted in the inset of (b).
    The inset of (c) shows the four-sublattice order on a triangular lattice with the numbers corresponding to the sublattice index.
    The spin texture was obtained by Monte Carlo simulations on a $50\times 50$ spin lattice with periodic boundary conditions. The parameters are set to $B_1/J_1=1$, $J_2/J_1=B_2/B_1=0.5$, $D_1/J_1=0.5$, and $b/J_1=0$. 
    }
	\label{fig1}
\end{figure}
%=========================================

Although skyrmions and $\mathbb{Z}_2$ vortices originate from different order parameter spaces, the recent discovery of three-sublattice SkX (AFM-SkX) provides an exciting opportunity to explore the interplay between these topological defects~\cite{rosalesThreesublatticeSkyrmionCrystal2015, diazTopologicalMagnonsEdge2019,gaoFractionalAntiferromagneticSkyrmion2020}. Introducing DM interactions into an AFM triangular magnet spatially modulates the 120$^\circ$ order, resulting in a three-sublattice helical phase and an AFM-SkX under magnetic fields. Furthermore, distorted chiral antiferromagnets with anisotropic DM interactions can generate pairs of $\mathbb{Z}_2$ vortices within AFM-SkXs, leading to the formation of $\mathbb{Z}_2$ vortex crystals~(VC)~\cite{Osorio2019}. Similarly, Kitaev interactions stabilize $\mathbb{Z}_2$ VCs in AFM triangular and honeycomb lattices~\cite{Rousoch2016, Yao2016}.
These studies suggest that a rich family of topological spin textures arises when weak perturbations break the $\mathrm{SO}(3)$ symmetry in AFM triangular magnets.

In this work, we propose the noncoplanar four-sublattice tetrahedral order~\cite{Martin2008} as a new building block for topological spin textures. 
This tetrahedral AFM order emerges as the ground state of a triangular lattice in the Kondo lattice model~\cite{Martin2008,katoStabilitySpontaneousQuantum2010, Akagi2012} and the AFM Heisenberg model with multiple-exchange interactions~\cite{Momoi1997}. Unlike the 120$^\circ$ order, it is characterized by a spatially uniform skyrmion number density~\cite{bergDefinitionStatisticalDistributions1981, Martin2008}, resulting in the quantum anomalous Hall effect. With an SO(3) order parameter space, the tetrahedral AFM order also hosts $\mathbb{Z}_2$ vortices~\cite{barrosEfficientLangevinSimulation2013a}.
Crucially, $\mathbb{Z}_2$ vortices introduce a fractional magnetic flux quantum to the quantum anomalous Hall state, giving rise to electronic fractionalization with Abelian anyonic statistics~\cite{Rahmani2013}.

Here, we reveal the spontaneous formation of a $\mathbb{Z}_2$ VC in the tetrahedral AFM order.
We construct a minimal spin-lattice model for AFM triangular magnets with positive biquadratic exchange interactions, where the tetrahedral AFM order emerges as the ground state.
Remarkably, we find that the uniform tetrahedral phase becomes unstable under small DM interactions, forming a $\mathbb{Z}_2$ VC in the absence of an external magnetic field.
The $\mathbb{Z}_2$ VC, shown in Fig.~\ref{fig1}, consists of four interpenetrating sublattices of triple-$q$ orders each characterized by the nontrivial skyrmion number.
We identify the $\mathbb{Z}_2$ VC phase across a broad range of magnetic fields and DM interactions, where each vortex is associated with a half-integer electric charge relative to the uniform background in the model of itinerant magnets.
Additionally, we find topologically nontrivial excitations that are coupled to spatially uniform temporally oscillating magnetic fields, highlighting the potential for spintronic applications.

\paragraph{Tetrahedral antiferromagnetic order:} 
We consider the following minimal classical spin-lattice model for stabilizing the tetrahedral AFM order:
\begin{align}
H_\text{BQ}
&=
\sum_{\braket{\vec{r},\vec{r}'}}J_1\vec{S}_{\vec{r}}\cdot\vec{S}_{\vec{r}'} +B_1(\vec{S}_{\vec{r}}\cdot\vec{S}_{\vec{r}'})^2
\notag
\\
&
+\sum_{\braket{\vec{r},\vec{r}''}_\textrm{nn}}
J_2\vec{S}_{\vec{r}}\cdot\vec{S}_{\vec{r}''} +B_2(\vec{S}_{\vec{r}}\cdot\vec{S}_{\vec{r}''})^2,
\label{eq:H_BQ}
\end{align}
where $\vec{S}_{\vec{r}}$ is a classical spin with a unit length, $\sum_{\braket{\vec{r},\vec{r}'}}$ and $\sum_{\braket{\vec{r},\vec{r}''}_\textrm{nn}}$ denote summation over nearest neighbors~(NN) and next-nearest neighbors~(NNN) on a triangular lattice, respectively. 
Following Ref.~\cite{Akagi2012}, we assume antiferromagnetic exchange $(J_1>0)$ and positive biquadratic interactions $(B_1>0)$.

\begin{figure}[tb]
	\centering
	\includegraphics[width=0.47\textwidth]{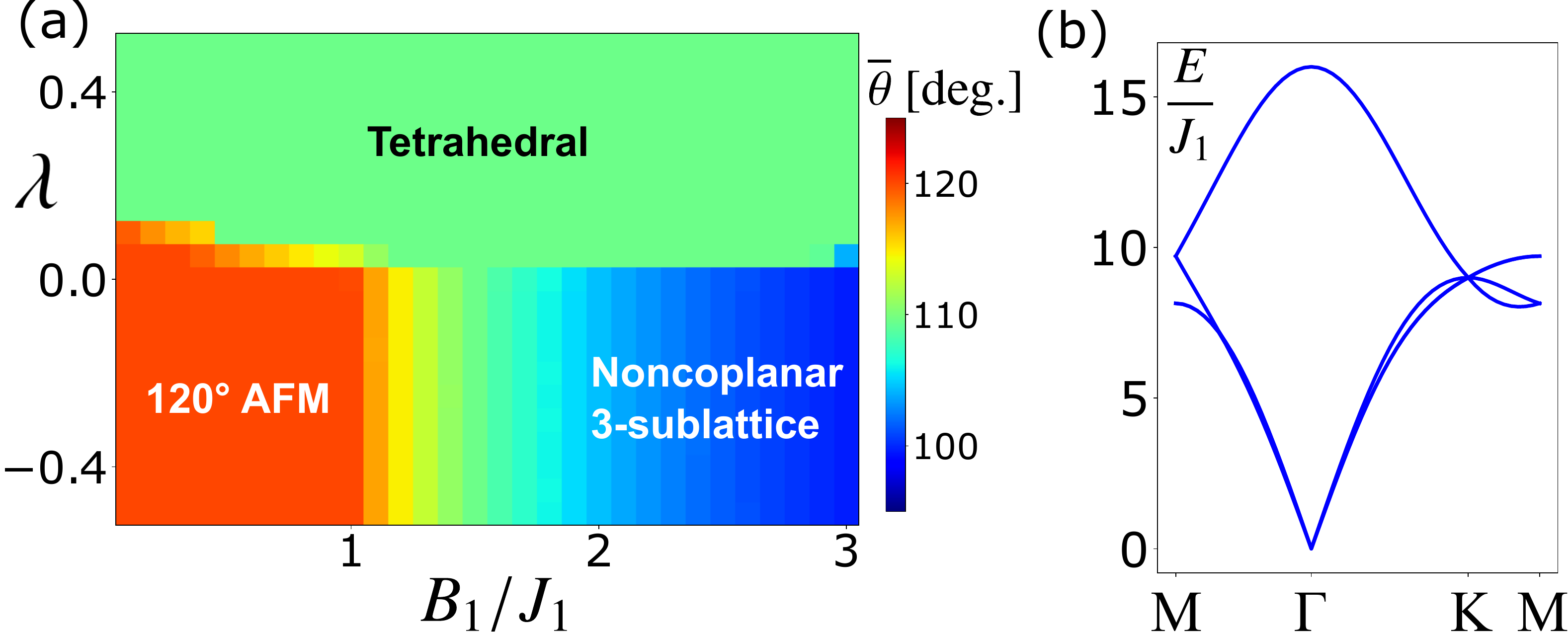}
	\caption{
    (a) Classical ground state phase diagram of $H_\text{BQ}$ in Eq.~\eqref{eq:H_BQ} spanned by the biquadratic interaction $B_1/J_1$ and the relative strength of next nearest coupling $\lambda=J_2/J_1=B_2/B_1$.
    This phase diagram is obtained by Monte Carlo simulations on a $30\times 30$ triangular lattice of spins.
    The color indicates the average angle $\overline{\theta}$ between nearest neighbor spins. 
    (b) Magnon band structure of the tetrahedral AFM order. The parameters are set as $B_1/J_1=1$ and $\lambda=0.5$. 
    }
	\label{fig2}
\end{figure}

Using Monte Carlo annealing simulations on a $30\times 30$ triangular lattice of spins with periodic boundary conditions, we investigate the stability of the tetrahedral AFM order within this model.
For simplicity, we assume the same ratio between NN and NNN interactions in both exchange and biquadratic terms, denoted as $\lambda=J_2/J_1=B_2/B_1$. 
Figure~\ref{fig2}(a) presents the magnetic phase diagram as a function of $B_1/J_1$ and $\lambda$, with color indicating the average angles ($\overline{\theta}$) between NN spins. 
As schematically shown in Fig.~\ref{fig1}(c), the tetrahedral AFM order features a noncoplanar four-sublattice configuration with the maximal angular separation between spins, given by $\overline{\theta}=\arctan(-1/3)\approx 109.5^\circ$~\cite{Martin2008}.
We find that the sign of the NNN interaction plays an essential role in stabilizing the tetrahedral AFM order.
For $\lambda>0$, the tetrahedral AFM order remains stable over a wide range of $B_1/J_1$. However, for $\lambda\le 0$, the ground state is the 120$^{\circ}$ order at $B_1/J_1\le 1$ and the three-sublattice noncoplanar order at $B_1/J_1>1$.

Since the Hamiltonian in Eq.~\eqref{eq:H_BQ} only depends on the dot product between spins, the angle $\overline{\theta}$ determines the energy of the classical spin ground state. 
Based on this observation, we find a simple ansatz explaining the phase diagram at $\lambda=0$, which is detailed in Sec.~IA and IB of the Supplemental Material (SM) \cite{SM}.
Assuming that the dot product is equal on all NN bonds, the total energy of the ground state is given by
\begin{equation}
    \frac{E}{M}=J_1\alpha+B_1\alpha^2=B_1\left(\alpha+\frac{J_1}{2B_1}\right)^2-\frac{J_1^2}{4B_1}
\end{equation}
where $M$ is the total number of bonds between NNs and $\alpha$ is the average value of the dot product between NN spins. 
The minimum value of the energy is $E_\textrm{min}/M=-J_1^2/\left(4B_1\right)$, when $\alpha=-J_1/(2B_1)$.
However, we cannot take $\alpha$ to be arbitrarily small. The smallest possible value in the triangular AFM magnet is $\alpha=-1/2$, corresponding to the coplanar 120$^{\circ}$ order. Therefore, the ground state is the 120$^{\circ}$ order for $B_1\le J_1$. For $B_1> J_1$, the energy is minimized by the noncoplanar three-sublattice orders, with a continuous variation in $\alpha$. Ultimately, for $B_1\gg J_1$, a $90^\circ$ order is stabilized~\cite{Akagi2012,groverNonAbelianSpinLiquid2011}.
This picture changes significantly upon introducing the NNN interaction with $\lambda>0$.
The NNN interaction penalizes parallel spin alignments of NNN spins, thus favoring the four-sublattice structure over the three-sublattice structure. 
As a result, the tetrahedral AFM order becomes the ground state.
In contrast, the NNN interaction with $\lambda<0$ favors the three-sublattice structure, leaving the magnetic phase diagram qualitatively unchanged. 
In the following, we fix $B_1/J_1=1$ and $\lambda=0.5$ to study the tetrahedral AFM order.

We perform the Holstein-Primakoff expansion~\cite{holsteinFieldDependenceIntrinsic1940, roldan-molinaTopologicalSpinWaves2016, diazChiralMagnonicEdge2020} to compute the magnon bands of the tetrahedral AFM order~(Sec.~IC in SM \cite{SM}).
%Due to the $\mathrm{SO}(3)$ symmetry of our model, the spectrum features three Goldstone modes~\cite{Akagi2013}, as illustrated in Fig.~\ref{fig2}(b).
In the tetrahedral AFM order, the spontaneous breaking of the full SO(3) spin-rotation symmetry gives rise to three linearly dispersing Goldstone modes~\cite{Akagi2013}, as illustrated in Fig.~\ref{fig2}(b).
Despite the nontrivial skyrmion number of the tetrahedral AFM order~(Sec.~ID in SM \cite{SM}), the magnon spectrum lacks a band gap, precluding the presence of topological magnon bands.

\paragraph{$Z_2$ vortex crystal phase:}
When perturbations weakly break the $\mathrm{SO}(3)$ symmetry, a novel long-range order can emerge due to spatial rotations of the $\mathrm{SO}(3)$ order~\cite{rodriguesDzyaloshinskiiMoriyaInducedSpintransfer2022}.
Here, we introduce the interfacial DM interaction as a symmetry-breaking perturbation. The Hamiltonian is given by
\begin{equation}
    H=H_\text{BQ}\,+\sum_{\braket{\vec{r},\vec{r}'}} \vec{D}_{1}^{\,\vec{r},\vec{r}'}\cdot\vec{S}_{\vec{r}}\times\vec{S}_{\vec{r}'}-b \sum_{\vec{r}}\vec{S}_{\vec{r}}\cdot\hat{\vec{z}},
    \label{eq:model}
\end{equation}
where the DM vectors are $\vec{D}_{1}^{\,\vec{r},\vec{r}'}=D_1\,\hat{\vec{z}}\times(\vec{r}-\vec{r}')/|\vec{r}-\vec{r}'|$, and the Zeeman coupling to external fields is parametrized by $b$. 
The DM interaction averages to zero in alternately canted spin configurations, leaving the energy of the tetrahedral AFM order unchanged.
However, the DM interaction has non-vanishing contributions to the spin wave Hamiltonian.
Crucially, the eigenvalues of the spin wave Hamiltonian become negative due to the DM interaction, indicating the instability of the uniform tetrahedral phase under even infinitesimal DM interactions~(Sec.~IIA in SM \cite{SM}).

Using Monte Carlo annealing, we find that the DM interaction stabilizes a long-range order by introducing spatial variations of the tetrahedral AFM order in the form of rotations, as illustrated in Fig.~\ref{fig1}. When decomposing the full texture---shown in Fig.~\ref{fig1}(a)---into its four sublattices, skyrmion-like configurations with six-fold rotational symmetry are identified, see Fig.~\ref{fig1}(b).
They are characterized by the nontrivial skyrmion number computed within a sublattice and the triple-$q$ structure in the static spin structure factor~(see Sec.~IIB in SM \cite{SM}).
In stark contrast to the three-sublattice AFM-SkXs~\cite{rosalesThreesublatticeSkyrmionCrystal2015, diazTopologicalMagnonsEdge2019, gaoFractionalAntiferromagneticSkyrmion2020}, this long-range order forms spontaneously even in the absence of magnetic fields.
Another difference from the three-sublattice AFM-SkXs is the presence of singular defects, as marked by the red circle in Fig.~\ref{fig1}(b).
While the three-sublattice AFM-SkXs can be described as a superposition of three ferromagnetic SkXs~\cite{rosalesThreesublatticeSkyrmionCrystal2015}, the singular defects prohibit decomposition into ferromagnetic SkXs.
%They also carry a large fractional skyrmion number~($\approx-\frac{1}{3}$), resulting in a nontrivial skyrmion number per sublattice with $Q_{\textrm{sub},p}=-1$ for $p=1,\ldots, 4$. We note the similarity with skyrmions in ferroelectric materials, where the integer-quantized skyrmion number is broken into equal fractions at the domain wall junctions~\cite{nahasDiscoveryStableSkyrmionic2015}.

\begin{figure}[tbp]
	\centering
	\includegraphics[width=0.45\textwidth]{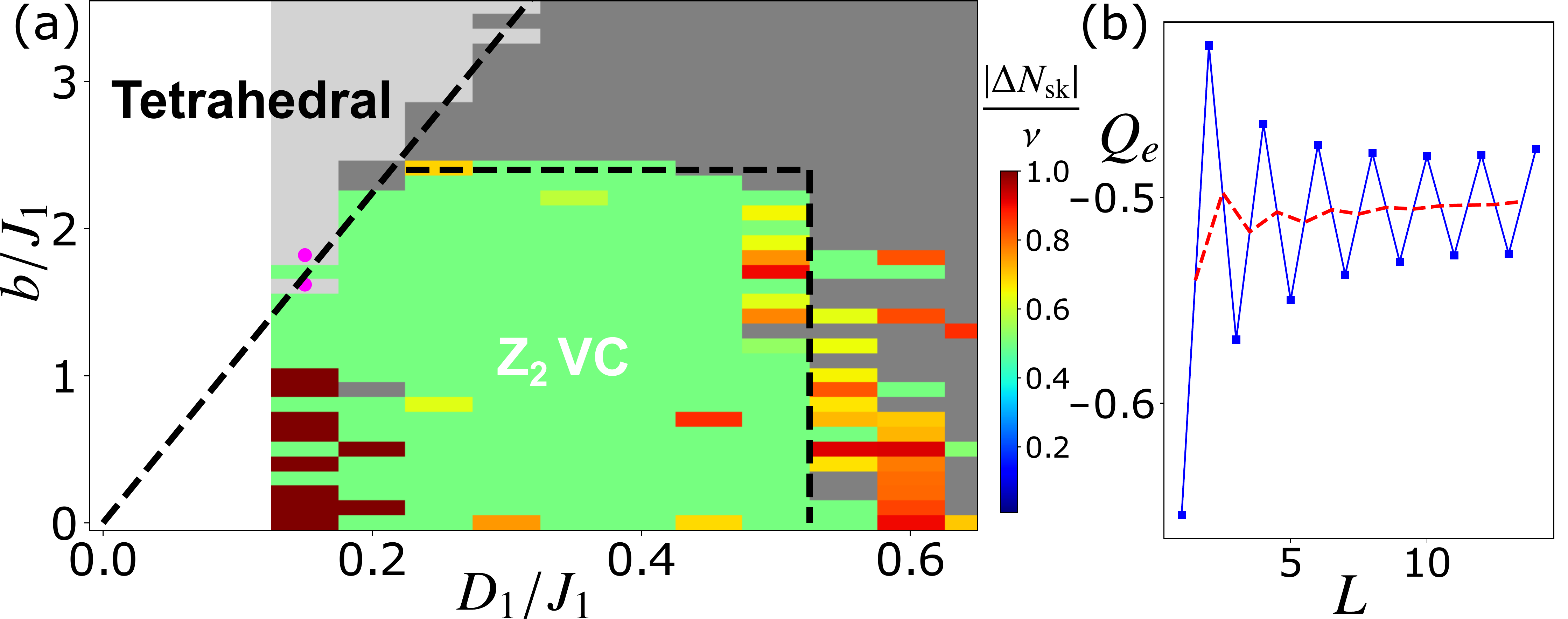}
	\caption{(a) The magnetic phase diagram for the model described in Eq.~\eqref{eq:model}, showing the stability region for the $\mathbb{Z}_2$ VC phase as a function of the relative strength of interfacial DM interactions ($D_1/J_1$) and external magnetic fields ($b/J_1$). This phase diagram is obtained by Monte Carlo simulations on a $120\times 120$ triangular lattice of spins. The color scale indicates the ratio between the total number of $\mathbb{Z}_2$ vortices~($\nu$) and the difference in the total skyrmion number compared to the uniform tetrahedral state ($\Delta N_\textrm{sk}=N_\textrm{sk}-N_\textrm{sk}^\textrm{tetra}$).
    Black dashed lines are included as visual guides. The tetrahedral phase denotes the uniform tetrahedral state, where $\nu=0$ (light gray region). The helical state with $\nu=0$ is also obtained as a metastable configuration~(magenta circles). The gray region indicates $|\Delta N_\textrm{sk}|/\nu>1.0$. The parameters are set as $B_1/J_1=1$ and $\lambda=0.5$. (b) The half-electric charge bound to a $\mathbb{Z}_2$ vortex in the Kondo lattice model, computed for a hexagon of length $L$. The red dashed line plots the average of two consecutive points. The magnetic unit cell of the $\mathbb{Z}_2$ VC in Fig.~\ref{fig1} is used for the calculation of the Kondo lattice model.
    }
	\label{fig3}
\end{figure}

To reveal the $\mathbb{Z}_2$ topology of the tetrahedral AFM order, a rigid regular tetrahedron is assigned to each four-sublattice plaquette, with its orientation described by the $\mathrm{SO}(3)$ group.
The first homotopy group for this $\mathrm{SO}(3)$ order is $\pi_1(\mathrm{SO}(3))=\mathbb{Z}_2$, with the corresponding topological invariant known as $\mathbb{Z}_2$ vorticity~\cite{Kawamura1984}. The calculation of the $\mathbb{Z}_2$ vorticity follows the framework outlined in Ref.~\cite{Kawamura1984} (see Sec.~IIC in SM\cite{SM}). 
Firstly,  to parameterize the $\mathrm{SO}(3)$ order for the $i$th tetrahedron, we define two orthogonal unit vectors: $\hat{\vec{a}}_{i}=\hat{\vec{S}}_{i1}$ and $\hat{\vec{b}}_{i}=\frac{\hat{\vec{S}}_{i1}\times (\hat{\vec{S}}_{i2}-\hat{\vec{S}}_{i3})}{|\,\hat{\vec{S}}_{i1}\times (\hat{\vec{S}}_{i2}-\hat{\vec{S}}_{i3})|}$~\cite{Barts2021}, where $\hat{\vec{S}}_{ip}$ denotes the spin in the $p$-th sublattice ($p=1,2,3,4$) of the $i$th tetrahedron.
The $\mathrm{SO}(3)$ rotation matrix is then constructed based on the rotations of $\hat{\vec{a}}$ and $\hat{\vec{b}}$ from the $i$th tetrahedron to the $j$th tetrahedron, characterized by the axis of rotation $\vec{\hat{n}}_{ij}$ and the angle $\omega_{ij}$. Finally, a link variable $U_{ij}$ is expressed as an $\mathrm{SU}(2)$ matrix,
\begin{equation}
    U_{ij}=\exp \left(\frac{\omega_{ij}}{2\mathrm{i}}\hat{\vec{n}}_{ij}\cdot\vec{\sigma}\right),
\end{equation}
where $-\pi<\omega_{ij}\le \pi$ and $\vec{\sigma}$ denotes a vector of Pauli matrices. We note that this constraint on $\omega_{ij}$ ensures a unique $\mathrm{SU}(2)$ representation of $\mathrm{SO}(3)$ rotations.
The $\mathbb{Z}_2$ vorticity is defined as~\cite{Kawamura1984}
\begin{equation}
    \nu_\mathcal{C}=\frac{1}{2}\textrm{Tr }\left[\prod_{(i,j) \in \mathcal{C}} \,U_{ij}\right], 
\end{equation}
where the product of link variables is computed over a closed loop $\mathcal{C}$. If a $\mathbb{Z}_2$ vortex exists inside the loop, $\nu_\mathcal{C}=-1$. Otherwise, $\nu_\mathcal{C}=1$.

Computing the $\mathbb{Z}_2$ vorticity, we identify four vortices per magnetic unit cell, which are marked in Fig.~\ref{fig1}(c).
Therefore, we conclude that a $\mathbb{Z}_2$ VC is realized.
To illustrate $\mathbb{Z}_2$ vortices, Fig.~\ref{fig1}(c) shows tetrahedra formed by the four-sublattice spins, with colors indicating the tetrahedron faces. 
Observing the colors and orientations of triangles, we find that the tetrahedral AFM order rotates by $2\pi$ around each vortex.
The tetrahedra are strongly distorted at $\mathbb{Z}_2$ vortex cores due to the mismatch between the local sixfold rotational symmetry and the four-sublattice tetrahedral order.
The $2\pi$ rotation of the tetrahedral AFM order is also observed in spin textures near the $\mathbb{Z}_2$ vortex, as shown in Fig.~\ref{fig1}(a).

We also study the stability of $\mathbb{Z}_2$ VCs against DM interactions and magnetic fields. 
Figure~\ref{fig3} presents the magnetic phase diagram obtained from Monte Carlo simulations of a $120\times 120$ spin lattice with periodic boundary conditions.
The color scale represents the skyrmion number associated with a $\mathbb{Z}_2$ vortex, which is estimated from the difference in the total skyrmion number compared to the uniform tetrahedral phase~($\Delta N_\textrm{sk}=N_\textrm{sk}-N_\textrm{sk}^\textrm{tetra}$) and the total number of $\mathbb{Z}_2$ vortices~($\nu$).
The $\mathbb{Z}_2$ VC phase is characterized by $\frac{|\Delta N_\textrm{sk}|}{\nu}=\frac{1}{2}$, leading to the emergent magnetic field with a half magnetic flux quantum per vortex~\cite{nagaosaTopologicalPropertiesDynamics2013}.
We note that the finite size effect results in $\frac{|\Delta N_\textrm{sk}|}{\nu}>\frac{1}{2}$ for $D_1/J_1\le0.2$ due to the large periodicity of the $\mathbb{Z}_2$ VC phase, which is approximately proportional to the inverse of $D_1/J_1$.
For small $D_1/J_1$, the $\mathbb{Z}_2$ VC phase remains robust up to a critical field. 
Beyond this critical magnetic field, a uniform four-sublattice structure emerges~(see Sec.~IID in SM\cite{SM}).
The critical magnetic fields are found to be proportional to $D_1/J_1$, as indicated by the left black dashed line in Fig.~\ref{fig3}.
We extrapolate the phase boundary between the $\mathbb{Z}_2$ VC and the uniform tetrahedral phase down to $b=D_1=0$, where the spin-wave calculation indicates the instability at $b=0$ and $D_1\ne0$.
For $D_1/J_1>0.3$ and $b/J_1>2.4$, the total skyrmion number deviates significantly from the uniform tetrahedral state, indicating that the tetrahedral description breaks down at large $D_1$ and $b$, although $\mathbb{Z}_2$ vortices are still obtained. The DM interactions are no longer perturbations for $D_1/J_1\ge 0.5$, introducing strong frustration and multiple metastable configurations.
We also obtain the helical state as a metastable configuration at the phase boundary between the $\mathbb{Z}_2$ VC and the uniform tetrahedral phase, as indicated by magenta circles in Fig.~\ref{fig3}~(see Sec.~IIE in SM\cite{SM}).

\paragraph{Fractional electric charge:}
A $\mathbb{Z}_2$ vortex of the tetrahedral order may confine a half-integer electric charge when local spins are coupled to conduction electrons~\cite{Rahmani2013}.
In the strong coupling limit, the low-energy model of the Kondo lattice model is given as~(see Sec. IIF in SM\cite{SM}) 
\begin{equation}
    H_\textrm{el}=-\sum_{\braket{\vec{r},\vec{r}'}}t\,\Big(\cos\frac{\theta_{\vec{r}}}{2}\cos\frac{\theta_{\vec{r}'}}{2}+\sin\frac{\theta_{\vec{r}}}{2}\sin\frac{\theta_{\vec{r}'}}{2}e^{-\textrm{i}(\phi_{\vec{r}}-\phi_{\vec{r}'})}\Big)d^\dagger_{\vec{r}}d^{\,}_{\vec{r}'},
\end{equation}
where $d^{\,}_{\vec{r}}$ is the annihilation operator of electrons with spins parallel to the the local spin moment of the $\mathbb{Z}_2$ VC, denoted as $\vec{S}_{\vec{r}}=(\sin \theta_{\vec{r}}\cos\phi_{\vec{r}},\sin \theta_{\vec{r}}\sin\phi_{\vec{r}},\cos\theta_{\vec{r}})$. The charge density is calculated as $\rho_{e,i}=\int_\textrm{BZ} \frac{d\vec{k}}{\Omega_\textrm{BZ}}\sum_{n\in\textrm{occ}}(\phi_{\vec{k},n,i}^*\phi_{\vec{k},n,i}-\frac{1}{2})$ in the half-filling condition, with $\Omega_\textrm{BZ}$ and $\phi_{\vec{k},n,i}$ representing the area of the 1st Brillouin zone and the $n$th energy eigenstate at $i$th sublattice, respectively. Here, the uniform background is subtracted. The bound charge of a $\mathbb{Z}_2$ vortex is estimated as $Q_e(L)=\sum_{i\in \textrm{Hex}(L)}\rho_{e,i}$, with $\textrm{Hex}(L)$ denoting a hexagon of size $L$ centered at the vortex core. Figure~\ref{fig3}(b) shows the bound charge as a function of $L$, gradually converging to $Q_e=\frac{1}{2}$.

\begin{figure}[tbp]
	\centering
	\includegraphics[width=0.5\textwidth]{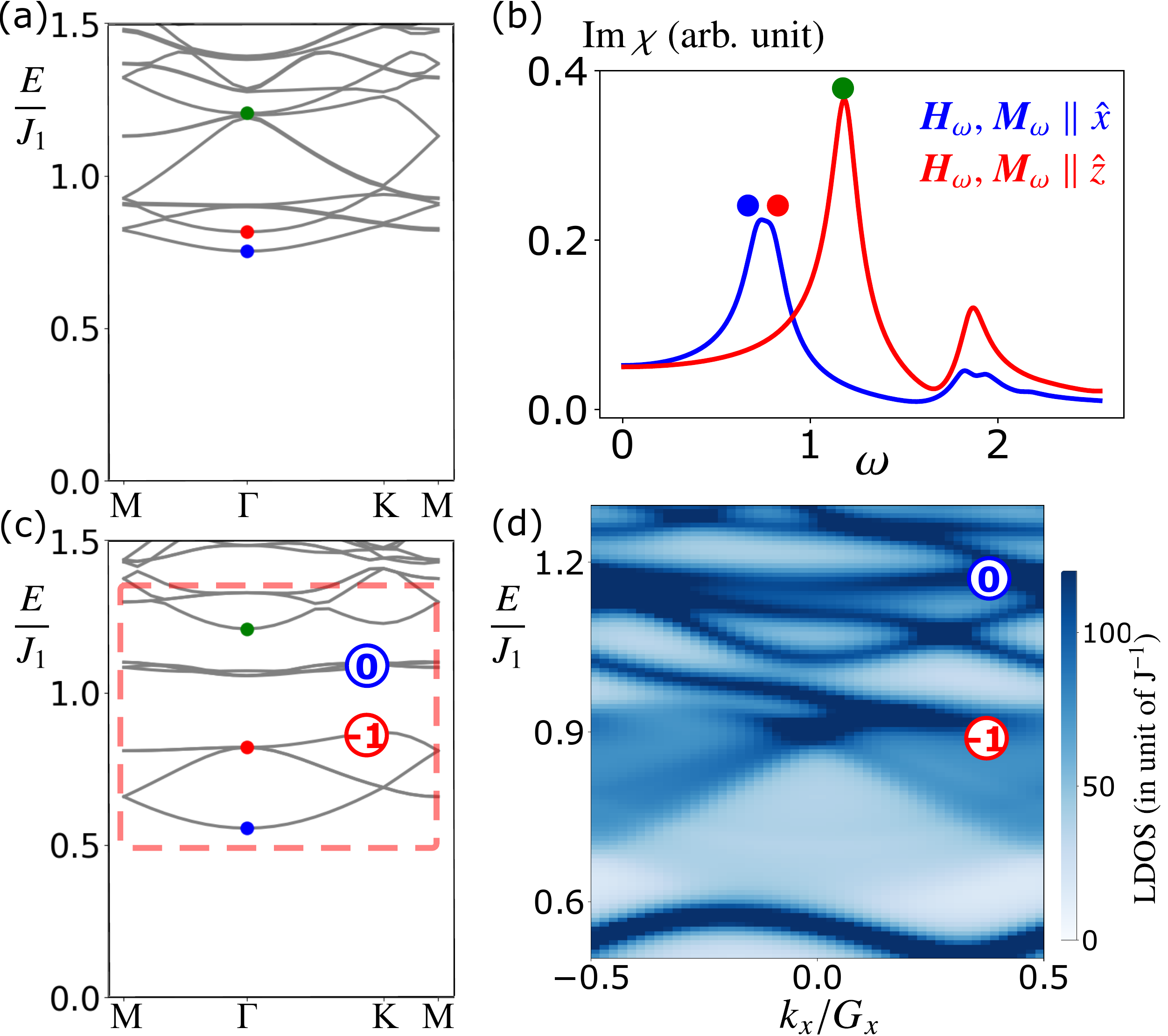}
	\caption{Magnetic activity and topological magnons in the $\mathbb{Z}_2$ VC phase.
    (a) Magnon band structure and (b) imaginary part of the dynamical susceptibility $\textrm{Im}\,\chi(\omega)$ at $b/J_1=0$. The blue (red) line shows the response of in-plane (out-of-plane) magnetization to in-plane (out-of-plane) fields.
    (c) Magnon band structure and (d) local density of states (LDOS) at the edge of a semi-infinite lattice at $b/J_1=0.5$. The LDOS is computed for regions enclosed by red dashed lines in (c). 
    Encircled numbers indicate Chern numbers of magnon bulk bands. 
    In (a)-(c), filled circles indicate magnetically active magnon modes at the $\Gamma$ point, corresponding to CCW~(blue), CW~(red), and breathing+polygon deformation mode~(green) from the lowest band upward. Parameters are consistent with those of Fig.~\ref{fig1} except for the applied magnetic field.
    }
	\label{fig4}
\end{figure}

\paragraph{Topological magnons:}
The magnon band structure of the $\mathbb{Z}_2$ VC without magnetic fields is shown in Fig.~\ref{fig4}(a), where magnetically active excitations are highlighted. 
These excitations are coupled with spatially uniform magnetic fields, and thus identified through the dynamical magnetization of magnons~\cite{hirosawaMagnetoelectricCavityMagnonics2022} (see Sec~IIG in SM\cite{SM}).
The lowest-energy magnon modes are the clockwise~(CW) and counter-clockwise~(CCW) rotation modes, as shown in SM Videos 1 and 2~\cite{SM}, both supporting large in-plane dynamical magnetization. 
These low-energy modes can be decomposed into the in-phase gyrotropic modes within each sublattice.
The CCW mode~(blue) has lower energy than the CW mode~(red) with the spin configurations of Fig.~\ref{fig1}. 
The breathing mode~(green) has higher energy than both CW and CCW modes in contrast to the behavior observed in ferromagnetic SkXs~\cite{mochizukiSpinWaveModesTheir2012}, characterized by twisting deformations that cause expansion and contraction~(SM video 3 \cite{SM}). 
Analysis of the magnon wave functions suggests this mode is a hybridization between the breathing mode and a sixth-order polygon deformation mode~\cite{schutteMagnonskyrmionScatteringChiral2014}.

We also conduct Landau-Lifshitz-Gilbert~(LLG) simulations to investigate magnetically active excitations. 
Figure~\ref{fig4}(b) presents the dynamical susceptibility,~$\textrm{Im}\chi_{xx}$~(blue) and $\textrm{Im}\chi_{zz}$~(red), where $\textrm{Im}\chi_{ab}(\omega)=M_a(\omega)/B_b(\omega)$ with $\vec{M}(\omega)$ and $\vec{B}(\omega)$ denoting magnetization and applied magnetic fields at frequency~$\omega$, respectively. 
The lowest peak in $\textrm{Im}\chi_{xx}$ corresponds to the CCW and CW modes, consistent with the spin wave analysis. 
Since their energy difference is small, both modes are excited simultaneously.
We also identify a peak for the breathing mode in $\textrm{Im}\chi_{zz}$ at $\omega\approx 1.3$ in agreement with the linear spin wave calculation. 

When a magnetic field is applied, the lowest four bands are separated by a gap from the higher-energy bands.
Crucially, they carry a nontrivial Chern number $C=-1$~\cite{fukuiChernNumbersDiscretized2005, shindouTopologicalChiralMagnonic2013}~(see Sec.~IIH in SM\cite{SM}).
Thus, for open boundary conditions, a magnonic chiral edge state arises inside the gap due to the bulk-boundary correspondence~\cite{hatsugaiChernNumberEdge1993, hatsugaiEdgeStatesInteger1993}. 
We employ the renormalization method to compute the local density of states~(LDOS) for a semi-infinite lattice~\cite{HENK199369, Mook2014edge}. The boundary spin textures are computed through Monte Carlo annealing with a specialized boundary condition~\cite{sunMonteCarloCalculations1996}.
Figure~\ref{fig4}(d) shows the LDOS at an edge of a semi-infinite $\mathbb{Z}_2$ VC, clearly showing the magnonic chiral edge state above the lowest magnon bands.

The tetrahedral AFM order has been recently reported in various materials, including Mn monolayers and Pd(Rh)/Mn bilayers on the Re(0001) surface~\cite{Spethmann2020, nickelCouplingTripleState2023}, intercalated van der Waals materials such as Co(Nb,Ta)$_3$S$_6$~\cite{Takagi2023,Park2023}, the van der Waals semiconductor GdGaI~\cite{okumaEmergentTopologicalMagnetism2024}, and has been predicted in CrSi$_2$P$_4$ monolayers~\cite{jiangAnomalousValleyHall2025}.
Our theory can be readily tested in these materials, as interfacial DM interactions can be engineered using heavy metal layers~\cite{fertSkyrmionsTrack2013}. 
The spin textures of $\mathbb{Z}_2$ VCs may be identified through periodic patterns of domain structures detected by spin-polarized scanning tunneling microscopy~\cite{Spethmann2020}.
Additionally, since the uniform tetrahedral phase does not support magnetically active modes at zero magnetic fields, the formation of $\mathbb{Z}_2$ VCs could be confirmed using ferromagnetic resonance techniques~\cite{onoseObservationMagneticExcitations2012}.  %\alex{Citation?} 
Furthermore, the magnonic chiral edge state could be measured with NV center magnetometry~\cite{duControlLocalMeasurement2017a, purserSpinwaveDetectionNitrogenvacancy2020}, or near-field Brillouin light scattering~\cite{jerschMappingLocalizedSpinwave2010}.

\paragraph{Conclusion}
We have discovered the formation of $\mathbb{Z}_2$ VCs within the noncoplanar four-sublattice tetrahedral AFM order, consisting of four interpenetrating triple-$q$ orders {characterized by the nontrivial skyrmion number in each respective sublattice.
In contrast to the previously reported three-sublattice AFM-SkXs, the $\mathbb{Z}_2$ VCs form spontaneously without the need for external magnetic fields. 
Using linear spin wave theory, we identified magnetically active excitations, including clockwise rotation and breathing modes.
When external magnetic fields are applied, a topological magnon gap emerges, enabling the formation of magnonic chiral edge states. 
Our findings reveal the robust formation of $\mathbb{Z}_2$ vortices with a half-integer electric bound charge relative to the uniform background, opening a path to novel topological phenomena and fractional excitations in triangular lattice antiferromagnets~\cite{Rahmani2013}.

\section{Acknowledgements}
We thank Yutaka Akagi, Oleksandr Pylypovskyi and Karin Everschor-Sitte for helpful discussions. 
This work was supported by the German Research Foundation (DFG) Project No.~320163632 (Emmy Noether Programme); Project No.~504261060 (Emmy Noether Programme); and Project No.~403233384 (SPP2137 Skyrmionics). T.~H. is supported by JSPS KAKENHI Grant Number JP23K13064 and Aoyama Gakuin University Research Institute “Early Eagle” grant program for promotion of research by early career researchers. M.~A.\ also acknowledges support from the UDE Postdoc Seed Funding.

\section{Data availability}
Data and computational codes are available upon reasonable request on Zenodo~\cite{hirosawa_2025_14978487}.


\begin{thebibliography}{68}%
\makeatletter
\providecommand \@ifxundefined [1]{%
 \@ifx{#1\undefined}
}%
\providecommand \@ifnum [1]{%
 \ifnum #1\expandafter \@firstoftwo
 \else \expandafter \@secondoftwo
 \fi
}%
\providecommand \@ifx [1]{%
 \ifx #1\expandafter \@firstoftwo
 \else \expandafter \@secondoftwo
 \fi
}%
\providecommand \natexlab [1]{#1}%
\providecommand \enquote  [1]{``#1''}%
\providecommand \bibnamefont  [1]{#1}%
\providecommand \bibfnamefont [1]{#1}%
\providecommand \citenamefont [1]{#1}%
\providecommand \href@noop [0]{\@secondoftwo}%
\providecommand \href [0]{\begingroup \@sanitize@url \@href}%
\providecommand \@href[1]{\@@startlink{#1}\@@href}%
\providecommand \@@href[1]{\endgroup#1\@@endlink}%
\providecommand \@sanitize@url [0]{\catcode `\\12\catcode `\$12\catcode
  `\&12\catcode `\#12\catcode `\^12\catcode `\_12\catcode `\%12\relax}%
\providecommand \@@startlink[1]{}%
\providecommand \@@endlink[0]{}%
\providecommand \url  [0]{\begingroup\@sanitize@url \@url }%
\providecommand \@url [1]{\endgroup\@href {#1}{\urlprefix }}%
\providecommand \urlprefix  [0]{URL }%
\providecommand \Eprint [0]{\href }%
\renewcommand{\doibase}[1]{https://dx.doi.org/\ifdefempty{#1}{}{#1}}%
\providecommand \selectlanguage [0]{\@gobble}%
\providecommand \bibinfo  [0]{\@secondoftwo}%
\providecommand \bibfield  [0]{\@secondoftwo}%
\providecommand \translation [1]{[#1]}%
\providecommand \BibitemOpen [0]{}%
\providecommand \bibitemStop [0]{}%
\providecommand \bibitemNoStop [0]{.\EOS\space}%
\providecommand \EOS [0]{\spacefactor3000\relax}%
\providecommand \BibitemShut  [1]{\csname bibitem#1\endcsname}%
\let\auto@bib@innerbib\@empty
%</preamble>
\bibitem [{\citenamefont {Zang}\ \emph {et~al.}(2018)\citenamefont {Zang},
  \citenamefont {Cros},\ and\ \citenamefont {Hoffmann}}]{zang2018topology}%
  \BibitemOpen
  \bibfield  {author} {\bibinfo {author} {\bibfnamefont {J.}~\bibnamefont
  {Zang}}, \bibinfo {author} {\bibfnamefont {V.}~\bibnamefont {Cros}}, \ and\
  \bibinfo {author} {\bibfnamefont {A.F.}\ \bibnamefont {Hoffmann}},\ }\href
  {https://books.google.co.jp/books?id=ThPOwQEACAAJ} {\emph {\bibinfo {title}
  {Topology in magnetism}}},\ Springer series in solid-state sciences\
  (\bibinfo  {publisher} {Springer},\ \bibinfo {year} {2018})\BibitemShut
  {NoStop}%
\bibitem [{\citenamefont {Nagaosa}\ and\ \citenamefont
  {Tokura}(2013)}]{nagaosaTopologicalPropertiesDynamics2013}%
  \BibitemOpen
  \bibfield  {author} {\bibinfo {author} {\bibfnamefont {Naoto}\ \bibnamefont
  {Nagaosa}}\ and\ \bibinfo {author} {\bibfnamefont {Yoshinori}\ \bibnamefont
  {Tokura}},\ }\bibfield  {title} {\enquote {\bibinfo {title} {Topological
  properties and dynamics of magnetic skyrmions},}\ }\href {\doibase
  10.1038/nnano.2013.243} {\bibfield  {journal} {\bibinfo  {journal} {Nat.
  Nanotechnol.}\ }\textbf {\bibinfo {volume} {8}},\ \bibinfo {pages} {899}
  (\bibinfo {year} {2013})}\BibitemShut {NoStop}%
\bibitem [{\citenamefont {Everschor-Sitte}\ \emph {et~al.}(2018)\citenamefont
  {Everschor-Sitte}, \citenamefont {Masell}, \citenamefont {Reeve},\ and\
  \citenamefont {Kl\"{a}ui}}]{everschorPerspectiveMagneticSkyrmions2018}%
  \BibitemOpen
  \bibfield  {author} {\bibinfo {author} {\bibfnamefont {K.}~\bibnamefont
  {Everschor-Sitte}}, \bibinfo {author} {\bibfnamefont {J.}~\bibnamefont
  {Masell}}, \bibinfo {author} {\bibfnamefont {R.~M.}\ \bibnamefont {Reeve}}, \
  and\ \bibinfo {author} {\bibfnamefont {M.}~\bibnamefont {Kl\"{a}ui}},\
  }\bibfield  {title} {\enquote {\bibinfo {title} {Perspective: {Magnetic}
  skyrmions—{Overview} of recent progress in an active research field},}\
  }\href {\doibase 10.1063/1.5048972} {\bibfield  {journal} {\bibinfo
  {journal} {J. Appl. Phys.}\ }\textbf {\bibinfo {volume} {124}},\ \bibinfo
  {pages} {240901} (\bibinfo {year} {2018})}\BibitemShut {NoStop}%
\bibitem [{\citenamefont {Bogdanov}\ and\ \citenamefont
  {Hubert}(1994)}]{bogdanovThermodynamicallyStableMagnetic1994}%
  \BibitemOpen
  \bibfield  {author} {\bibinfo {author} {\bibfnamefont {A.}~\bibnamefont
  {Bogdanov}}\ and\ \bibinfo {author} {\bibfnamefont {A.}~\bibnamefont
  {Hubert}},\ }\bibfield  {title} {\enquote {\bibinfo {title}
  {Thermodynamically stable magnetic vortex states in magnetic crystals},}\
  }\href {\doibase 10.1016/0304-8853(94)90046-9} {\bibfield  {journal}
  {\bibinfo  {journal} {J. Magn. Magn. Mater.}\ }\textbf {\bibinfo {volume}
  {138}},\ \bibinfo {pages} {255} (\bibinfo {year} {1994})}\BibitemShut
  {NoStop}%
\bibitem [{\citenamefont {R\"{o}{\ss}ler}\ \emph {et~al.}(2006)\citenamefont
  {R\"{o}{\ss}ler}, \citenamefont {Bogdanov},\ and\ \citenamefont
  {Pfleiderer}}]{rosslerSpontaneousSkyrmionGround2006}%
  \BibitemOpen
  \bibfield  {author} {\bibinfo {author} {\bibfnamefont {U.~K.}\ \bibnamefont
  {R\"{o}{\ss}ler}}, \bibinfo {author} {\bibfnamefont {A.~N.}\ \bibnamefont
  {Bogdanov}}, \ and\ \bibinfo {author} {\bibfnamefont {C.}~\bibnamefont
  {Pfleiderer}},\ }\bibfield  {title} {\enquote {\bibinfo {title} {Spontaneous
  skyrmion ground states in magnetic metals},}\ }\href {\doibase
  10.1038/nature05056} {\bibfield  {journal} {\bibinfo  {journal} {Nature}\
  }\textbf {\bibinfo {volume} {442}},\ \bibinfo {pages} {797} (\bibinfo {year}
  {2006})}\BibitemShut {NoStop}%
\bibitem [{\citenamefont {M\"{u}hlbauer}\ \emph {et~al.}(2009)\citenamefont
  {M\"{u}hlbauer}, \citenamefont {Binz}, \citenamefont {Jonietz}, \citenamefont
  {Pfleiderer}, \citenamefont {Rosch}, \citenamefont {Neubauer}, \citenamefont
  {Georgii},\ and\ \citenamefont {Böni}}]{muhlbauerSkyrmionLatticeChiral2009}%
  \BibitemOpen
  \bibfield  {author} {\bibinfo {author} {\bibfnamefont {S.}~\bibnamefont
  {M\"{u}hlbauer}}, \bibinfo {author} {\bibfnamefont {B.}~\bibnamefont {Binz}},
  \bibinfo {author} {\bibfnamefont {F.}~\bibnamefont {Jonietz}}, \bibinfo
  {author} {\bibfnamefont {C.}~\bibnamefont {Pfleiderer}}, \bibinfo {author}
  {\bibfnamefont {A.}~\bibnamefont {Rosch}}, \bibinfo {author} {\bibfnamefont
  {A.}~\bibnamefont {Neubauer}}, \bibinfo {author} {\bibfnamefont
  {R.}~\bibnamefont {Georgii}}, \ and\ \bibinfo {author} {\bibfnamefont
  {P.}~\bibnamefont {Böni}},\ }\bibfield  {title} {\enquote {\bibinfo {title}
  {Skyrmion {Lattice} in a {Chiral} {Magnet}},}\ }\href {\doibase
  10.1126/science.1166767} {\bibfield  {journal} {\bibinfo  {journal}
  {Science}\ }\textbf {\bibinfo {volume} {323}},\ \bibinfo {pages} {915}
  (\bibinfo {year} {2009})}\BibitemShut {NoStop}%
\bibitem [{\citenamefont {Yu}\ \emph {et~al.}(2010)\citenamefont {Yu},
  \citenamefont {Onose}, \citenamefont {Kanazawa}, \citenamefont {Park},
  \citenamefont {Han}, \citenamefont {Matsui}, \citenamefont {Nagaosa},\ and\
  \citenamefont {Tokura}}]{yuRealspaceObservationTwodimensional2010}%
  \BibitemOpen
  \bibfield  {author} {\bibinfo {author} {\bibfnamefont {X.~Z.}\ \bibnamefont
  {Yu}}, \bibinfo {author} {\bibfnamefont {Y.}~\bibnamefont {Onose}}, \bibinfo
  {author} {\bibfnamefont {N.}~\bibnamefont {Kanazawa}}, \bibinfo {author}
  {\bibfnamefont {J.~H.}\ \bibnamefont {Park}}, \bibinfo {author}
  {\bibfnamefont {J.~H.}\ \bibnamefont {Han}}, \bibinfo {author} {\bibfnamefont
  {Y.}~\bibnamefont {Matsui}}, \bibinfo {author} {\bibfnamefont
  {N.}~\bibnamefont {Nagaosa}}, \ and\ \bibinfo {author} {\bibfnamefont
  {Y.}~\bibnamefont {Tokura}},\ }\bibfield  {title} {\enquote {\bibinfo {title}
  {Real-space observation of a two-dimensional skyrmion crystal},}\ }\href
  {\doibase 10.1038/nature09124} {\bibfield  {journal} {\bibinfo  {journal}
  {Nature}\ }\textbf {\bibinfo {volume} {465}},\ \bibinfo {pages} {901}
  (\bibinfo {year} {2010})}\BibitemShut {NoStop}%
\bibitem [{\citenamefont {Okubo}\ \emph {et~al.}(2012)\citenamefont {Okubo},
  \citenamefont {Chung},\ and\ \citenamefont {Kawamura}}]{Okubo2012}%
  \BibitemOpen
  \bibfield  {author} {\bibinfo {author} {\bibfnamefont {Tsuyoshi}\
  \bibnamefont {Okubo}}, \bibinfo {author} {\bibfnamefont {Sungki}\
  \bibnamefont {Chung}}, \ and\ \bibinfo {author} {\bibfnamefont {Hikaru}\
  \bibnamefont {Kawamura}},\ }\bibfield  {title} {\enquote {\bibinfo {title}
  {Multiple-$q$ states and the skyrmion lattice of the triangular-lattice
  heisenberg antiferromagnet under magnetic fields},}\ }\href {\doibase
  10.1103/PhysRevLett.108.017206} {\bibfield  {journal} {\bibinfo  {journal}
  {Phys. Rev. Lett.}\ }\textbf {\bibinfo {volume} {108}},\ \bibinfo {pages}
  {017206} (\bibinfo {year} {2012})}\BibitemShut {NoStop}%
\bibitem [{\citenamefont {Leonov}\ and\ \citenamefont
  {Mostovoy}(2015)}]{Leonov2015}%
  \BibitemOpen
  \bibfield  {author} {\bibinfo {author} {\bibfnamefont {A.~O.}\ \bibnamefont
  {Leonov}}\ and\ \bibinfo {author} {\bibfnamefont {M.}~\bibnamefont
  {Mostovoy}},\ }\bibfield  {title} {\enquote {\bibinfo {title} {Multiply
  periodic states and isolated skyrmions in an anisotropic frustrated
  magnet},}\ }\href {\doibase 10.1038/ncomms9275} {\bibfield  {journal}
  {\bibinfo  {journal} {Nat. Commun.}\ }\textbf {\bibinfo {volume} {6}},\
  \bibinfo {pages} {8275} (\bibinfo {year} {2015})}\BibitemShut {NoStop}%
\bibitem [{\citenamefont {Hayami}\ \emph {et~al.}(2017)\citenamefont {Hayami},
  \citenamefont {Ozawa},\ and\ \citenamefont
  {Motome}}]{hayamiEffectiveBilinearbiquadraticModel2017a}%
  \BibitemOpen
  \bibfield  {author} {\bibinfo {author} {\bibfnamefont {Satoru}\ \bibnamefont
  {Hayami}}, \bibinfo {author} {\bibfnamefont {Ryo}\ \bibnamefont {Ozawa}}, \
  and\ \bibinfo {author} {\bibfnamefont {Yukitoshi}\ \bibnamefont {Motome}},\
  }\bibfield  {title} {\enquote {\bibinfo {title} {Effective
  bilinear-biquadratic model for noncoplanar ordering in itinerant magnets},}\
  }\href {\doibase 10.1103/PhysRevB.95.224424} {\bibfield  {journal} {\bibinfo
  {journal} {Phys. Rev. B}\ }\textbf {\bibinfo {volume} {95}},\ \bibinfo
  {pages} {224424} (\bibinfo {year} {2017})}\BibitemShut {NoStop}%
\bibitem [{\citenamefont {Kurumaji}\ \emph {et~al.}(2019)\citenamefont
  {Kurumaji}, \citenamefont {Nakajima}, \citenamefont {Hirschberger},
  \citenamefont {Kikkawa}, \citenamefont {Yamasaki}, \citenamefont {Sagayama},
  \citenamefont {Nakao}, \citenamefont {Taguchi}, \citenamefont {Arima},\ and\
  \citenamefont {Tokura}}]{kurumajiSkyrmionLatticeGiant2019}%
  \BibitemOpen
  \bibfield  {author} {\bibinfo {author} {\bibfnamefont {Takashi}\ \bibnamefont
  {Kurumaji}}, \bibinfo {author} {\bibfnamefont {Taro}\ \bibnamefont
  {Nakajima}}, \bibinfo {author} {\bibfnamefont {Max}\ \bibnamefont
  {Hirschberger}}, \bibinfo {author} {\bibfnamefont {Akiko}\ \bibnamefont
  {Kikkawa}}, \bibinfo {author} {\bibfnamefont {Yuichi}\ \bibnamefont
  {Yamasaki}}, \bibinfo {author} {\bibfnamefont {Hajime}\ \bibnamefont
  {Sagayama}}, \bibinfo {author} {\bibfnamefont {Hironori}\ \bibnamefont
  {Nakao}}, \bibinfo {author} {\bibfnamefont {Yasujiro}\ \bibnamefont
  {Taguchi}}, \bibinfo {author} {\bibfnamefont {Taka-hisa}\ \bibnamefont
  {Arima}}, \ and\ \bibinfo {author} {\bibfnamefont {Yoshinori}\ \bibnamefont
  {Tokura}},\ }\bibfield  {title} {\enquote {\bibinfo {title} {Skyrmion lattice
  with a giant topological {Hall} effect in a frustrated triangular-lattice
  magnet},}\ }\href {\doibase 10.1126/science.aau0968} {\bibfield  {journal}
  {\bibinfo  {journal} {Science}\ }\textbf {\bibinfo {volume} {365}},\ \bibinfo
  {pages} {914} (\bibinfo {year} {2019})}\BibitemShut {NoStop}%
\bibitem [{\citenamefont {Ezawa}(2010)}]{ezawaGiantSkyrmionsStabilized2010}%
  \BibitemOpen
  \bibfield  {author} {\bibinfo {author} {\bibfnamefont {Motohiko}\
  \bibnamefont {Ezawa}},\ }\bibfield  {title} {\enquote {\bibinfo {title}
  {Giant {Skyrmions} {Stabilized} by {Dipole}-{Dipole} {Interactions} in {Thin}
  {Ferromagnetic} {Films}},}\ }\href {\doibase 10.1103/PhysRevLett.105.197202}
  {\bibfield  {journal} {\bibinfo  {journal} {Phys. Rev. Lett.}\ }\textbf
  {\bibinfo {volume} {105}},\ \bibinfo {pages} {197202} (\bibinfo {year}
  {2010})}\BibitemShut {NoStop}%
\bibitem [{\citenamefont {Yu}\ \emph {et~al.}(2012)\citenamefont {Yu},
  \citenamefont {Mostovoy}, \citenamefont {Tokunaga}, \citenamefont {Zhang},
  \citenamefont {Kimoto}, \citenamefont {Matsui}, \citenamefont {Kaneko},
  \citenamefont {Nagaosa},\ and\ \citenamefont
  {Tokura}}]{yuMagneticStripesSkyrmions2012}%
  \BibitemOpen
  \bibfield  {author} {\bibinfo {author} {\bibfnamefont {Xiuzhen}\ \bibnamefont
  {Yu}}, \bibinfo {author} {\bibfnamefont {Maxim}\ \bibnamefont {Mostovoy}},
  \bibinfo {author} {\bibfnamefont {Yusuke}\ \bibnamefont {Tokunaga}}, \bibinfo
  {author} {\bibfnamefont {Weizhu}\ \bibnamefont {Zhang}}, \bibinfo {author}
  {\bibfnamefont {Koji}\ \bibnamefont {Kimoto}}, \bibinfo {author}
  {\bibfnamefont {Yoshio}\ \bibnamefont {Matsui}}, \bibinfo {author}
  {\bibfnamefont {Yoshio}\ \bibnamefont {Kaneko}}, \bibinfo {author}
  {\bibfnamefont {Naoto}\ \bibnamefont {Nagaosa}}, \ and\ \bibinfo {author}
  {\bibfnamefont {Yoshinori}\ \bibnamefont {Tokura}},\ }\bibfield  {title}
  {\enquote {\bibinfo {title} {Magnetic stripes and skyrmions with helicity
  reversals},}\ }\href {\doibase 10.1073/pnas.1118496109} {\bibfield  {journal}
  {\bibinfo  {journal} {Proc. Natl. Acad. Sci.}\ }\textbf {\bibinfo {volume}
  {109}},\ \bibinfo {pages} {8856} (\bibinfo {year} {2012})}\BibitemShut
  {NoStop}%
\bibitem [{\citenamefont {Heinze}\ \emph {et~al.}(2011)\citenamefont {Heinze},
  \citenamefont {von Bergmann}, \citenamefont {Menzel}, \citenamefont {Brede},
  \citenamefont {Kubetzka}, \citenamefont {Wiesendanger}, \citenamefont
  {Bihlmayer},\ and\ \citenamefont {Bl{\"u}gel}}]{Heinze2011}%
  \BibitemOpen
  \bibfield  {author} {\bibinfo {author} {\bibfnamefont {Stefan}\ \bibnamefont
  {Heinze}}, \bibinfo {author} {\bibfnamefont {Kirsten}\ \bibnamefont {von
  Bergmann}}, \bibinfo {author} {\bibfnamefont {Matthias}\ \bibnamefont
  {Menzel}}, \bibinfo {author} {\bibfnamefont {Jens}\ \bibnamefont {Brede}},
  \bibinfo {author} {\bibfnamefont {Andr{\'e}}\ \bibnamefont {Kubetzka}},
  \bibinfo {author} {\bibfnamefont {Roland}\ \bibnamefont {Wiesendanger}},
  \bibinfo {author} {\bibfnamefont {Gustav}\ \bibnamefont {Bihlmayer}}, \ and\
  \bibinfo {author} {\bibfnamefont {Stefan}\ \bibnamefont {Bl{\"u}gel}},\
  }\bibfield  {title} {\enquote {\bibinfo {title} {Spontaneous atomic-scale
  magnetic skyrmion lattice in two dimensions},}\ }\href {\doibase
  10.1038/nphys2045} {\bibfield  {journal} {\bibinfo  {journal} {Nat. Phys.}\
  }\textbf {\bibinfo {volume} {7}},\ \bibinfo {pages} {713} (\bibinfo {year}
  {2011})}\BibitemShut {NoStop}%
\bibitem [{\citenamefont {Ohgushi}\ \emph {et~al.}(2000)\citenamefont
  {Ohgushi}, \citenamefont {Murakami},\ and\ \citenamefont
  {Nagaosa}}]{ohgushiSpinAnisotropyQuantum2000}%
  \BibitemOpen
  \bibfield  {author} {\bibinfo {author} {\bibfnamefont {Kenya}\ \bibnamefont
  {Ohgushi}}, \bibinfo {author} {\bibfnamefont {Shuichi}\ \bibnamefont
  {Murakami}}, \ and\ \bibinfo {author} {\bibfnamefont {Naoto}\ \bibnamefont
  {Nagaosa}},\ }\bibfield  {title} {\enquote {\bibinfo {title} {Spin anisotropy
  and quantum {Hall} effect in the kagom\'e lattice: {Chiral} spin state based
  on a ferromagnet},}\ }\href {\doibase 10.1103/PhysRevB.62.R6065} {\bibfield
  {journal} {\bibinfo  {journal} {Phys. Rev. B}\ }\textbf {\bibinfo {volume}
  {62}},\ \bibinfo {pages} {R6065} (\bibinfo {year} {2000})}\BibitemShut
  {NoStop}%
\bibitem [{\citenamefont {Nagaosa}(2006)}]{nagaosaAnomalousHallEffect2006}%
  \BibitemOpen
  \bibfield  {author} {\bibinfo {author} {\bibfnamefont {Naoto}\ \bibnamefont
  {Nagaosa}},\ }\bibfield  {title} {\enquote {\bibinfo {title} {Anomalous
  {Hall} {Effect} –{A} {New} {Perspective}–},}\ }\href {\doibase
  10.1143/JPSJ.75.042001} {\bibfield  {journal} {\bibinfo  {journal} {J. Phys.
  Soc. Jpn.}\ }\textbf {\bibinfo {volume} {75}},\ \bibinfo {pages} {042001}
  (\bibinfo {year} {2006})}\BibitemShut {NoStop}%
\bibitem [{\citenamefont {Martin}\ and\ \citenamefont
  {Batista}(2008)}]{Martin2008}%
  \BibitemOpen
  \bibfield  {author} {\bibinfo {author} {\bibfnamefont {Ivar}\ \bibnamefont
  {Martin}}\ and\ \bibinfo {author} {\bibfnamefont {C.~D.}\ \bibnamefont
  {Batista}},\ }\bibfield  {title} {\enquote {\bibinfo {title} {Itinerant
  electron-driven chiral magnetic ordering and spontaneous quantum hall effect
  in triangular lattice models},}\ }\href {\doibase
  10.1103/PhysRevLett.101.156402} {\bibfield  {journal} {\bibinfo  {journal}
  {Phys. Rev. Lett.}\ }\textbf {\bibinfo {volume} {101}},\ \bibinfo {pages}
  {156402} (\bibinfo {year} {2008})}\BibitemShut {NoStop}%
\bibitem [{\citenamefont {Neubauer}\ \emph {et~al.}(2009)\citenamefont
  {Neubauer}, \citenamefont {Pfleiderer}, \citenamefont {Binz}, \citenamefont
  {Rosch}, \citenamefont {Ritz}, \citenamefont {Niklowitz},\ and\ \citenamefont
  {Böni}}]{neubauerTopologicalHallEffect2009}%
  \BibitemOpen
  \bibfield  {author} {\bibinfo {author} {\bibfnamefont {A.}~\bibnamefont
  {Neubauer}}, \bibinfo {author} {\bibfnamefont {C.}~\bibnamefont
  {Pfleiderer}}, \bibinfo {author} {\bibfnamefont {B.}~\bibnamefont {Binz}},
  \bibinfo {author} {\bibfnamefont {A.}~\bibnamefont {Rosch}}, \bibinfo
  {author} {\bibfnamefont {R.}~\bibnamefont {Ritz}}, \bibinfo {author}
  {\bibfnamefont {P.~G.}\ \bibnamefont {Niklowitz}}, \ and\ \bibinfo {author}
  {\bibfnamefont {P.}~\bibnamefont {Böni}},\ }\bibfield  {title} {\enquote
  {\bibinfo {title} {Topological {Hall} {Effect} in the {$A$} {Phase} of
  {MnSi}},}\ }\href {\doibase 10.1103/PhysRevLett.102.186602} {\bibfield
  {journal} {\bibinfo  {journal} {Phys. Rev. Lett.}\ }\textbf {\bibinfo
  {volume} {102}},\ \bibinfo {pages} {186602} (\bibinfo {year}
  {2009})}\BibitemShut {NoStop}%
\bibitem [{\citenamefont {Jonietz}\ \emph {et~al.}(2010)\citenamefont
  {Jonietz}, \citenamefont {M\"{u}hlbauer}, \citenamefont {Pfleiderer},
  \citenamefont {Neubauer}, \citenamefont {M\"{u}nzer}, \citenamefont {Bauer},
  \citenamefont {Adams}, \citenamefont {Georgii}, \citenamefont {B\"{o}ni},
  \citenamefont {Duine}, \citenamefont {Everschor}, \citenamefont {Garst},\
  and\ \citenamefont {Rosch}}]{jonietzSpinTransferTorques2010}%
  \BibitemOpen
  \bibfield  {author} {\bibinfo {author} {\bibfnamefont {F.}~\bibnamefont
  {Jonietz}}, \bibinfo {author} {\bibfnamefont {S.}~\bibnamefont
  {M\"{u}hlbauer}}, \bibinfo {author} {\bibfnamefont {C.}~\bibnamefont
  {Pfleiderer}}, \bibinfo {author} {\bibfnamefont {A.}~\bibnamefont
  {Neubauer}}, \bibinfo {author} {\bibfnamefont {W.}~\bibnamefont
  {M\"{u}nzer}}, \bibinfo {author} {\bibfnamefont {A.}~\bibnamefont {Bauer}},
  \bibinfo {author} {\bibfnamefont {T.}~\bibnamefont {Adams}}, \bibinfo
  {author} {\bibfnamefont {R.}~\bibnamefont {Georgii}}, \bibinfo {author}
  {\bibfnamefont {P.}~\bibnamefont {B\"{o}ni}}, \bibinfo {author}
  {\bibfnamefont {R.~A.}\ \bibnamefont {Duine}}, \bibinfo {author}
  {\bibfnamefont {K.}~\bibnamefont {Everschor}}, \bibinfo {author}
  {\bibfnamefont {M.}~\bibnamefont {Garst}}, \ and\ \bibinfo {author}
  {\bibfnamefont {A.}~\bibnamefont {Rosch}},\ }\bibfield  {title} {\enquote
  {\bibinfo {title} {Spin {Transfer} {Torques} in {MnSi} at {Ultralow}
  {Current} {Densities}},}\ }\href {\doibase 10.1126/science.1195709}
  {\bibfield  {journal} {\bibinfo  {journal} {Science}\ }\textbf {\bibinfo
  {volume} {330}},\ \bibinfo {pages} {1648} (\bibinfo {year}
  {2010})}\BibitemShut {NoStop}%
\bibitem [{\citenamefont {Schulz}\ \emph {et~al.}(2012)\citenamefont {Schulz},
  \citenamefont {Ritz}, \citenamefont {Bauer}, \citenamefont {Halder},
  \citenamefont {Wagner}, \citenamefont {Franz}, \citenamefont {Pfleiderer},
  \citenamefont {Everschor}, \citenamefont {Garst},\ and\ \citenamefont
  {Rosch}}]{schulzEmergentElectrodynamicsSkyrmions2012}%
  \BibitemOpen
  \bibfield  {author} {\bibinfo {author} {\bibfnamefont {T.}~\bibnamefont
  {Schulz}}, \bibinfo {author} {\bibfnamefont {R.}~\bibnamefont {Ritz}},
  \bibinfo {author} {\bibfnamefont {A.}~\bibnamefont {Bauer}}, \bibinfo
  {author} {\bibfnamefont {M.}~\bibnamefont {Halder}}, \bibinfo {author}
  {\bibfnamefont {M.}~\bibnamefont {Wagner}}, \bibinfo {author} {\bibfnamefont
  {C.}~\bibnamefont {Franz}}, \bibinfo {author} {\bibfnamefont
  {C.}~\bibnamefont {Pfleiderer}}, \bibinfo {author} {\bibfnamefont
  {K.}~\bibnamefont {Everschor}}, \bibinfo {author} {\bibfnamefont
  {M.}~\bibnamefont {Garst}}, \ and\ \bibinfo {author} {\bibfnamefont
  {A.}~\bibnamefont {Rosch}},\ }\bibfield  {title} {\enquote {\bibinfo {title}
  {Emergent electrodynamics of skyrmions in a chiral magnet},}\ }\href
  {\doibase 10.1038/nphys2231} {\bibfield  {journal} {\bibinfo  {journal} {Nat.
  Phys.}\ }\textbf {\bibinfo {volume} {8}},\ \bibinfo {pages} {301} (\bibinfo
  {year} {2012})}\BibitemShut {NoStop}%
\bibitem [{\citenamefont {Rold\`{a}n-Molina}\ \emph {et~al.}(2016)\citenamefont
  {Rold\`{a}n-Molina}, \citenamefont {Nunez},\ and\ \citenamefont
  {Fern\'{a}ndez-Rossier}}]{roldan-molinaTopologicalSpinWaves2016}%
  \BibitemOpen
  \bibfield  {author} {\bibinfo {author} {\bibfnamefont {A.}~\bibnamefont
  {Rold\`{a}n-Molina}}, \bibinfo {author} {\bibfnamefont {A.~S.}\ \bibnamefont
  {Nunez}}, \ and\ \bibinfo {author} {\bibfnamefont {J.}~\bibnamefont
  {Fern\'{a}ndez-Rossier}},\ }\bibfield  {title} {\enquote {\bibinfo {title}
  {Topological spin waves in the atomic-scale magnetic skyrmion crystal},}\
  }\href {\doibase 10.1088/1367-2630/18/4/045015} {\bibfield  {journal}
  {\bibinfo  {journal} {New J. Phys.}\ }\textbf {\bibinfo {volume} {18}},\
  \bibinfo {pages} {045015} (\bibinfo {year} {2016})}\BibitemShut {NoStop}%
\bibitem [{\citenamefont {Garst}\ \emph {et~al.}(2017)\citenamefont {Garst},
  \citenamefont {Waizner},\ and\ \citenamefont
  {Grundler}}]{garstCollectiveSpinExcitations2017}%
  \BibitemOpen
  \bibfield  {author} {\bibinfo {author} {\bibfnamefont {Markus}\ \bibnamefont
  {Garst}}, \bibinfo {author} {\bibfnamefont {Johannes}\ \bibnamefont
  {Waizner}}, \ and\ \bibinfo {author} {\bibfnamefont {Dirk}\ \bibnamefont
  {Grundler}},\ }\bibfield  {title} {\enquote {\bibinfo {title} {Collective
  spin excitations of helices and magnetic skyrmions: review and perspectives
  of magnonics in non-centrosymmetric magnets},}\ }\href {\doibase
  10.1088/1361-6463/aa7573} {\bibfield  {journal} {\bibinfo  {journal} {J.
  Phys. D Appl. Phys.}\ }\textbf {\bibinfo {volume} {50}},\ \bibinfo {pages}
  {293002} (\bibinfo {year} {2017})}\BibitemShut {NoStop}%
\bibitem [{\citenamefont {D\'{i}az}\ \emph {et~al.}(2020)\citenamefont
  {D\'{i}az}, \citenamefont {Hirosawa}, \citenamefont {Klinovaja},\ and\
  \citenamefont {Loss}}]{diazChiralMagnonicEdge2020}%
  \BibitemOpen
  \bibfield  {author} {\bibinfo {author} {\bibfnamefont {Sebasti\'{a}n~A.}\
  \bibnamefont {D\'{i}az}}, \bibinfo {author} {\bibfnamefont {Tomoki}\
  \bibnamefont {Hirosawa}}, \bibinfo {author} {\bibfnamefont {Jelena}\
  \bibnamefont {Klinovaja}}, \ and\ \bibinfo {author} {\bibfnamefont {Daniel}\
  \bibnamefont {Loss}},\ }\bibfield  {title} {\enquote {\bibinfo {title}
  {Chiral magnonic edge states in ferromagnetic skyrmion crystals controlled by
  magnetic fields},}\ }\href {\doibase 10.1103/PhysRevResearch.2.013231}
  {\bibfield  {journal} {\bibinfo  {journal} {Phys. Rev. Research}\ }\textbf
  {\bibinfo {volume} {2}},\ \bibinfo {pages} {013231} (\bibinfo {year}
  {2020})}\BibitemShut {NoStop}%
\bibitem [{\citenamefont {Mook}\ \emph {et~al.}(2020)\citenamefont {Mook},
  \citenamefont {Klinovaja},\ and\ \citenamefont
  {Loss}}]{mookQuantumDampingSkyrmion2020}%
  \BibitemOpen
  \bibfield  {author} {\bibinfo {author} {\bibfnamefont {Alexander}\
  \bibnamefont {Mook}}, \bibinfo {author} {\bibfnamefont {Jelena}\ \bibnamefont
  {Klinovaja}}, \ and\ \bibinfo {author} {\bibfnamefont {Daniel}\ \bibnamefont
  {Loss}},\ }\bibfield  {title} {\enquote {\bibinfo {title} {Quantum damping of
  skyrmion crystal eigenmodes due to spontaneous quasiparticle decay},}\ }\href
  {\doibase 10.1103/PhysRevResearch.2.033491} {\bibfield  {journal} {\bibinfo
  {journal} {Phys. Rev. Research}\ }\textbf {\bibinfo {volume} {2}},\ \bibinfo
  {pages} {033491} (\bibinfo {year} {2020})}\BibitemShut {NoStop}%
\bibitem [{\citenamefont {Hirosawa}\ \emph
  {et~al.}(2022{\natexlab{a}})\citenamefont {Hirosawa}, \citenamefont
  {Klinovaja}, \citenamefont {Loss},\ and\ \citenamefont
  {D\'{i}az}}]{hirosawaLaserControlledRealReciprocalSpace2022}%
  \BibitemOpen
  \bibfield  {author} {\bibinfo {author} {\bibfnamefont {Tomoki}\ \bibnamefont
  {Hirosawa}}, \bibinfo {author} {\bibfnamefont {Jelena}\ \bibnamefont
  {Klinovaja}}, \bibinfo {author} {\bibfnamefont {Daniel}\ \bibnamefont
  {Loss}}, \ and\ \bibinfo {author} {\bibfnamefont {Sebasti\'{a}n~A.}\
  \bibnamefont {D\'{i}az}},\ }\bibfield  {title} {\enquote {\bibinfo {title}
  {Laser-{Controlled} {Real}- and {Reciprocal}-{Space} {Topology} in
  {Multiferroic} {Insulators}},}\ }\href {\doibase
  10.1103/PhysRevLett.128.037201} {\bibfield  {journal} {\bibinfo  {journal}
  {Phys. Rev. Lett.}\ }\textbf {\bibinfo {volume} {128}},\ \bibinfo {pages}
  {037201} (\bibinfo {year} {2022}{\natexlab{a}})}\BibitemShut {NoStop}%
\bibitem [{\citenamefont {Hirosawa}\ \emph {et~al.}(2020)\citenamefont
  {Hirosawa}, \citenamefont {D\'{i}az}, \citenamefont {Klinovaja},\ and\
  \citenamefont {Loss}}]{hirosawaMagnonicQuadrupoleTopological2020}%
  \BibitemOpen
  \bibfield  {author} {\bibinfo {author} {\bibfnamefont {Tomoki}\ \bibnamefont
  {Hirosawa}}, \bibinfo {author} {\bibfnamefont {Sebasti\'{a}n~A.}\
  \bibnamefont {D\'{i}az}}, \bibinfo {author} {\bibfnamefont {Jelena}\
  \bibnamefont {Klinovaja}}, \ and\ \bibinfo {author} {\bibfnamefont {Daniel}\
  \bibnamefont {Loss}},\ }\bibfield  {title} {\enquote {\bibinfo {title}
  {Magnonic {Quadrupole} {Topological} {Insulator} in {Antiskyrmion}
  {Crystals}},}\ }\href {\doibase 10.1103/PhysRevLett.125.207204} {\bibfield
  {journal} {\bibinfo  {journal} {Phys. Rev. Lett.}\ }\textbf {\bibinfo
  {volume} {125}},\ \bibinfo {pages} {207204} (\bibinfo {year}
  {2020})}\BibitemShut {NoStop}%
\bibitem [{\citenamefont {Kawamura}\ and\ \citenamefont
  {Miyashita}(1984)}]{Kawamura1984}%
  \BibitemOpen
  \bibfield  {author} {\bibinfo {author} {\bibfnamefont {Hikaru}\ \bibnamefont
  {Kawamura}}\ and\ \bibinfo {author} {\bibfnamefont {Seiji}\ \bibnamefont
  {Miyashita}},\ }\bibfield  {title} {\enquote {\bibinfo {title} {Phase
  transition of the two-dimensional $\mathrm{H}$eisenberg antiferromagnet on
  the triangular lattice},}\ }\href {\doibase 10.1143/JPSJ.53.4138} {\bibfield
  {journal} {\bibinfo  {journal} {J. Phys. Soc. Jpn.}\ }\textbf {\bibinfo
  {volume} {53}},\ \bibinfo {pages} {4138} (\bibinfo {year}
  {1984})}\BibitemShut {NoStop}%
\bibitem [{\citenamefont {Tomiyasu}\ \emph {et~al.}(2022)\citenamefont
  {Tomiyasu}, \citenamefont {Mizuta}, \citenamefont {Matsuura}, \citenamefont
  {Aoyama},\ and\ \citenamefont
  {Kawamura}}]{tomiyasuObservationTopological2022}%
  \BibitemOpen
  \bibfield  {author} {\bibinfo {author} {\bibfnamefont {K.}~\bibnamefont
  {Tomiyasu}}, \bibinfo {author} {\bibfnamefont {Y.~P.}\ \bibnamefont
  {Mizuta}}, \bibinfo {author} {\bibfnamefont {M.}~\bibnamefont {Matsuura}},
  \bibinfo {author} {\bibfnamefont {K.}~\bibnamefont {Aoyama}}, \ and\ \bibinfo
  {author} {\bibfnamefont {H.}~\bibnamefont {Kawamura}},\ }\bibfield  {title}
  {\enquote {\bibinfo {title} {Observation of topological $\mathbb{Z}_2$ vortex
  fluctuations in the frustrated {{Heisenberg}} magnet {NaCrO}$_2$},}\ }\href
  {\doibase 10.1103/PhysRevB.106.054407} {\bibfield  {journal} {\bibinfo
  {journal} {Phys. Rev. B}\ }\textbf {\bibinfo {volume} {106}},\ \bibinfo
  {pages} {054407} (\bibinfo {year} {2022})}\BibitemShut {NoStop}%
\bibitem [{\citenamefont {Rosales}\ \emph {et~al.}(2015)\citenamefont
  {Rosales}, \citenamefont {Cabra},\ and\ \citenamefont
  {Pujol}}]{rosalesThreesublatticeSkyrmionCrystal2015}%
  \BibitemOpen
  \bibfield  {author} {\bibinfo {author} {\bibfnamefont {H.~D.}\ \bibnamefont
  {Rosales}}, \bibinfo {author} {\bibfnamefont {D.~C.}\ \bibnamefont {Cabra}},
  \ and\ \bibinfo {author} {\bibfnamefont {Pierre}\ \bibnamefont {Pujol}},\
  }\bibfield  {title} {\enquote {\bibinfo {title} {Three-sublattice skyrmion
  crystal in the antiferromagnetic triangular lattice},}\ }\href {\doibase
  10.1103/PhysRevB.92.214439} {\bibfield  {journal} {\bibinfo  {journal} {Phys.
  Rev. B}\ }\textbf {\bibinfo {volume} {92}},\ \bibinfo {pages} {214439}
  (\bibinfo {year} {2015})}\BibitemShut {NoStop}%
\bibitem [{\citenamefont {Díaz}\ \emph {et~al.}(2019)\citenamefont {Díaz},
  \citenamefont {Klinovaja},\ and\ \citenamefont
  {Loss}}]{diazTopologicalMagnonsEdge2019}%
  \BibitemOpen
  \bibfield  {author} {\bibinfo {author} {\bibfnamefont {Sebastián~A.}\
  \bibnamefont {Díaz}}, \bibinfo {author} {\bibfnamefont {Jelena}\
  \bibnamefont {Klinovaja}}, \ and\ \bibinfo {author} {\bibfnamefont {Daniel}\
  \bibnamefont {Loss}},\ }\bibfield  {title} {\enquote {\bibinfo {title}
  {Topological {Magnons} and {Edge} {States} in {Antiferromagnetic} {Skyrmion}
  {Crystals}},}\ }\href {\doibase 10.1103/PhysRevLett.122.187203} {\bibfield
  {journal} {\bibinfo  {journal} {Phys. Rev. Lett.}\ }\textbf {\bibinfo
  {volume} {122}},\ \bibinfo {pages} {187203} (\bibinfo {year}
  {2019})}\BibitemShut {NoStop}%
\bibitem [{\citenamefont {Gao}\ \emph {et~al.}(2020)\citenamefont {Gao},
  \citenamefont {Rosales}, \citenamefont {G\'{o}mez~Albarrac\'{i}n},
  \citenamefont {Tsurkan}, \citenamefont {Kaur}, \citenamefont {Fennell},
  \citenamefont {Steffens}, \citenamefont {Boehm}, \citenamefont
  {\v{C}erm\'{a}k}, \citenamefont {Schneidewind}, \citenamefont {Ressouche},
  \citenamefont {Cabra}, \citenamefont {R\"{u}egg},\ and\ \citenamefont
  {Zaharko}}]{gaoFractionalAntiferromagneticSkyrmion2020}%
  \BibitemOpen
  \bibfield  {author} {\bibinfo {author} {\bibfnamefont {Shang}\ \bibnamefont
  {Gao}}, \bibinfo {author} {\bibfnamefont {H.~Diego}\ \bibnamefont {Rosales}},
  \bibinfo {author} {\bibfnamefont {Flavia~A.}\ \bibnamefont
  {G\'{o}mez~Albarrac\'{i}n}}, \bibinfo {author} {\bibfnamefont {Vladimir}\
  \bibnamefont {Tsurkan}}, \bibinfo {author} {\bibfnamefont {Guratinder}\
  \bibnamefont {Kaur}}, \bibinfo {author} {\bibfnamefont {Tom}\ \bibnamefont
  {Fennell}}, \bibinfo {author} {\bibfnamefont {Paul}\ \bibnamefont
  {Steffens}}, \bibinfo {author} {\bibfnamefont {Martin}\ \bibnamefont
  {Boehm}}, \bibinfo {author} {\bibfnamefont {Petr}\ \bibnamefont
  {\v{C}erm\'{a}k}}, \bibinfo {author} {\bibfnamefont {Astrid}\ \bibnamefont
  {Schneidewind}}, \bibinfo {author} {\bibfnamefont {Eric}\ \bibnamefont
  {Ressouche}}, \bibinfo {author} {\bibfnamefont {Daniel~C.}\ \bibnamefont
  {Cabra}}, \bibinfo {author} {\bibfnamefont {Christian}\ \bibnamefont
  {R\"{u}egg}}, \ and\ \bibinfo {author} {\bibfnamefont {Oksana}\ \bibnamefont
  {Zaharko}},\ }\bibfield  {title} {\enquote {\bibinfo {title} {Fractional
  antiferromagnetic skyrmion lattice induced by anisotropic couplings},}\
  }\href {\doibase 10.1038/s41586-020-2716-8} {\bibfield  {journal} {\bibinfo
  {journal} {Nature}\ }\textbf {\bibinfo {volume} {586}},\ \bibinfo {pages}
  {37} (\bibinfo {year} {2020})}\BibitemShut {NoStop}%
\bibitem [{\citenamefont {Osorio}\ \emph {et~al.}(2019)\citenamefont {Osorio},
  \citenamefont {Sturla}, \citenamefont {Rosales},\ and\ \citenamefont
  {Cabra}}]{Osorio2019}%
  \BibitemOpen
  \bibfield  {author} {\bibinfo {author} {\bibfnamefont {S.~A.}\ \bibnamefont
  {Osorio}}, \bibinfo {author} {\bibfnamefont {M.~B.}\ \bibnamefont {Sturla}},
  \bibinfo {author} {\bibfnamefont {H.~D.}\ \bibnamefont {Rosales}}, \ and\
  \bibinfo {author} {\bibfnamefont {D.~C.}\ \bibnamefont {Cabra}},\ }\bibfield
  {title} {\enquote {\bibinfo {title} {From skyrmions to {$\mathbb{Z}_{2}$}
  vortices in distorted chiral antiferromagnets},}\ }\href {\doibase
  10.1103/PhysRevB.100.220404} {\bibfield  {journal} {\bibinfo  {journal}
  {Phys. Rev. B}\ }\textbf {\bibinfo {volume} {100}},\ \bibinfo {pages}
  {220404} (\bibinfo {year} {2019})}\BibitemShut {NoStop}%
\bibitem [{\citenamefont {Rousochatzakis}\ \emph {et~al.}(2016)\citenamefont
  {Rousochatzakis}, \citenamefont {R\"ossler}, \citenamefont {van~den Brink},\
  and\ \citenamefont {Daghofer}}]{Rousoch2016}%
  \BibitemOpen
  \bibfield  {author} {\bibinfo {author} {\bibfnamefont {Ioannis}\ \bibnamefont
  {Rousochatzakis}}, \bibinfo {author} {\bibfnamefont {Ulrich~K.}\ \bibnamefont
  {R\"ossler}}, \bibinfo {author} {\bibfnamefont {Jeroen}\ \bibnamefont
  {van~den Brink}}, \ and\ \bibinfo {author} {\bibfnamefont {Maria}\
  \bibnamefont {Daghofer}},\ }\bibfield  {title} {\enquote {\bibinfo {title}
  {Kitaev anisotropy induces mesoscopic {$\mathbb{Z}_{2}$} vortex crystals in
  frustrated hexagonal antiferromagnets},}\ }\href {\doibase
  10.1103/PhysRevB.93.104417} {\bibfield  {journal} {\bibinfo  {journal} {Phys.
  Rev. B}\ }\textbf {\bibinfo {volume} {93}},\ \bibinfo {pages} {104417}
  (\bibinfo {year} {2016})}\BibitemShut {NoStop}%
\bibitem [{\citenamefont {Yao}\ and\ \citenamefont {Dong}(2016)}]{Yao2016}%
  \BibitemOpen
  \bibfield  {author} {\bibinfo {author} {\bibfnamefont {Xiaoyan}\ \bibnamefont
  {Yao}}\ and\ \bibinfo {author} {\bibfnamefont {Shuai}\ \bibnamefont {Dong}},\
  }\bibfield  {title} {\enquote {\bibinfo {title} {Topological triple-vortex
  lattice stabilized by mixed frustration in expanded honeycomb
  kitaev-heisenberg model},}\ }\href {\doibase 10.1038/srep26750} {\bibfield
  {journal} {\bibinfo  {journal} {Sci. Rep.}\ }\textbf {\bibinfo {volume}
  {6}},\ \bibinfo {pages} {26750} (\bibinfo {year} {2016})}\BibitemShut
  {NoStop}%
\bibitem [{\citenamefont {Kato}\ \emph {et~al.}(2010)\citenamefont {Kato},
  \citenamefont {Martin},\ and\ \citenamefont
  {Batista}}]{katoStabilitySpontaneousQuantum2010}%
  \BibitemOpen
  \bibfield  {author} {\bibinfo {author} {\bibfnamefont {Yasuyuki}\
  \bibnamefont {Kato}}, \bibinfo {author} {\bibfnamefont {Ivar}\ \bibnamefont
  {Martin}}, \ and\ \bibinfo {author} {\bibfnamefont {C.~D.}\ \bibnamefont
  {Batista}},\ }\bibfield  {title} {\enquote {\bibinfo {title} {Stability of
  the {Spontaneous} {Quantum} {Hall} {State} in the {Triangular}
  {Kondo}-{Lattice} {Model}},}\ }\href {\doibase
  10.1103/PhysRevLett.105.266405} {\bibfield  {journal} {\bibinfo  {journal}
  {Phys. Rev. Lett.}\ }\textbf {\bibinfo {volume} {105}},\ \bibinfo {pages}
  {266405} (\bibinfo {year} {2010})}\BibitemShut {NoStop}%
\bibitem [{\citenamefont {Akagi}\ \emph {et~al.}(2012)\citenamefont {Akagi},
  \citenamefont {Udagawa},\ and\ \citenamefont {Motome}}]{Akagi2012}%
  \BibitemOpen
  \bibfield  {author} {\bibinfo {author} {\bibfnamefont {Yutaka}\ \bibnamefont
  {Akagi}}, \bibinfo {author} {\bibfnamefont {Masafumi}\ \bibnamefont
  {Udagawa}}, \ and\ \bibinfo {author} {\bibfnamefont {Yukitoshi}\ \bibnamefont
  {Motome}},\ }\bibfield  {title} {\enquote {\bibinfo {title} {Hidden
  multiple-spin interactions as an origin of spin scalar chiral order in
  frustrated kondo lattice models},}\ }\href {\doibase
  10.1103/PhysRevLett.108.096401} {\bibfield  {journal} {\bibinfo  {journal}
  {Phys. Rev. Lett.}\ }\textbf {\bibinfo {volume} {108}},\ \bibinfo {pages}
  {096401} (\bibinfo {year} {2012})}\BibitemShut {NoStop}%
\bibitem [{\citenamefont {Momoi}\ \emph {et~al.}(1997)\citenamefont {Momoi},
  \citenamefont {Kubo},\ and\ \citenamefont {Niki}}]{Momoi1997}%
  \BibitemOpen
  \bibfield  {author} {\bibinfo {author} {\bibfnamefont {Tsutomu}\ \bibnamefont
  {Momoi}}, \bibinfo {author} {\bibfnamefont {Kenn}\ \bibnamefont {Kubo}}, \
  and\ \bibinfo {author} {\bibfnamefont {Koji}\ \bibnamefont {Niki}},\
  }\bibfield  {title} {\enquote {\bibinfo {title} {Possible chiral phase
  transition in two-dimensional solid ${}^{3}\mathrm{He}$},}\ }\href {\doibase
  10.1103/PhysRevLett.79.2081} {\bibfield  {journal} {\bibinfo  {journal}
  {Phys. Rev. Lett.}\ }\textbf {\bibinfo {volume} {79}},\ \bibinfo {pages}
  {2081--2084} (\bibinfo {year} {1997})}\BibitemShut {NoStop}%
\bibitem [{\citenamefont {Berg}\ and\ \citenamefont
  {L\"{u}scher}(1981)}]{bergDefinitionStatisticalDistributions1981}%
  \BibitemOpen
  \bibfield  {author} {\bibinfo {author} {\bibfnamefont {B.}~\bibnamefont
  {Berg}}\ and\ \bibinfo {author} {\bibfnamefont {M.}~\bibnamefont
  {L\"{u}scher}},\ }\bibfield  {title} {\enquote {\bibinfo {title} {Definition
  and statistical distributions of a topological number in the lattice {O}(3)
  $\sigma$-model},}\ }\href {\doibase 10.1016/0550-3213(81)90568-X} {\bibfield
  {journal} {\bibinfo  {journal} {Nucl. Phys. B}\ }\textbf {\bibinfo {volume}
  {190}},\ \bibinfo {pages} {412} (\bibinfo {year} {1981})}\BibitemShut
  {NoStop}%
\bibitem [{\citenamefont {Barros}\ and\ \citenamefont
  {Kato}(2013)}]{barrosEfficientLangevinSimulation2013a}%
  \BibitemOpen
  \bibfield  {author} {\bibinfo {author} {\bibfnamefont {Kipton}\ \bibnamefont
  {Barros}}\ and\ \bibinfo {author} {\bibfnamefont {Yasuyuki}\ \bibnamefont
  {Kato}},\ }\bibfield  {title} {\enquote {\bibinfo {title} {Efficient
  {Langevin} simulation of coupled classical fields and fermions},}\ }\href
  {\doibase 10.1103/PhysRevB.88.235101} {\bibfield  {journal} {\bibinfo
  {journal} {Phys. Rev. B}\ }\textbf {\bibinfo {volume} {88}},\ \bibinfo
  {pages} {235101} (\bibinfo {year} {2013})}\BibitemShut {NoStop}%
\bibitem [{\citenamefont {Rahmani}\ \emph {et~al.}(2013)\citenamefont
  {Rahmani}, \citenamefont {Muniz},\ and\ \citenamefont
  {Martin}}]{Rahmani2013}%
  \BibitemOpen
  \bibfield  {author} {\bibinfo {author} {\bibfnamefont {Armin}\ \bibnamefont
  {Rahmani}}, \bibinfo {author} {\bibfnamefont {Rodrigo~A.}\ \bibnamefont
  {Muniz}}, \ and\ \bibinfo {author} {\bibfnamefont {Ivar}\ \bibnamefont
  {Martin}},\ }\bibfield  {title} {\enquote {\bibinfo {title} {Anyons in
  {Integer} {Quantum} {Hall} {Magnets}},}\ }\href {\doibase
  10.1103/PhysRevX.3.031008} {\bibfield  {journal} {\bibinfo  {journal} {Phys.
  Rev. X}\ }\textbf {\bibinfo {volume} {3}},\ \bibinfo {pages} {031008}
  (\bibinfo {year} {2013})}\BibitemShut {NoStop}%
\bibitem [{SM()}]{SM}%
  \BibitemOpen
  \href@noop {} {}\bibinfo {note} {See Supplemental Material
  at...,}\BibitemShut {NoStop}%
\bibitem [{\citenamefont {Grover}\ and\ \citenamefont
  {Senthil}(2011)}]{groverNonAbelianSpinLiquid2011}%
  \BibitemOpen
  \bibfield  {author} {\bibinfo {author} {\bibfnamefont {Tarun}\ \bibnamefont
  {Grover}}\ and\ \bibinfo {author} {\bibfnamefont {T.}~\bibnamefont
  {Senthil}},\ }\bibfield  {title} {\enquote {\bibinfo {title} {Non-{Abelian}
  {Spin} {Liquid} in a {Spin}-{One} {Quantum} {Magnet}},}\ }\href {\doibase
  10.1103/PhysRevLett.107.077203} {\bibfield  {journal} {\bibinfo  {journal}
  {Phys. Rev. Lett.}\ }\textbf {\bibinfo {volume} {107}},\ \bibinfo {pages}
  {077203} (\bibinfo {year} {2011})}\BibitemShut {NoStop}%
\bibitem [{\citenamefont {Holstein}\ and\ \citenamefont
  {Primakoff}(1940)}]{holsteinFieldDependenceIntrinsic1940}%
  \BibitemOpen
  \bibfield  {author} {\bibinfo {author} {\bibfnamefont {T.}~\bibnamefont
  {Holstein}}\ and\ \bibinfo {author} {\bibfnamefont {H.}~\bibnamefont
  {Primakoff}},\ }\bibfield  {title} {\enquote {\bibinfo {title} {Field
  {{Dependence}} of the {{Intrinsic Domain Magnetization}} of a
  {{Ferromagnet}}},}\ }\href {\doibase 10.1103/PhysRev.58.1098} {\bibfield
  {journal} {\bibinfo  {journal} {Physical Review}\ }\textbf {\bibinfo {volume}
  {58}},\ \bibinfo {pages} {1098} (\bibinfo {year} {1940})}\BibitemShut
  {NoStop}%
\bibitem [{\citenamefont {Akagi}\ \emph {et~al.}(2013)\citenamefont {Akagi},
  \citenamefont {Udagawa},\ and\ \citenamefont {Motome}}]{Akagi2013}%
  \BibitemOpen
  \bibfield  {author} {\bibinfo {author} {\bibfnamefont {Yutaka}\ \bibnamefont
  {Akagi}}, \bibinfo {author} {\bibfnamefont {Masafumi}\ \bibnamefont
  {Udagawa}}, \ and\ \bibinfo {author} {\bibfnamefont {Yukitoshi}\ \bibnamefont
  {Motome}},\ }\bibfield  {title} {\enquote {\bibinfo {title} {Effect of
  quantum spin fluctuation on scalar chiral ordering in the kondo lattice model
  on a triangular lattice},}\ }\href {\doibase 10.7566/JPSJ.82.123709}
  {\bibfield  {journal} {\bibinfo  {journal} {J. Phys. Soc. Jpn.}\ }\textbf
  {\bibinfo {volume} {82}},\ \bibinfo {pages} {123709} (\bibinfo {year}
  {2013})}\BibitemShut {NoStop}%
\bibitem [{\citenamefont {Rodrigues}\ \emph {et~al.}(2022)\citenamefont
  {Rodrigues}, \citenamefont {Salimath}, \citenamefont {Everschor-Sitte},\ and\
  \citenamefont {Hals}}]{rodriguesDzyaloshinskiiMoriyaInducedSpintransfer2022}%
  \BibitemOpen
  \bibfield  {author} {\bibinfo {author} {\bibfnamefont {Davi~R.}\ \bibnamefont
  {Rodrigues}}, \bibinfo {author} {\bibfnamefont {Akshaykumar}\ \bibnamefont
  {Salimath}}, \bibinfo {author} {\bibfnamefont {Karin}\ \bibnamefont
  {Everschor-Sitte}}, \ and\ \bibinfo {author} {\bibfnamefont {Kjetil M.~D.}\
  \bibnamefont {Hals}},\ }\bibfield  {title} {\enquote {\bibinfo {title}
  {Dzyaloshinskii-{Moriya} induced spin-transfer torques in kagome
  antiferromagnets},}\ }\href {\doibase 10.1103/PhysRevB.105.174401} {\bibfield
   {journal} {\bibinfo  {journal} {Phys. Rev. B}\ }\textbf {\bibinfo {volume}
  {105}},\ \bibinfo {pages} {174401} (\bibinfo {year} {2022})}\BibitemShut
  {NoStop}%
\bibitem [{\citenamefont {Barts}\ and\ \citenamefont
  {Mostovoy}(2021)}]{Barts2021}%
  \BibitemOpen
  \bibfield  {author} {\bibinfo {author} {\bibfnamefont {Evgenii}\ \bibnamefont
  {Barts}}\ and\ \bibinfo {author} {\bibfnamefont {Maxim}\ \bibnamefont
  {Mostovoy}},\ }\bibfield  {title} {\enquote {\bibinfo {title} {Magnetic
  particles and strings in iron langasite},}\ }\href {\doibase
  10.1038/s41535-021-00408-4} {\bibfield  {journal} {\bibinfo  {journal} {npj
  Quantum Mater.}\ }\textbf {\bibinfo {volume} {6}},\ \bibinfo {pages} {104}
  (\bibinfo {year} {2021})}\BibitemShut {NoStop}%
\bibitem [{\citenamefont {Hirosawa}\ \emph
  {et~al.}(2022{\natexlab{b}})\citenamefont {Hirosawa}, \citenamefont {Mook},
  \citenamefont {Klinovaja},\ and\ \citenamefont
  {Loss}}]{hirosawaMagnetoelectricCavityMagnonics2022}%
  \BibitemOpen
  \bibfield  {author} {\bibinfo {author} {\bibfnamefont {Tomoki}\ \bibnamefont
  {Hirosawa}}, \bibinfo {author} {\bibfnamefont {Alexander}\ \bibnamefont
  {Mook}}, \bibinfo {author} {\bibfnamefont {Jelena}\ \bibnamefont
  {Klinovaja}}, \ and\ \bibinfo {author} {\bibfnamefont {Daniel}\ \bibnamefont
  {Loss}},\ }\bibfield  {title} {\enquote {\bibinfo {title} {Magnetoelectric
  {Cavity} {Magnonics} in {Skyrmion} {Crystals}},}\ }\href {\doibase
  10.1103/PRXQuantum.3.040321} {\bibfield  {journal} {\bibinfo  {journal} {PRX
  Quantum}\ }\textbf {\bibinfo {volume} {3}},\ \bibinfo {pages} {040321}
  (\bibinfo {year} {2022}{\natexlab{b}})}\BibitemShut {NoStop}%
\bibitem [{\citenamefont {Mochizuki}(2012)}]{mochizukiSpinWaveModesTheir2012}%
  \BibitemOpen
  \bibfield  {author} {\bibinfo {author} {\bibfnamefont {Masahito}\
  \bibnamefont {Mochizuki}},\ }\bibfield  {title} {\enquote {\bibinfo {title}
  {Spin-{Wave} {Modes} and {Their} {Intense} {Excitation} {Effects} in
  {Skyrmion} {Crystals}},}\ }\href {\doibase 10.1103/PhysRevLett.108.017601}
  {\bibfield  {journal} {\bibinfo  {journal} {Phys. Rev. Lett.}\ }\textbf
  {\bibinfo {volume} {108}},\ \bibinfo {pages} {017601} (\bibinfo {year}
  {2012})}\BibitemShut {NoStop}%
\bibitem [{\citenamefont {Sch{\"u}tte}\ and\ \citenamefont
  {Garst}(2014)}]{schutteMagnonskyrmionScatteringChiral2014}%
  \BibitemOpen
  \bibfield  {author} {\bibinfo {author} {\bibfnamefont {Christoph}\
  \bibnamefont {Sch{\"u}tte}}\ and\ \bibinfo {author} {\bibfnamefont {Markus}\
  \bibnamefont {Garst}},\ }\bibfield  {title} {\enquote {\bibinfo {title}
  {Magnon-skyrmion scattering in chiral magnets},}\ }\href {\doibase
  10.1103/PhysRevB.90.094423} {\bibfield  {journal} {\bibinfo  {journal} {Phys.
  Rev. B}\ }\textbf {\bibinfo {volume} {90}},\ \bibinfo {pages} {094423}
  (\bibinfo {year} {2014})}\BibitemShut {NoStop}%
\bibitem [{\citenamefont {Fukui}\ \emph {et~al.}(2005)\citenamefont {Fukui},
  \citenamefont {Hatsugai},\ and\ \citenamefont
  {Suzuki}}]{fukuiChernNumbersDiscretized2005}%
  \BibitemOpen
  \bibfield  {author} {\bibinfo {author} {\bibfnamefont {Takahiro}\
  \bibnamefont {Fukui}}, \bibinfo {author} {\bibfnamefont {Yasuhiro}\
  \bibnamefont {Hatsugai}}, \ and\ \bibinfo {author} {\bibfnamefont {Hiroshi}\
  \bibnamefont {Suzuki}},\ }\bibfield  {title} {\enquote {\bibinfo {title}
  {Chern {{Numbers}} in {{Discretized Brillouin Zone}}: {{Efficient Method}} of
  {{Computing}} ({{Spin}}) {{Hall Conductances}}},}\ }\href {\doibase
  10.1143/JPSJ.74.1674} {\bibfield  {journal} {\bibinfo  {journal} {J. Phys.
  Soc. Jpn.}\ }\textbf {\bibinfo {volume} {74}},\ \bibinfo {pages} {1674--1677}
  (\bibinfo {year} {2005})}\BibitemShut {NoStop}%
\bibitem [{\citenamefont {Shindou}\ \emph {et~al.}(2013)\citenamefont
  {Shindou}, \citenamefont {Matsumoto}, \citenamefont {Murakami},\ and\
  \citenamefont {Ohe}}]{shindouTopologicalChiralMagnonic2013}%
  \BibitemOpen
  \bibfield  {author} {\bibinfo {author} {\bibfnamefont {Ryuichi}\ \bibnamefont
  {Shindou}}, \bibinfo {author} {\bibfnamefont {Ryo}\ \bibnamefont
  {Matsumoto}}, \bibinfo {author} {\bibfnamefont {Shuichi}\ \bibnamefont
  {Murakami}}, \ and\ \bibinfo {author} {\bibfnamefont {Jun-ichiro}\
  \bibnamefont {Ohe}},\ }\bibfield  {title} {\enquote {\bibinfo {title}
  {Topological chiral magnonic edge mode in a magnonic crystal},}\ }\href
  {\doibase 10.1103/PhysRevB.87.174427} {\bibfield  {journal} {\bibinfo
  {journal} {Phys. Rev. B}\ }\textbf {\bibinfo {volume} {87}},\ \bibinfo
  {pages} {174427} (\bibinfo {year} {2013})}\BibitemShut {NoStop}%
\bibitem [{\citenamefont
  {Hatsugai}(1993{\natexlab{a}})}]{hatsugaiChernNumberEdge1993}%
  \BibitemOpen
  \bibfield  {author} {\bibinfo {author} {\bibfnamefont {Yasuhiro}\
  \bibnamefont {Hatsugai}},\ }\bibfield  {title} {\enquote {\bibinfo {title}
  {Chern number and edge states in the integer quantum {Hall} effect},}\ }\href
  {\doibase 10.1103/PhysRevLett.71.3697} {\bibfield  {journal} {\bibinfo
  {journal} {Phys. Rev. Lett.}\ }\textbf {\bibinfo {volume} {71}},\ \bibinfo
  {pages} {3697} (\bibinfo {year} {1993}{\natexlab{a}})}\BibitemShut {NoStop}%
\bibitem [{\citenamefont
  {Hatsugai}(1993{\natexlab{b}})}]{hatsugaiEdgeStatesInteger1993}%
  \BibitemOpen
  \bibfield  {author} {\bibinfo {author} {\bibfnamefont {Yasuhiro}\
  \bibnamefont {Hatsugai}},\ }\bibfield  {title} {\enquote {\bibinfo {title}
  {Edge states in the integer quantum {Hall} effect and the {Riemann} surface
  of the {Bloch} function},}\ }\href {\doibase 10.1103/PhysRevB.48.11851}
  {\bibfield  {journal} {\bibinfo  {journal} {Phys. Rev. B}\ }\textbf {\bibinfo
  {volume} {48}},\ \bibinfo {pages} {11851} (\bibinfo {year}
  {1993}{\natexlab{b}})}\BibitemShut {NoStop}%
\bibitem [{\citenamefont {Henk}\ and\ \citenamefont
  {Schattke}(1993)}]{HENK199369}%
  \BibitemOpen
  \bibfield  {author} {\bibinfo {author} {\bibfnamefont {J.}~\bibnamefont
  {Henk}}\ and\ \bibinfo {author} {\bibfnamefont {W.}~\bibnamefont
  {Schattke}},\ }\bibfield  {title} {\enquote {\bibinfo {title} {A subroutine
  package for computing {Green's} functions of relaxed surfaces by the
  renormalization method},}\ }\href {\doibase
  https://doi.org/10.1016/0010-4655(93)90038-E} {\bibfield  {journal} {\bibinfo
   {journal} {Comput. Phys. Commun.}\ }\textbf {\bibinfo {volume} {77}},\
  \bibinfo {pages} {69} (\bibinfo {year} {1993})}\BibitemShut {NoStop}%
\bibitem [{\citenamefont {Mook}\ \emph {et~al.}(2014)\citenamefont {Mook},
  \citenamefont {Henk},\ and\ \citenamefont {Mertig}}]{Mook2014edge}%
  \BibitemOpen
  \bibfield  {author} {\bibinfo {author} {\bibfnamefont {Alexander}\
  \bibnamefont {Mook}}, \bibinfo {author} {\bibfnamefont {J\"urgen}\
  \bibnamefont {Henk}}, \ and\ \bibinfo {author} {\bibfnamefont {Ingrid}\
  \bibnamefont {Mertig}},\ }\bibfield  {title} {\enquote {\bibinfo {title}
  {Edge states in topological magnon insulators},}\ }\href {\doibase
  10.1103/PhysRevB.90.024412} {\bibfield  {journal} {\bibinfo  {journal} {Phys.
  Rev. B}\ }\textbf {\bibinfo {volume} {90}},\ \bibinfo {pages} {024412}
  (\bibinfo {year} {2014})}\BibitemShut {NoStop}%
\bibitem [{\citenamefont {Sun}\ \emph {et~al.}(1996)\citenamefont {Sun},
  \citenamefont {Zhang},\ and\ \citenamefont
  {Li}}]{sunMonteCarloCalculations1996}%
  \BibitemOpen
  \bibfield  {author} {\bibinfo {author} {\bibfnamefont {Gang}\ \bibnamefont
  {Sun}}, \bibinfo {author} {\bibfnamefont {Xiang-Dong}\ \bibnamefont {Zhang}},
  \ and\ \bibinfo {author} {\bibfnamefont {Bo-Zang}\ \bibnamefont {Li}},\
  }\bibfield  {title} {\enquote {\bibinfo {title} {Monte--{{Carlo
  Calculations}} for the {{Surface}} on a {{Semi-Infinite XY Model}}},}\ }\href
  {\doibase 10.1088/0253-6102/25/1/115} {\bibfield  {journal} {\bibinfo
  {journal} {Commun. Theor. Phys.}\ }\textbf {\bibinfo {volume} {25}},\
  \bibinfo {pages} {115} (\bibinfo {year} {1996})}\BibitemShut {NoStop}%
\bibitem [{\citenamefont {Spethmann}\ \emph {et~al.}(2020)\citenamefont
  {Spethmann}, \citenamefont {Meyer}, \citenamefont {von Bergmann},
  \citenamefont {Wiesendanger}, \citenamefont {Heinze},\ and\ \citenamefont
  {Kubetzka}}]{Spethmann2020}%
  \BibitemOpen
  \bibfield  {author} {\bibinfo {author} {\bibfnamefont {Jonas}\ \bibnamefont
  {Spethmann}}, \bibinfo {author} {\bibfnamefont {Sebastian}\ \bibnamefont
  {Meyer}}, \bibinfo {author} {\bibfnamefont {Kirsten}\ \bibnamefont {von
  Bergmann}}, \bibinfo {author} {\bibfnamefont {Roland}\ \bibnamefont
  {Wiesendanger}}, \bibinfo {author} {\bibfnamefont {Stefan}\ \bibnamefont
  {Heinze}}, \ and\ \bibinfo {author} {\bibfnamefont {Andr\'e}\ \bibnamefont
  {Kubetzka}},\ }\bibfield  {title} {\enquote {\bibinfo {title} {Discovery of
  magnetic single- and triple-$\mathbf{q}$ states in
  $\mathrm{Mn}/\mathrm{Re}(0001)$},}\ }\href {\doibase
  10.1103/PhysRevLett.124.227203} {\bibfield  {journal} {\bibinfo  {journal}
  {Phys. Rev. Lett.}\ }\textbf {\bibinfo {volume} {124}},\ \bibinfo {pages}
  {227203} (\bibinfo {year} {2020})}\BibitemShut {NoStop}%
\bibitem [{\citenamefont {Nickel}\ \emph {et~al.}(2023)\citenamefont {Nickel},
  \citenamefont {Kubetzka}, \citenamefont {Haldar}, \citenamefont
  {Wiesendanger}, \citenamefont {Heinze},\ and\ \citenamefont {von
  Bergmann}}]{nickelCouplingTripleState2023}%
  \BibitemOpen
  \bibfield  {author} {\bibinfo {author} {\bibfnamefont {Felix}\ \bibnamefont
  {Nickel}}, \bibinfo {author} {\bibfnamefont {Andr\'{e}}\ \bibnamefont
  {Kubetzka}}, \bibinfo {author} {\bibfnamefont {Soumyajyoti}\ \bibnamefont
  {Haldar}}, \bibinfo {author} {\bibfnamefont {Roland}\ \bibnamefont
  {Wiesendanger}}, \bibinfo {author} {\bibfnamefont {Stefan}\ \bibnamefont
  {Heinze}}, \ and\ \bibinfo {author} {\bibfnamefont {Kirsten}\ \bibnamefont
  {von Bergmann}},\ }\bibfield  {title} {\enquote {\bibinfo {title} {Coupling
  of the triple-${q}$ state to the atomic lattice by anisotropic symmetric
  exchange},}\ }\href {\doibase 10.1103/PhysRevB.108.L180411} {\bibfield
  {journal} {\bibinfo  {journal} {Phys. Rev. B}\ }\textbf {\bibinfo {volume}
  {108}},\ \bibinfo {pages} {L180411} (\bibinfo {year} {2023})}\BibitemShut
  {NoStop}%
\bibitem [{\citenamefont {Takagi}\ \emph {et~al.}(2023)\citenamefont {Takagi},
  \citenamefont {Takagi}, \citenamefont {Minami}, \citenamefont {Nomoto},
  \citenamefont {Ohishi}, \citenamefont {Suzuki}, \citenamefont {Yanagi},
  \citenamefont {Hirayama}, \citenamefont {Khanh}, \citenamefont {Karube},
  \citenamefont {Saito}, \citenamefont {Hashizume}, \citenamefont {Kiyanagi},
  \citenamefont {Tokura}, \citenamefont {Arita}, \citenamefont {Nakajima},\
  and\ \citenamefont {Seki}}]{Takagi2023}%
  \BibitemOpen
  \bibfield  {author} {\bibinfo {author} {\bibfnamefont {H.}~\bibnamefont
  {Takagi}}, \bibinfo {author} {\bibfnamefont {R.}~\bibnamefont {Takagi}},
  \bibinfo {author} {\bibfnamefont {S.}~\bibnamefont {Minami}}, \bibinfo
  {author} {\bibfnamefont {T.}~\bibnamefont {Nomoto}}, \bibinfo {author}
  {\bibfnamefont {K.}~\bibnamefont {Ohishi}}, \bibinfo {author} {\bibfnamefont
  {M.-T.}\ \bibnamefont {Suzuki}}, \bibinfo {author} {\bibfnamefont
  {Y.}~\bibnamefont {Yanagi}}, \bibinfo {author} {\bibfnamefont
  {M.}~\bibnamefont {Hirayama}}, \bibinfo {author} {\bibfnamefont {N.~D.}\
  \bibnamefont {Khanh}}, \bibinfo {author} {\bibfnamefont {K.}~\bibnamefont
  {Karube}}, \bibinfo {author} {\bibfnamefont {H.}~\bibnamefont {Saito}},
  \bibinfo {author} {\bibfnamefont {D.}~\bibnamefont {Hashizume}}, \bibinfo
  {author} {\bibfnamefont {R.}~\bibnamefont {Kiyanagi}}, \bibinfo {author}
  {\bibfnamefont {Y.}~\bibnamefont {Tokura}}, \bibinfo {author} {\bibfnamefont
  {R.}~\bibnamefont {Arita}}, \bibinfo {author} {\bibfnamefont
  {T.}~\bibnamefont {Nakajima}}, \ and\ \bibinfo {author} {\bibfnamefont
  {S.}~\bibnamefont {Seki}},\ }\bibfield  {title} {\enquote {\bibinfo {title}
  {Spontaneous topological hall effect induced by non-coplanar
  antiferromagnetic order in intercalated van der waals materials},}\ }\href
  {\doibase 10.1038/s41567-023-02017-3} {\bibfield  {journal} {\bibinfo
  {journal} {Nat. Phys.}\ }\textbf {\bibinfo {volume} {19}},\ \bibinfo {pages}
  {961} (\bibinfo {year} {2023})}\BibitemShut {NoStop}%
\bibitem [{\citenamefont {Park}\ \emph {et~al.}(2023)\citenamefont {Park},
  \citenamefont {Cho}, \citenamefont {Kim}, \citenamefont {An}, \citenamefont
  {Kang}, \citenamefont {Avdeev}, \citenamefont {Sibille}, \citenamefont
  {Iida}, \citenamefont {Kajimoto}, \citenamefont {Lee}, \citenamefont {Ju},
  \citenamefont {Cho}, \citenamefont {Noh}, \citenamefont {Han}, \citenamefont
  {Zhang}, \citenamefont {Batista},\ and\ \citenamefont {Park}}]{Park2023}%
  \BibitemOpen
  \bibfield  {author} {\bibinfo {author} {\bibfnamefont {Pyeongjae}\
  \bibnamefont {Park}}, \bibinfo {author} {\bibfnamefont {Woonghee}\
  \bibnamefont {Cho}}, \bibinfo {author} {\bibfnamefont {Chaebin}\ \bibnamefont
  {Kim}}, \bibinfo {author} {\bibfnamefont {Yeochan}\ \bibnamefont {An}},
  \bibinfo {author} {\bibfnamefont {Yoon-Gu}\ \bibnamefont {Kang}}, \bibinfo
  {author} {\bibfnamefont {Maxim}\ \bibnamefont {Avdeev}}, \bibinfo {author}
  {\bibfnamefont {Romain}\ \bibnamefont {Sibille}}, \bibinfo {author}
  {\bibfnamefont {Kazuki}\ \bibnamefont {Iida}}, \bibinfo {author}
  {\bibfnamefont {Ryoichi}\ \bibnamefont {Kajimoto}}, \bibinfo {author}
  {\bibfnamefont {Ki~Hoon}\ \bibnamefont {Lee}}, \bibinfo {author}
  {\bibfnamefont {Woori}\ \bibnamefont {Ju}}, \bibinfo {author} {\bibfnamefont
  {En-Jin}\ \bibnamefont {Cho}}, \bibinfo {author} {\bibfnamefont {Han-Jin}\
  \bibnamefont {Noh}}, \bibinfo {author} {\bibfnamefont {Myung~Joon}\
  \bibnamefont {Han}}, \bibinfo {author} {\bibfnamefont {Shang-Shun}\
  \bibnamefont {Zhang}}, \bibinfo {author} {\bibfnamefont {Cristian~D.}\
  \bibnamefont {Batista}}, \ and\ \bibinfo {author} {\bibfnamefont {Je-Geun}\
  \bibnamefont {Park}},\ }\bibfield  {title} {\enquote {\bibinfo {title}
  {Tetrahedral triple-{Q} magnetic ordering and large spontaneous hall
  conductivity in the metallic triangular antiferromagnet
  $\mathrm{Co}_{1/3}\mathrm{TaS}_2$},}\ }\href {\doibase
  10.1038/s41467-023-43853-4} {\bibfield  {journal} {\bibinfo  {journal} {Nat.
  Commun.}\ }\textbf {\bibinfo {volume} {14}},\ \bibinfo {pages} {8346}
  (\bibinfo {year} {2023})}\BibitemShut {NoStop}%
\bibitem [{\citenamefont {Okuma}\ \emph {et~al.}(2024)\citenamefont {Okuma},
  \citenamefont {Yamagami}, \citenamefont {Fujisawa}, \citenamefont {Hsu},
  \citenamefont {Obata}, \citenamefont {Tomoda}, \citenamefont {Dronova},
  \citenamefont {Kuroda}, \citenamefont {Ishikawa}, \citenamefont {Kawaguchi},
  \citenamefont {Aido}, \citenamefont {Kindo}, \citenamefont {Chan},
  \citenamefont {Lin}, \citenamefont {Ihara}, \citenamefont {Kondo},\ and\
  \citenamefont {Okada}}]{okumaEmergentTopologicalMagnetism2024}%
  \BibitemOpen
  \bibfield  {author} {\bibinfo {author} {\bibfnamefont {R.}~\bibnamefont
  {Okuma}}, \bibinfo {author} {\bibfnamefont {K.}~\bibnamefont {Yamagami}},
  \bibinfo {author} {\bibfnamefont {Y.}~\bibnamefont {Fujisawa}}, \bibinfo
  {author} {\bibfnamefont {C.~H.}\ \bibnamefont {Hsu}}, \bibinfo {author}
  {\bibfnamefont {Y.}~\bibnamefont {Obata}}, \bibinfo {author} {\bibfnamefont
  {N.}~\bibnamefont {Tomoda}}, \bibinfo {author} {\bibfnamefont
  {M.}~\bibnamefont {Dronova}}, \bibinfo {author} {\bibfnamefont
  {K.}~\bibnamefont {Kuroda}}, \bibinfo {author} {\bibfnamefont
  {H.}~\bibnamefont {Ishikawa}}, \bibinfo {author} {\bibfnamefont
  {K.}~\bibnamefont {Kawaguchi}}, \bibinfo {author} {\bibfnamefont
  {K.}~\bibnamefont {Aido}}, \bibinfo {author} {\bibfnamefont {K.}~\bibnamefont
  {Kindo}}, \bibinfo {author} {\bibfnamefont {Y.~H.}\ \bibnamefont {Chan}},
  \bibinfo {author} {\bibfnamefont {H.}~\bibnamefont {Lin}}, \bibinfo {author}
  {\bibfnamefont {Y.}~\bibnamefont {Ihara}}, \bibinfo {author} {\bibfnamefont
  {T.}~\bibnamefont {Kondo}}, \ and\ \bibinfo {author} {\bibfnamefont
  {Y.}~\bibnamefont {Okada}},\ }\bibfield  {title} {\enquote {\bibinfo {title}
  {Emergent topological magnetism in {Hund}'s excitonic insulator},}\ }\href
  {http://arxiv.org/abs/2405.16781} {\bibfield  {journal} {\bibinfo  {journal}
  {arXiv:2405.16781 [cond-mat.str-el]}\ } (\bibinfo {year} {2024})}\BibitemShut
  {NoStop}%
\bibitem [{\citenamefont {Jiang}\ \emph {et~al.}(2025)\citenamefont {Jiang},
  \citenamefont {Qiao},\ and\ \citenamefont
  {Zhang}}]{jiangAnomalousValleyHall2025}%
  \BibitemOpen
  \bibfield  {author} {\bibinfo {author} {\bibfnamefont {Xiu-Cai}\ \bibnamefont
  {Jiang}}, \bibinfo {author} {\bibfnamefont {Li-Ya}\ \bibnamefont {Qiao}}, \
  and\ \bibinfo {author} {\bibfnamefont {Yu-Zhong}\ \bibnamefont {Zhang}},\
  }\bibfield  {title} {\enquote {\bibinfo {title} {Anomalous valley {{Hall}}
  effect in monolayer chromium-based triple-${Q}$ magnets},}\ }\href {\doibase
  10.1103/PhysRevB.111.L140416} {\bibfield  {journal} {\bibinfo  {journal}
  {Phys. Rev. B}\ }\textbf {\bibinfo {volume} {111}},\ \bibinfo {pages}
  {L140416} (\bibinfo {year} {2025})}\BibitemShut {NoStop}%
\bibitem [{\citenamefont {Fert}\ \emph {et~al.}(2013)\citenamefont {Fert},
  \citenamefont {Cros},\ and\ \citenamefont
  {Sampaio}}]{fertSkyrmionsTrack2013}%
  \BibitemOpen
  \bibfield  {author} {\bibinfo {author} {\bibfnamefont {Albert}\ \bibnamefont
  {Fert}}, \bibinfo {author} {\bibfnamefont {Vincent}\ \bibnamefont {Cros}}, \
  and\ \bibinfo {author} {\bibfnamefont {Jo\~{a}o}\ \bibnamefont {Sampaio}},\
  }\bibfield  {title} {\enquote {\bibinfo {title} {Skyrmions on the track},}\
  }\href {\doibase 10.1038/nnano.2013.29} {\bibfield  {journal} {\bibinfo
  {journal} {Nat. Nanotechnol.}\ }\textbf {\bibinfo {volume} {8}},\ \bibinfo
  {pages} {152} (\bibinfo {year} {2013})}\BibitemShut {NoStop}%
\bibitem [{\citenamefont {Onose}\ \emph {et~al.}(2012)\citenamefont {Onose},
  \citenamefont {Okamura}, \citenamefont {Seki}, \citenamefont {Ishiwata},\
  and\ \citenamefont {Tokura}}]{onoseObservationMagneticExcitations2012}%
  \BibitemOpen
  \bibfield  {author} {\bibinfo {author} {\bibfnamefont {Y.}~\bibnamefont
  {Onose}}, \bibinfo {author} {\bibfnamefont {Y.}~\bibnamefont {Okamura}},
  \bibinfo {author} {\bibfnamefont {S.}~\bibnamefont {Seki}}, \bibinfo {author}
  {\bibfnamefont {S.}~\bibnamefont {Ishiwata}}, \ and\ \bibinfo {author}
  {\bibfnamefont {Y.}~\bibnamefont {Tokura}},\ }\bibfield  {title} {\enquote
  {\bibinfo {title} {Observation of {{Magnetic Excitations}} of {{Skyrmion
  Crystal}} in a {{Helimagnetic Insulator}} {Cu}$_2${OSeO}$_3$},}\ }\href
  {\doibase 10.1103/PhysRevLett.109.037603} {\bibfield  {journal} {\bibinfo
  {journal} {Phys. Rev. Lett.}\ }\textbf {\bibinfo {volume} {109}},\ \bibinfo
  {pages} {037603} (\bibinfo {year} {2012})}\BibitemShut {NoStop}%
\bibitem [{\citenamefont {Du}\ \emph {et~al.}(2017)\citenamefont {Du},
  \citenamefont {Sar}, \citenamefont {Zhou}, \citenamefont {Upadhyaya},
  \citenamefont {Casola}, \citenamefont {Zhang}, \citenamefont {Onbasli},
  \citenamefont {Ross}, \citenamefont {Walsworth}, \citenamefont
  {Tserkovnyak},\ and\ \citenamefont
  {Yacoby}}]{duControlLocalMeasurement2017a}%
  \BibitemOpen
  \bibfield  {author} {\bibinfo {author} {\bibfnamefont {Chunhui}\ \bibnamefont
  {Du}}, \bibinfo {author} {\bibfnamefont {Toeno van~der}\ \bibnamefont {Sar}},
  \bibinfo {author} {\bibfnamefont {Tony~X.}\ \bibnamefont {Zhou}}, \bibinfo
  {author} {\bibfnamefont {Pramey}\ \bibnamefont {Upadhyaya}}, \bibinfo
  {author} {\bibfnamefont {Francesco}\ \bibnamefont {Casola}}, \bibinfo
  {author} {\bibfnamefont {Huiliang}\ \bibnamefont {Zhang}}, \bibinfo {author}
  {\bibfnamefont {Mehmet~C.}\ \bibnamefont {Onbasli}}, \bibinfo {author}
  {\bibfnamefont {Caroline~A.}\ \bibnamefont {Ross}}, \bibinfo {author}
  {\bibfnamefont {Ronald~L.}\ \bibnamefont {Walsworth}}, \bibinfo {author}
  {\bibfnamefont {Yaroslav}\ \bibnamefont {Tserkovnyak}}, \ and\ \bibinfo
  {author} {\bibfnamefont {Amir}\ \bibnamefont {Yacoby}},\ }\bibfield  {title}
  {\enquote {\bibinfo {title} {Control and local measurement of the spin
  chemical potential in a magnetic insulator},}\ }\href {\doibase
  10.1126/science.aak9611} {\bibfield  {journal} {\bibinfo  {journal}
  {Science}\ }\textbf {\bibinfo {volume} {357}},\ \bibinfo {pages} {195}
  (\bibinfo {year} {2017})}\BibitemShut {NoStop}%
\bibitem [{\citenamefont {Purser}\ \emph {et~al.}(2020)\citenamefont {Purser},
  \citenamefont {Bhallamudi}, \citenamefont {Guo}, \citenamefont {Page},
  \citenamefont {Guo}, \citenamefont {Fuchs},\ and\ \citenamefont
  {Hammel}}]{purserSpinwaveDetectionNitrogenvacancy2020}%
  \BibitemOpen
  \bibfield  {author} {\bibinfo {author} {\bibfnamefont {C.~M.}\ \bibnamefont
  {Purser}}, \bibinfo {author} {\bibfnamefont {V.~P.}\ \bibnamefont
  {Bhallamudi}}, \bibinfo {author} {\bibfnamefont {F.}~\bibnamefont {Guo}},
  \bibinfo {author} {\bibfnamefont {M.~R.}\ \bibnamefont {Page}}, \bibinfo
  {author} {\bibfnamefont {Q.}~\bibnamefont {Guo}}, \bibinfo {author}
  {\bibfnamefont {G.~D.}\ \bibnamefont {Fuchs}}, \ and\ \bibinfo {author}
  {\bibfnamefont {P.~C.}\ \bibnamefont {Hammel}},\ }\bibfield  {title}
  {\enquote {\bibinfo {title} {Spinwave detection by nitrogen-vacancy centers
  in diamond as a function of probe–sample separation},}\ }\href {\doibase
  10.1063/1.5141921} {\bibfield  {journal} {\bibinfo  {journal} {Appl. Phys.
  Lett.}\ }\textbf {\bibinfo {volume} {116}},\ \bibinfo {pages} {202401}
  (\bibinfo {year} {2020})}\BibitemShut {NoStop}%
\bibitem [{\citenamefont {Jersch}\ \emph {et~al.}(2010)\citenamefont {Jersch},
  \citenamefont {Demidov}, \citenamefont {Fuchs}, \citenamefont {Rott},
  \citenamefont {Krzysteczko}, \citenamefont {Münchenberger}, \citenamefont
  {Reiss},\ and\ \citenamefont
  {Demokritov}}]{jerschMappingLocalizedSpinwave2010}%
  \BibitemOpen
  \bibfield  {author} {\bibinfo {author} {\bibfnamefont {J.}~\bibnamefont
  {Jersch}}, \bibinfo {author} {\bibfnamefont {V.~E.}\ \bibnamefont {Demidov}},
  \bibinfo {author} {\bibfnamefont {H.}~\bibnamefont {Fuchs}}, \bibinfo
  {author} {\bibfnamefont {K.}~\bibnamefont {Rott}}, \bibinfo {author}
  {\bibfnamefont {P.}~\bibnamefont {Krzysteczko}}, \bibinfo {author}
  {\bibfnamefont {J.}~\bibnamefont {Münchenberger}}, \bibinfo {author}
  {\bibfnamefont {G.}~\bibnamefont {Reiss}}, \ and\ \bibinfo {author}
  {\bibfnamefont {S.~O.}\ \bibnamefont {Demokritov}},\ }\bibfield  {title}
  {\enquote {\bibinfo {title} {Mapping of localized spin-wave excitations by
  near-field {Brillouin} light scattering},}\ }\href {\doibase
  10.1063/1.3502599} {\bibfield  {journal} {\bibinfo  {journal} {Appl. Phys.
  Lett.}\ }\textbf {\bibinfo {volume} {97}},\ \bibinfo {pages} {152502}
  (\bibinfo {year} {2010})}\BibitemShut {NoStop}%
\bibitem [{\citenamefont {Hirosawa}\ \emph {et~al.}(2025)\citenamefont
  {Hirosawa}, \citenamefont {Mook},\ and\ \citenamefont
  {Azhar}}]{hirosawa_2025_14978487}%
  \BibitemOpen
  \bibfield  {author} {\bibinfo {author} {\bibfnamefont {Tomoki}\ \bibnamefont
  {Hirosawa}}, \bibinfo {author} {\bibfnamefont {Alexander}\ \bibnamefont
  {Mook}}, \ and\ \bibinfo {author} {\bibfnamefont {Maria}\ \bibnamefont
  {Azhar}},\ }\href {\doibase 10.5281/zenodo.14978487} {\enquote {\bibinfo
  {title} {Z2 vortex crystals and topological magnons in a tetrahedral
  antiferromagnet},}\ } (\bibinfo {year} {2025})\BibitemShut {NoStop}%
\end{thebibliography}
\end{document}

\title{
Supplementary Material: \\ $\mathbb{Z}_2$ Vortex Crystals in Tetrahedral Antiferromagnets: Fractional Charges and Topological Magnons
%$\mathbb{Z}_2$ Vortex Crystals and Topological Magnons in a Tetrahedral Antiferromagnet
}
\author{Tomoki Hirosawa}
\affiliation{Department of Physical Sciences, Aoyama Gakuin University, Sagamihara, Kanagawa 252-5258, Japan}
\author{Alexander Mook}
\affiliation{University of Münster, Institute of Solid State Theory, 48149 Münster, Germany}
\affiliation{Institute of Physics, Johannes Gutenberg University Mainz, Mainz D-55128, Germany}
\author{Maria Azhar}
\affiliation{Faculty of Physics and Center for Nanointegration Duisburg-Essen (CENIDE), University of Duisburg-Essen, 47057 Duisburg, Germany}

\maketitle

\tableofcontents

\section{The uniform tetrahedral phase}

\subsection{Energies calculated using a simple spin model}

We consider a simple model defined on a triangular lattice that includes only nearest-neighbor interactions. The Hamiltonian is expressed as

\begin{equation}
H_\text{BQ}=
\sum_{\braket{\vec{r},\vec{r}'}}J_1\vec{S}_{\vec{r}}\cdot\vec{S}_{\vec{r}'} +B_1(\vec{S}_{\vec{r}}\cdot\vec{S}_{\vec{r}'})^2.
\end{equation}
	
\begin{figure}[htbp]
\includegraphics[width=0.6\textwidth]{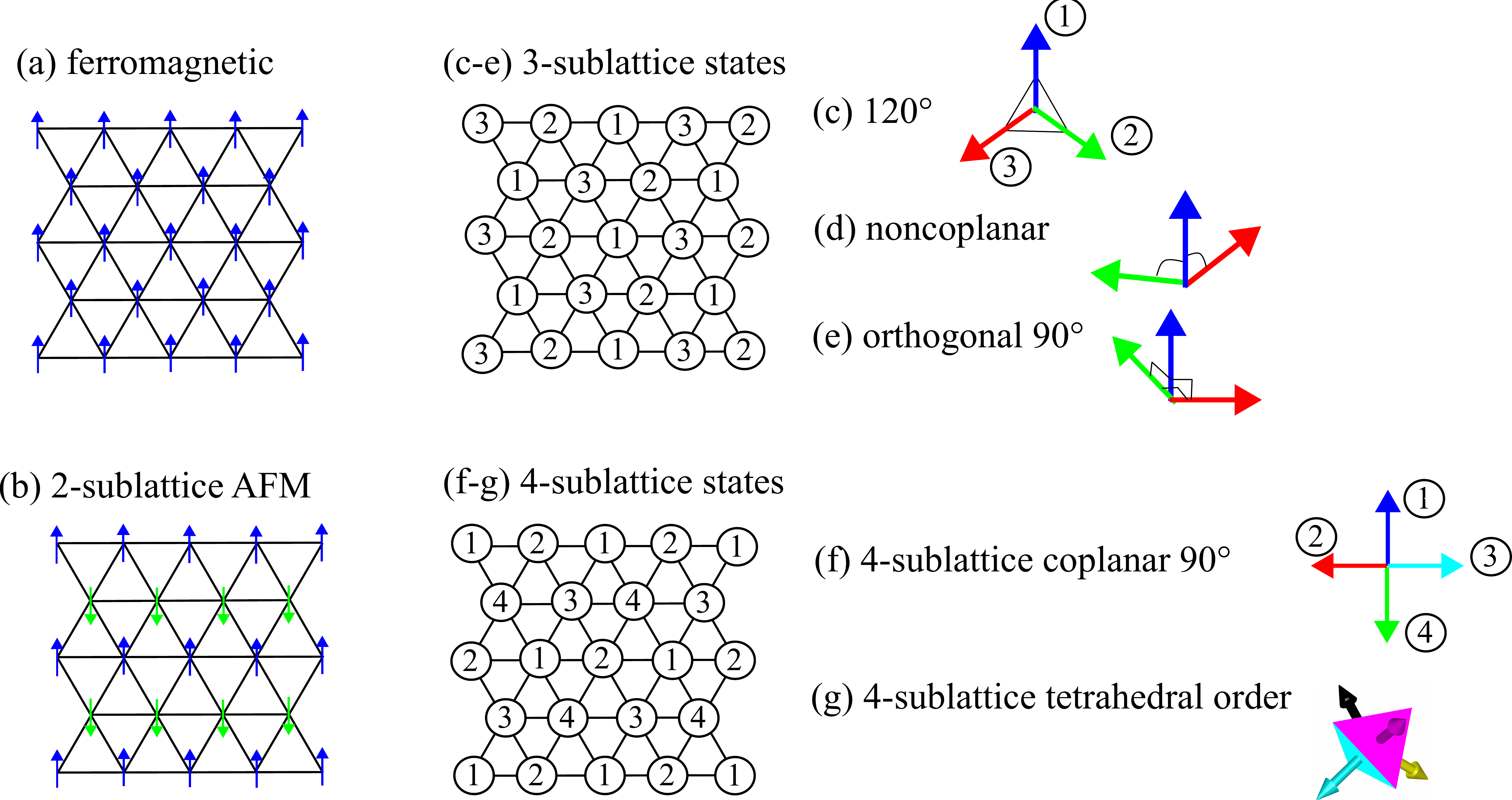}
	\caption{Visualization of the ordered spin states considered in our energy calculations. These include (a) the ferromagnetic state, (b) the two-sublattice collinear antiferromagnetic order, (c) the three-sublattice coplanar \(120^\circ\) order, (d) the three-sublattice noncoplanar order with a uniform nearest-neighbor dot product \(\mathbf{S}_r \cdot \mathbf{S}_{r'}\), (e) the three-sublattice orthogonal 90$^\circ$ order (g) the four-sublattice tetrahedral order, and (f) the four-sublattice 90° order. The spin configurations are represented in a schematic manner, highlighting their respective lattice structures and orientations. These states correspond to those whose energies are computed and analyzed in Table \ref{table:table1} and Fig. \ref{fig:figs2}.}

	\label{fig:figs2a}
\end{figure}

\begin{figure}[htbp]
	\includegraphics[width=0.7\textwidth]{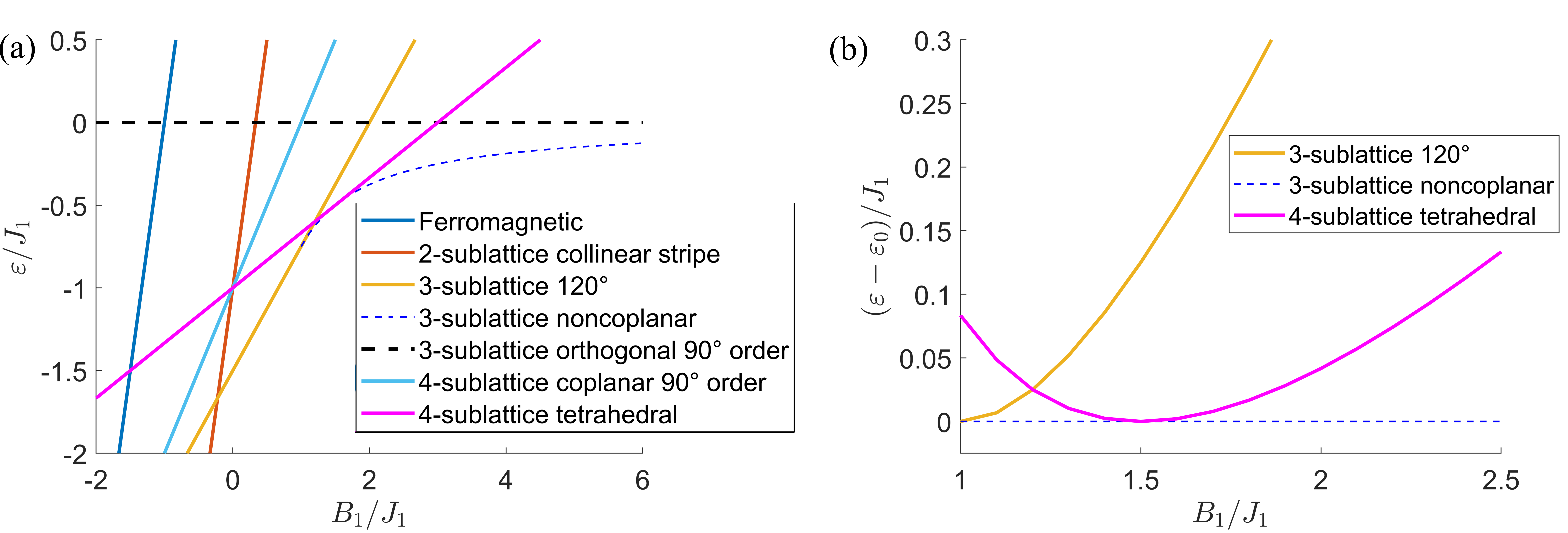}
	\caption{
		(a) Energy per lattice site normalized by $J_1$ ($\varepsilon/J_1$) plotted using the expressions from Table \ref{table:table1}. (b) Comparison of the energies of three states: the tetrahedral state, the $120^\circ$ state, and the 3-sublattice noncoplanar state (respectively states (g), (c) and (d)) measured relative to the energy of state (d). 
	}
	\label{fig:figs2}
\end{figure}

The energy per lattice site, denoted as $\varepsilon$, is computed for various ordered states and summarized in Table \ref{table:table1}. The ordered states under consideration include: (a) the ferromagnetic state, (b) the 2-sublattice collinear antiferromagnetic order, (c) the 3-sublattice coplanar 120$^\circ$ order, (d) the  3-sublattice noncoplanar order with a uniform value of $\bm{S}_{\vec{r}}\cdot\bm{S}_{\vec{r}'}$, (e) the  3-sublattice orthogonal 90$^\circ$ order, (f) the 4-sublattice 90$^\circ$ order and (g) the 4-sublattice tetrahedral order. These states are illustrated in Fig. \ref{fig:figs2a}, and their corresponding energy per site is plotted in Fig. \ref{fig:figs2}.

\begin{table}[h]
	\begin{tabular}{| l |l |}
		\hline
		State &  $\varepsilon$ \\ 
		\hline
		(a) Ferromagnetic & $3J_1+3B_1$  \\  
 		(b) 2-sublattice collinear stripe   & $-J_1+3B_1$ \\       
		(c) 3-sublattice 120$^{\circ}$   & $-3J_1/2+3B_1/4$ \\        
		(d) 3-sublattice noncoplanar ($\bm{S}_{\vec{r}}\cdot\bm{S}_{\vec{r}'}=\mathrm{uniform}$) & $-3J_1^2/4B_1$ \\
 		(e) 3-sublattice orthogonal 90$^\circ$ order & 0 \\       
		(f) 4-sublattice coplanar 90$^\circ$ order & $-J_1+B_1$ \\
		(g) 4-sublattice tetrahedral &  $-J_1+B_1/3$ \\

		\hline        
	\end{tabular}
      \caption{Energy per lattice site $\varepsilon$ for various states. }  
           \label{table:table1}
\end{table}

For $B_1>J_1$, the 3-sublattice noncoplanar order (d)  becomes the ground state. To compute the energy of this state, without loss of generality, we assume the spin configuration $\vec{S}_p=(\cos\phi_p\sin\Theta,\sin\phi_p\sin\Theta,\cos\Theta)$ with $\phi_p=2\pi p/3$, where $p$ is an index labeling the sublattice, and $p=1,2,3$. The dot product between any pair of nearest neighbors is then given by $\bm{S}_i\cdot\bm{S}_j=-\frac{1}{2}(1-3\cos^2\Theta)$. Minimizing the energy yields $\bm{S}_i\cdot\bm{S}_j=-J_1/2B_1$, resulting in $\Theta=\arccos\sqrt{(-J_1/B_1+1)/3}$. In the limit $B_1\gg J_1 $, the 90 degree 3-sublattice order (e) forms with $\Theta=\arccos\sqrt{1/3}$.  At $B_1=1.5J_1$, the tetrahedral state (g) has the lowest energy. For $B_1= J_1$, the 120$^\circ$ ordering (c) is the ground state.

Furthermore, the introduction of a next-nearest-neighbor interaction, parameterized by $\lambda$, extends the stability range of the tetrahedral phase, as demonstrated in Fig.~2 of the main text.

\subsection{Monte Carlo simulations}
\begin{figure}[tb]
	\centering
	\includegraphics[width=0.9\textwidth]{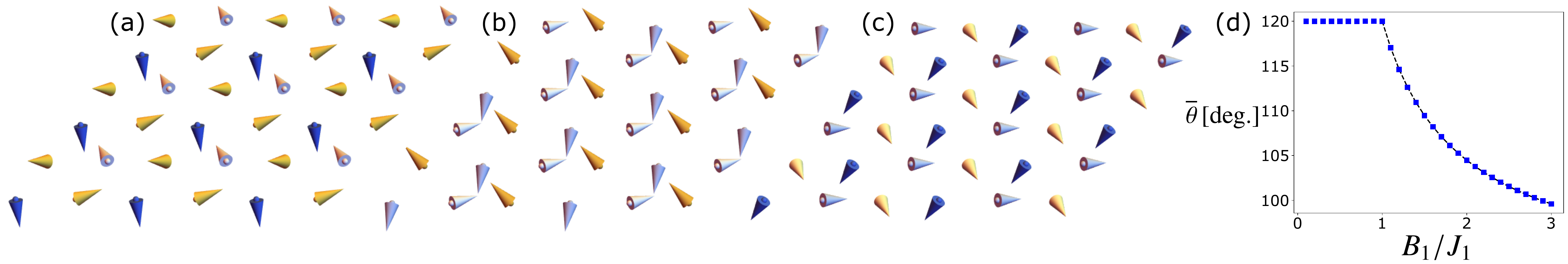}
	\caption{
    (a-c) Classical ground-state spin configurations of Eq.~\eqref{eq:biquadratic_Hamil} are plotted. They are obtained by Monte Carlo annealing of $30\times 30$ spins under periodic boundary conditions (not showing all spins). 
    The parameters are (a) $B_1/J_1=1$ and $\lambda=0.5$, (b) $B_1/J_1=1$ and $\lambda=-0.5$, and (c) $B_1/J_1=3$ and $\lambda=-0.5$.
    (d) The average angle $\overline{\theta}$ between nearest neighbor spins is plotted as a function of $B_1/J_1$ at $\lambda=-0.5$. The dashed line shows the ansatz solution $\overline{\theta}=\arccos(-J_1/2B_1)$, which fits the numerical data points well for $B_1/J_1>1$.  
    }
	\label{fig_MC1}
\end{figure}

We employ the following minimal spin lattice model on a triangular lattice~\cite{Martin2008,Akagi2012}:
\begin{align}
H
=
\sum_{\braket{\vec{r},\vec{r}'}}\left[J_1\vec{S}_{\vec{r}}\cdot\vec{S}_{\vec{r}'} 
+B_1
(\vec{S}_{\vec{r}}\cdot\vec{S}_{\vec{r}'})^2\right]
+\sum_{\braket{\vec{r},\vec{r}''}_\textrm{nn}}
\left[J_2\vec{S}_{\vec{r}}\cdot\vec{S}_{\vec{r}''} 
+B_2(\vec{S}_{\vec{r}}\cdot\vec{S}_{\vec{r}''})^2\right]-b \sum_{\vec{r}}\vec{S}_{\vec{r}}\cdot\hat{\vec{z}},
\label{eq:biquadratic_Hamil}
\end{align}
where the exchange coupling and the biquadratic interaction are included up to the next-nearest neighbor.
Setting parameters as $J_2/J_1=B_2/B_1=\lambda$ and $b=0$, we performed Monte Carlo simulations on a $30\times 30$ spin lattice under periodic boundary conditions for various values of $\lambda$ and $B_1/J_1$.
As discussed in the main text, the four-sublattice tetrahedral order is obtained as the ground state for $\lambda>0$.
Figure~\ref{fig_MC1}(a) shows the spin texture of the tetrahedral order.
In contrast, the 3-sublattice orders are stable for $\lambda\leq 0$. 
Specifically, the $120^\circ$ order is obtained as the ground state for $B_1/J_1\leq 1$, while the noncoplanar 3-sublattice order becomes stable for $B_1/J_1>1$.
Figures~\ref{fig_MC1}(b) and \ref{fig_MC1}(c) show the spin textures of the 120$^\circ$ order and the noncoplanar 3-sublattice order, respectively.
We also confirm that the variation in the average angle $\overline{\theta}$ between nearest-neighbor spins with respect to $B_1/J_1$ is consistent with the analytical ansatz $\overline{\theta}=\arccos(-J_1/2B_1)$ for $B_1/J_1>1$, as shown in Fig.~\ref{fig_MC1}(d).

\subsection{Magnetic excitations}
\label{sec: magnetic_excitations}
To obtain spin wave excitations of the uniform tetrahedral phase, we employ the Holstein-Primakoff expansion. For this purpose, we have rewritten the Hamiltonian as
\begin{equation}
	H
	=
	\sum_{\braket{\vec{r},\vec{r}'}}\left[\frac{J_1}{S^2}\hat{\vec{S}}_{\vec{r}}\cdot\hat{\vec{S}}_{\vec{r}'} 
	+\frac{B_1}{S^4}
	(\hat{\vec{S}}_{\vec{r}}\cdot\hat{\vec{S}}_{\vec{r}'})^2\right]
	+\sum_{\braket{\vec{r},\vec{r}''}_\textrm{nn}}
	\left[\frac{J_2}{S^2}\hat{\vec{S}}_{\vec{r}}\cdot\hat{\vec{S}}_{\vec{r}''} 
	+\frac{B_2}{S^4}(\hat{\vec{S}}_{\vec{r}}\cdot\hat{\vec{S}}_{\vec{r}''})^2\right]-\frac{b}{S} \sum_{\vec{r}}\hat{\vec{S}}_{\vec{r}}\cdot\hat{\vec{z}},
    \label{eq:biquadratic_Hamil2}
\end{equation}
with quantum spin operators $\hat{\vec{S}}_{\vec{r}}$. We note that this Hamiltonian becomes equivalent to the classical spin model in the limit of infinite $S$.
We write the spin operator as $\hat{\vec{S}}_{\vec{r}}=\textbf{S}_{\vec{r}}^1\hat{\vec{e}}_{\vec{r}}^{1}+\textbf{S}_{\vec{r}}^2\hat{\vec{e}}_{\vec{r}}^{2}+\textbf{S}_{\vec{r}}^3\hat{\vec{e}}_{\vec{r}}^{3}$.
Here, we define the local orthonormal basis $(\hat{\vec{e}}_{\vec{r}}^{1},\hat{\vec{e}}_{\vec{r}}^{2},\hat{\vec{e}}_{\vec{r}}^{3})$, where $\hat{\vec{e}}_{\vec{r}}^{3}$ is a unit vector parallel to $\vec{S}_{\vec{r}}$. 
Performing the Holstein-Primakoff transformation and truncating higher-order terms~\cite{holsteinFieldDependenceIntrinsic1940, roldan-molinaTopologicalSpinWaves2016, diazChiralMagnonicEdge2020}, the spin operators transform as
\begin{align}
\textbf{S}_{\vec{r}}^{+} \approx \sqrt{2S} a_{\vec{r}},\quad
\textbf{S}_{\vec{r}}^{-} \approx \sqrt{2S} a_{\vec{r}}^\dagger,\quad
\textbf{S}_{\vec{r}}^{3} = S - a_{\vec{r}}^\dagger a_{\vec{r}},
\end{align}
where $\textbf{S}_{\vec{r}}^{\pm} = \textbf{S}_{\vec{r}}^{1} \pm \textrm{i}\textbf{S}_{\vec{r}}^{2}$, and $a_{\vec{r}}$, $a_{\vec{r}}^\dag$ are the magnon creation and annihilation operators.
Expanding each term of Eq.~\eqref{eq:biquadratic_Hamil2} up to quadratic terms in magnon operators yields the spin wave Hamiltonian, 
\begin{align}
\mathcal{H} = \frac{1}{2S}\sum_{\vec{k}}\sum_{ij}\psi_{\vec{k} i}^\dagger H_{ij}^\text{sw}(\vec{k}) \psi_{\vec{k} j} \,,
\label{eq:sw_tetra}
\end{align}
with
\begin{align}
H_{ij}^\text{sw}(\vec{k}) =
\begin{pmatrix}
\Omega_{ij}(\vec{k}) & \Delta_{ij}(\vec{k})\\
\Delta_{ij}^*(-\vec{k}) &\Omega_{ij}^* (-\vec{k})
\end{pmatrix}\,,
\end{align}
where $\psi_{\vec{k} i}=(a_{\vec{k} i}, a^\dagger_{-\vec{k} i})^T$.
We define $a_{\vec{k} i}=1/\sqrt{N_\textrm{u}}\sum_{\vec{R}}e^{-\mathrm{i}\vec{k}\cdot(\vec{R}+\vec{r}_i)}a_{\vec{r}}$, where $N_\textrm{u}$ is the number of magnetic unit cells and $\vec{R}$ is a Bravais lattice vector.
The elements of the Hamilton matrix $H_{ij}^\text{sw}(\vec{k})$ are given by
\begin{align*}
\Omega_{ij}(\vec{k})&
=\delta_{ij}\Lambda_i(\vec{k})+B_{ij}(\vec{k})
\left\{
F_{ij}^{13}F_{ij}^{31}+\textrm{i}F_{ij}^{23}F_{ij}^{31}-\textrm{i}F_{ij}^{13}F_{ij}^{32}+F_{ij}^{23}F_{ij}^{32}+F_{ij}^{33}(F_{ij}^{11}-\textrm{i}F_{ij}^{12}+\textrm{i}F_{ij}^{21}+F_{ij}^{22})
\right\}\\
&+\frac{J_{ij}(\vec{k})}{2}(F_{ij}^{11}-\textrm{i}F_{ij}^{12}+\textrm{i}F_{ij}^{21}+F_{ij}^{22}),
\\
\Lambda_i(\vec{k})
&=\sum_{j\ne i}B_{ij}(\vec{k}=0)\left\{(F_{ij}^{13})^2+(F_{ij}^{23})^2-2(F_{ij}^{33})^2\right\}-J_{ij}(\vec{k}=0)F_{ij}^{33}+b\hat{\vec{e}}^3_{\vec{r}_i}\cdot\hat{\vec{z}}\,
\\
\Delta_{ij}(\vec{k})
&=
\delta_{ij}\Lambda'_i(\vec{k})
+B_{ij}(\vec{k})
\left\{F_{ij}^{13}F_{ij}^{31}+\textrm{i}F_{ij}^{23}F_{ij}^{31}+\textrm{i}F_{ij}^{13}F_{ij}^{32}-F_{ij}^{23}F_{ij}^{32}+F_{ij}^{33}(F_{ij}^{11}+\textrm{i}F_{ij}^{12}+\textrm{i}F_{ij}^{21}-F_{ij}^{22})\right\}\\
&+\frac{J_{ij}(\vec{k})}{2}(F_{ij}^{11}+\textrm{i}F_{ij}^{12}+\textrm{i}F_{ij}^{21}-F_{ij}^{22}),
\\
\Lambda'_i(\vec{k})
&=\sum_{j\ne i}B_{ij}(\vec{k}=0)
\left\{(F_{ij}^{13})^2+2\textrm{i}F_{ij}^{13}F_{ij}^{23}-(F_{ij}^{23})^2\right\},
\end{align*}
where $F_{ij}^{ab}=\hat{\vec{e}}^a_{\vec{r}_i}\cdot\hat{\vec{e}}^b_{\vec{r}_j}$. We define the Fourier transforms of the exchange and biquadratic interactions as
\begin{align*}
J_{ij}(\vec{k})
&=
\sum_{\vec{R}}\Big[
J_1\Big(\delta_{\vec{R}+\delta\vec{r}_{ij},\,\pm\vec{a}_1}+\delta_{\vec{R}+\delta\vec{r}_{ij},\,\pm\vec{a}_2}+\delta_{\vec{R}+\delta\vec{r}_{ij},\,\pm(\vec{a}_1-\vec{a}_2})\Big)\\
&+J_2\Big(\delta_{\vec{R}+\delta\vec{r}_{ij},\,\pm(\vec{a}_1+\vec{a}_2)}+\delta_{\vec{R}+\delta\vec{r}_{ij},\,\pm(-\vec{a}_1+2\vec{a}_2)}+\delta_{\vec{R}+\delta\vec{r}_{ij},\,\pm(-2\vec{a}_1+\vec{a}_2})\Big)
\Big]e^{-\mathrm{i}\vec{k}\cdot(\vec{R}+\delta\vec{r}_{ij})}
,
\\
B_{ij}(\vec{k})
&=
\sum_{\vec{R}}\Big[
B_1\Big(\delta_{\vec{R}+\delta\vec{r}_{ij},\,\pm\vec{a}_1}+\delta_{\vec{R}+\delta\vec{r}_{ij},\,\pm\vec{a}_2}+\delta_{\vec{R}+\delta\vec{r}_{ij},\,\pm(\vec{a}_1-\vec{a}_2})\Big)\\
&+B_2\Big(\delta_{\vec{R}+\delta\vec{r}_{ij},\,\pm(\vec{a}_1+\vec{a}_2)}+\delta_{\vec{R}+\delta\vec{r}_{ij},\,\pm(-\vec{a}_1+2\vec{a}_2)}+\delta_{\vec{R}+\delta\vec{r}_{ij},\,\pm(-2\vec{a}_1+\vec{a}_2})\Big)
\Big]e^{-\mathrm{i}\vec{k}\cdot(\vec{R}+\delta\vec{r}_{ij})}
,
\end{align*}
where $\vec{a}_1$ and $\vec{a}_2$ are primitive lattice vectors of the triangular lattice, and $\delta\vec{r}_{ij}=\vec{r}_{i}-\vec{r}_{j}$.

The spin-wave Hamiltonian of Eq.~\eqref{eq:sw_tetra} is diagonalized using a paraunitary matrix $T_{\vec{k}}$~\cite{colpaDiagonalizationQuadraticBoson1978}, which satisfies the bosonic commutation relations $T^\dagger_{\vec{k}}\Sigma T_{\vec{k}}=T_{\vec{k}}\Sigma T^\dagger_{\vec{k}}=\Sigma$ where
\begin{equation}
    \Sigma=
    \begin{pmatrix}
        I_{N_\textrm{b}\times N_\textrm{b}}&0
        \\
        0 & -I_{N_\textrm{b}\times N_\textrm{b}}
    \end{pmatrix},
\end{equation}
with $N_\textrm{b}$ representing the number of spins within a magnetic unit cell. The total number of spins is given by $N=N_\textrm{u}N_\textrm{b}$.

\begin{figure}[t]
	\centering
	\includegraphics[width=0.9\textwidth]{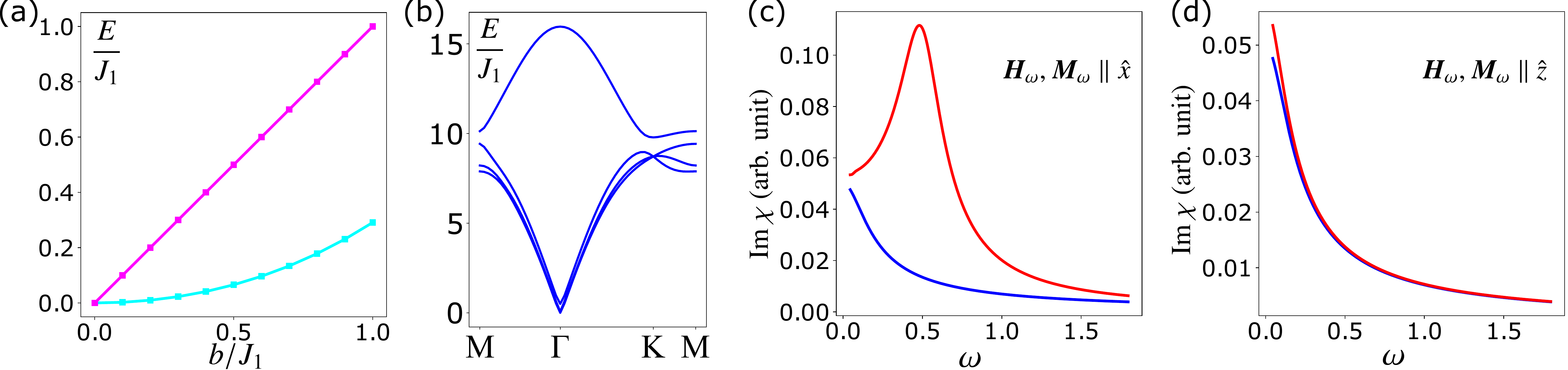}
	\caption{
    (a) The energy of the second and third lowest-energy magnon modes of the uniform tetrahedral phase is plotted against the out-of-plane magnetic field~$b$.
    (b) The magnon band structure is plotted at $b/J_1=0.5$. 
    (c,d) The imaginary part of the dynamical susceptibility $\textrm{Im}\,\chi(\omega)$ at $b/J_1=0$~(blue) and $b/J_1=0.5$~(red). 
    The response of in-plane magnetization to in-plane fields is plotted in (c), while the response of out-of-plane magnetization to out-of-plane fields is plotted in (d). In (a-d), we set $B_1/J_1 =1$ and $\lambda=0.5$.
    }
	\label{fig_SWtetra1}
\end{figure}

We find three Goldstone modes and one optical branch without magnetic fields, consistent with previously reported magnon dispersions obtained using the Kondo lattice model~\cite{Akagi2013}.
When an out-of-plane magnetic field is applied, the $\mathrm{SO}(3)$ symmetry is reduced to $\mathrm{SO}(2)$ symmetry. 
As a result, two of the Goldstone modes acquire energy gaps.
Figure~\ref{fig_SWtetra1}(a) shows the energy of these two modes as a function of the magnetic field strength, showing linear and quadratic dependencies on $b$.
The magnon mode with a linear dependence on $b$ is characterized by an intrinsic magnetic moment~\cite{mookQuantumDampingSkyrmion2020}, while the quadratic dependence implies a field-induced magnetic moment.

Since the topological spin textures often lead to the nontrivial magnon band topology~\cite{dugaevBerryPhaseMagnons2005a, vanhoogdalemMagneticTextureinducedThermal2013}, we investigate the topological properties of magnon bands in the uniform tetrahedral phase.
Denoting the magnon wave function as $\ket{u_{\vec{k}}^n}=T_{\vec{k}}\vec{v}_n$ with $v_n^j=\delta_{j,n}$, the orthogonality relation holds as $\braket{u_{\vec{k}}^m |u_{\vec{k}}^n}_\textrm{para}=\xi \vec{v}_m^T T^\dagger_{\vec{k}}\Sigma T_{\vec{k}}\vec{v}_n =\delta_{mn}$ with $\xi=\pm1$ for particle/hole bands.
The Berry curvature and Chern number of the $n$th magnon band are defined as~\cite{fukuiChernNumbersDiscretized2005, shindouTopologicalChiralMagnonic2013}
\begin{align}
   \Omega_{\vec{k}}^n
   &=\textrm{i}\left[\Braket{\frac{\partial u_{\vec{k}}^n }{\partial k_x} |\frac{\partial u_{\vec{k}}^n}{\partial k_y} }_\textrm{para}-\Braket{\frac{\partial u_{\vec{k}}^n }{\partial k_y} |\frac{\partial u_{\vec{k}}^n}{\partial k_x} }_\textrm{para}\right],\quad\quad
   C_n
   =
    \int_\textrm{BZ} \frac{d\vec{k}}{2\pi}\Omega_{\vec{k}}^n.  
\end{align}
We observe a field-induced gap between the third and fourth bands, as shown in Fig.~\ref{fig_SWtetra1}(b).
However, the Chern number below this band gap vanishes. 
Thus, it does not support topological magnons despite the tetrahedral order's topologically nontrivial spin textures.

We also investigate the magnetic activity of magnons using the Landau-Lifshitz-Gilbert~(LLG) equation:
\begin{equation}
    \frac{d\vec{S}_{\vec{r}}}{dt}=-\frac{\vec{S}_{\vec{r}}}{1+\alpha^2}\times\left[ \vec{H}_{\vec{r}}^\textrm{eff}+ \alpha\vec{S}_{\vec{r}}\times\vec{H}_{\vec{r}}^\textrm{eff} \right] \,, 
    \label{eq: general_LLG}
\end{equation}
where $\vec{H}_{\vec{r}}^\textrm{eff} = -J_1^{-1}\partial H_\textrm{BQ}/\partial \vec{S}_{\vec{r}}$, and $\alpha=0.01$ is the Gilbert damping constant.
The dimensionless time is measured in units of $J_1^{-1}$.
We apply a spatially uniform magnetic pulse and compute the imaginary part of the dynamical susceptibility~$\textrm{Im}\chi_{ab}(\omega)$~\cite{mochizukiSpinWaveModesTheir2012}.
It is defined as $\textrm{Im}\chi_{ab}(\omega)=M_a(\omega)/B_b(\omega)$, where $M_a(\omega)$ is given by the Fourier transform of $M_a(t)=\sum_{\vec{r}}S_{\vec{r},i}(t)$ and $B_b(\omega)$ is a constant for pulse fields. 
Figure~\ref{fig_SWtetra1}(c) and \ref{fig_SWtetra1}(d) show $\textrm{Im}\chi_{xx}(\omega)$ and $\textrm{Im}\chi_{zz}(\omega)$ at $b/J_1=0$~(blue) and $b/J_1=0.5$~(red), respectively.
Both $\textrm{Im}\chi_{xx}(\omega)$ and $\textrm{Im}\chi_{zz}(\omega)$ show no peak at $b/J_1=0$, indicating a lack of magnetically active excitations.
However, with the application of an external magnetic field~$b$, the third-lowest magnon mode (with energy proportional to $b$) becomes magnetically active under in-plane pulse fields, as shown in Fig~\ref{fig_SWtetra1}(c).

\subsection{Skyrmion number}
\label{sec: topo_charge}

The noncollinear spin texture is characterized by the skyrmion number, also known as the topological charge.
In a two-dimensional continuum model, it is defined as~\cite{TopologicalPropertiesDynamics2013}
\begin{equation}
N_\textrm{sk}
=
\frac{1}{4\pi} \int dr^2 \,\vec{n}(\vec{r})\cdot
\left[
\frac{\partial \vec{n}(\vec{r})}{\partial x}\times \frac{\partial\vec{n}(\vec{r})}{\partial y}
\right],
\label{eq: topo_charge1}
\end{equation}
where $\vec{n}(\vec{r})$ is a unit vector parallel to the local spin.
The skyrmion number is equal to the number of times $\vec{n}(\vec{r})$ covers the surface of a unit sphere, and is quantized to an integer under periodic boundary conditions.

In a discrete lattice system, we employ the method of Ref.~\cite{bergDefinitionStatisticalDistributions1981} to discretize Eq.~\eqref{eq: topo_charge1}. In this scheme, the skyrmion number density on the $i^{\mathrm{th}}$ triangular plaquette is defined as
\begin{equation}
\rho_{\textrm{sk},i}=\frac{\sigma S(\vec{s}^i_1,\vec{s}^i_2,\vec{s}^i_3)}{4\pi},
\label{eq: topo_charge2}
\end{equation}
where $\sigma S(\vec{s}^i_1,\vec{s}^i_2,\vec{s}^i_3)$ is the signed area of the spherical triangle formed by the three normalized spins ordered clockwise, and the sign is determined by $\sigma=\textrm{sign}[\vec{s}^i_1\cdot(\vec{s}^i_2\times \vec{s}^i_3)]$.
The total skyrmion number is defined as $N_\textrm{sk}=\sum_i \rho_{\textrm{sk},i}$, where the sum runs over all triangle plaquettes. 
We confirmed that this definition (Eq.~\eqref{eq: topo_charge2}) always takes an integer value within numerical precision.
We also define the sublattice skyrmion number as the sum of skyrmion number density of triangle plaquettes within sublattice sites, denoted as $ N_\textrm{sk}^\textrm{sub}=\sum_{i\in \textrm{sub}} \rho_{\textrm{sk},i}^\textrm{sub}$.

For the uniform tetrahedral phase, the skyrmion number density is spatially uniform and equal to $\pm \frac{1}{4}$. Since a magnetic unit cell of the tetrahedral order consists of four triangular plaquettes, the total skyrmion number is quantized to be $\pm 1$ per magnetic unit cell.

\section{$\mathbb{Z}_2$ vortex crystal phase}
\subsection{DMI-induced instability of uniform tetrahedral phase}

We introduce the interfacial Dzyaloshinskii-Moriya~(DM) interaction as a perturbation to the uniform tetrahedral phase:
\begin{equation}
    H_\textrm{DM}=\sum_{\braket{\vec{r},\vec{r}'}} \frac{D_1\,\hat{\vec{z}}\times(\vec{r}-\vec{r}')}{|\vec{r}-\vec{r}'|}\cdot(\vec{S}_{\vec{r}}\times\vec{S}_{\vec{r}'}).
\end{equation}
Since the DM interaction is antisymmetric with respect to $\vec{r}-\vec{r}'$, it is canceled in alternately canted spin configurations.
However, it has a finite contribution to the spin wave Hamiltonian.
Performing the Holstein-Primakoff expansion as detailed in Sec.~\ref{sec: magnetic_excitations}, we derive biquadratic magnon terms from the DM interaction as
\begin{align}
\mathcal{H}^\textrm{DM}_{ij} =  
\begin{pmatrix}
\Omega^\textrm{DM}_{ij}(\vec{k}) & \Delta^\textrm{DM}_{ij}(\vec{k})\\
\Delta^\textrm{DM *}_{ij}(-\vec{k}) &\Omega^\textrm{DM\,*}_{ij} (-\vec{k})
\end{pmatrix},
\label{eq:sw_tetraDM}
\end{align}
where
\begin{align}
\Omega^\textrm{DM}_{ij}(\vec{k})&
=\frac{D_{ij}^c(\vec{k})}{2}(V_{ij}^{11c}-\textrm{i}V_{ij}^{12c}+\textrm{i}V_{ij}^{21c}+V_{ij}^{22c}),
\end{align}
\begin{align}
\Delta^\textrm{DM}_{ij}(\vec{k})
&=
\frac{D_{ij}^c(\vec{k})}{2}(V_{ij}^{11c}+\textrm{i}V_{ij}^{12c}+\textrm{i}V_{ij}^{21c}-V_{ij}^{22c}),
\end{align}
\begin{equation}
\vec{D}_{ij}(\vec{k})=\sum_{\vec{R}}D_1\vec{\hat{z}}\times(\vec{R}+\delta\vec{r}_{ij})\Big(\delta_{\vec{R}+\delta\vec{r}_{ij},\,\pm\vec{a}_1}+\delta_{\vec{R}+\delta\vec{r}_{ij},\,\pm\vec{a}_2}\Big)e^{-\mathrm{i}\vec{k}\cdot(\vec{R}+\delta\vec{r}_{ij})},
\end{equation}
and $V_{ij}^{abc}=(\hat{e}_{\vec{r}_i}^a\times \hat{e}^b_{\vec{r}_j})^c$.

The spin wave Hamiltonian $H^\text{sw}(\vec{k})$ must be a positive semi-definite matrix for well-defined spin wave excitations, ensuring all eigenvalues are non-negative. 
A negative eigenvalue of $H^\text{sw}(\vec{k})$ indicates the instability of the uniform tetrahedral phase.
Here, we use perturbation theory to analyze the zero mode of $H^\text{sw}(\vec{k})$.
For simplicity, we assume $b=0$ in the following.
From Eq.~\eqref{eq:sw_tetra}, the block matrices of the spin wave Hamiltonian at $\vec{k}=0$ are:
\begin{align*}
\Omega_{ij}(\vec{k}=0)
=(10B-3J)\begin{pmatrix}
%first
\frac{4B+2J}{10B-3J}&
\frac{-1-\sqrt{3}\textrm{i}}{9}
&
\frac{2}{9}
&
\frac{-1-\sqrt{3}\textrm{i}}{9}
\\
%second
\frac{-1+\sqrt{3}\textrm{i}}{9}
&
\frac{4B+2J}{10B-3J}
&
\frac{-1+\sqrt{3}\textrm{i}}{9}
&
\frac{2}{9}
\\
%third
\frac{2}{9}
&
\frac{-1-\sqrt{3}\textrm{i}}{9}
&
\frac{4B+2J}{10B-3J}
&
\frac{-1-\sqrt{3}\textrm{i}}{9}
\\
%fourth
\frac{-1+\sqrt{3}\textrm{i}}{9}
&
\frac{2}{9}
&
\frac{-1+\sqrt{3}\textrm{i}}{9}
&
\frac{4B+2J}{10B-3J}
\end{pmatrix},
\end{align*}
\begin{align*}
\Delta_{ij}(\vec{k}=0)
=(2B+3J)
\begin{pmatrix}
%first
0
&
\frac{2(1-\sqrt{3}\textrm{i})}{9}
&
\frac{2(1+\sqrt{3}\textrm{i})}{9}
&
\frac{2(1+\sqrt{3}\textrm{i})}{9}
\\
%second
\frac{2(1-\sqrt{3}\textrm{i})}{9}
&
0
&
\frac{2(1+\sqrt{3}\textrm{i})}{9}
&
\frac{2(1-\sqrt{3}\textrm{i})}{9}
\\
%third
\frac{2(1+\sqrt{3}\textrm{i})}{9}
&
\frac{2(1+\sqrt{3}\textrm{i})}{9}
&
0
&
\frac{2(1-\sqrt{3}\textrm{i})}{9}
\\
%fourth
\frac{2(1+\sqrt{3}\textrm{i})}{9}
&
\frac{2(1-\sqrt{3}\textrm{i})}{9}
&
\frac{2(1-\sqrt{3}\textrm{i})}{9}
&
0
\end{pmatrix},
\end{align*}
where $J=J_1+J_2$ and $B=B_1+B_2$. 
One of the zero modes is given by 
\begin{equation*}
\vec{v}_{0}=\frac{1}{\sqrt{6}}\left(\frac{1}{2},\frac{-1+\sqrt{3}\textrm{i}}{4},\frac{1}{2},\frac{3(1-\sqrt{3}\textrm{i})}{4},\frac{1+\sqrt{3}\textrm{i}}{2},-1,-1,0\right).
\end{equation*}
Although $\vec{v}_{0}$ does not satisfy the bosonic orthogonality relation (as $\vec{v}_{0}^\dagger\Sigma\vec{v}_{0}=0$), an orthogonal basis can be constructed using $\vec{v}_0$ and $\vec{\nu}_0$, that satisfies $\Sigma H^\textrm{sw}\vec{\nu}_{0}=-\textrm{i}\vec{v}_{0}/\mu$ ($\mu>0$)~\cite{alma991025119919703276}.

While the exchange and biquadratic interactions are symmetric in $\vec{k}$, the DM interaction leads to antisymmetric terms in the spin wave Hamiltonian.
Thus, the DM interaction contributes the lowest order term in the Taylor expansion at $k=0$.
The first order perturbation correction to the eigenvalue of the zero mode near $k=0$ is: 
\begin{equation}
    \Delta \beta=\vec{v}_{0}^\dagger \mathcal{H}^\textrm{sw}(\vec{k})\vec{v}_{0}\approx\frac{D_1\big[(3+\sqrt{3})k_y-2k_x\big]}{\sqrt{3}}.
    \label{eq_deltaE_DM}
\end{equation}
We note that $\Delta \beta$ depends on the directions of spins of the uniform tetrahedral phase, which is degenerate under the $\mathrm{SO}(3)$ rotation. 
The above expression is obtained with $\vec{S}_1=(1,1,1)/\sqrt{3}$, $\vec{S}_2=(1,-1,-1)/\sqrt{3}$, $\vec{S}_3=(-1,1,-1)/\sqrt{3}$, and $\vec{S}_4=(-1,-1,1)/\sqrt{3}$.
The linear terms in $k_x$ and $k_y$ give antisymmetric corrections to the eigenvalue of zero modes. 
Consequently, one eigenvalue of $\mathcal{H}^\textrm{sw}$ becomes negative at $k\ne 0$ under infinitesimal DM interaction, indicating instability in the uniform tetrahedral phase.

\subsection{Comparison of the sublattice spin texture with the ferromagnetic skyrmion crystal}

\begin{figure}[t]
	\centering
	\includegraphics[width=0.9\textwidth]{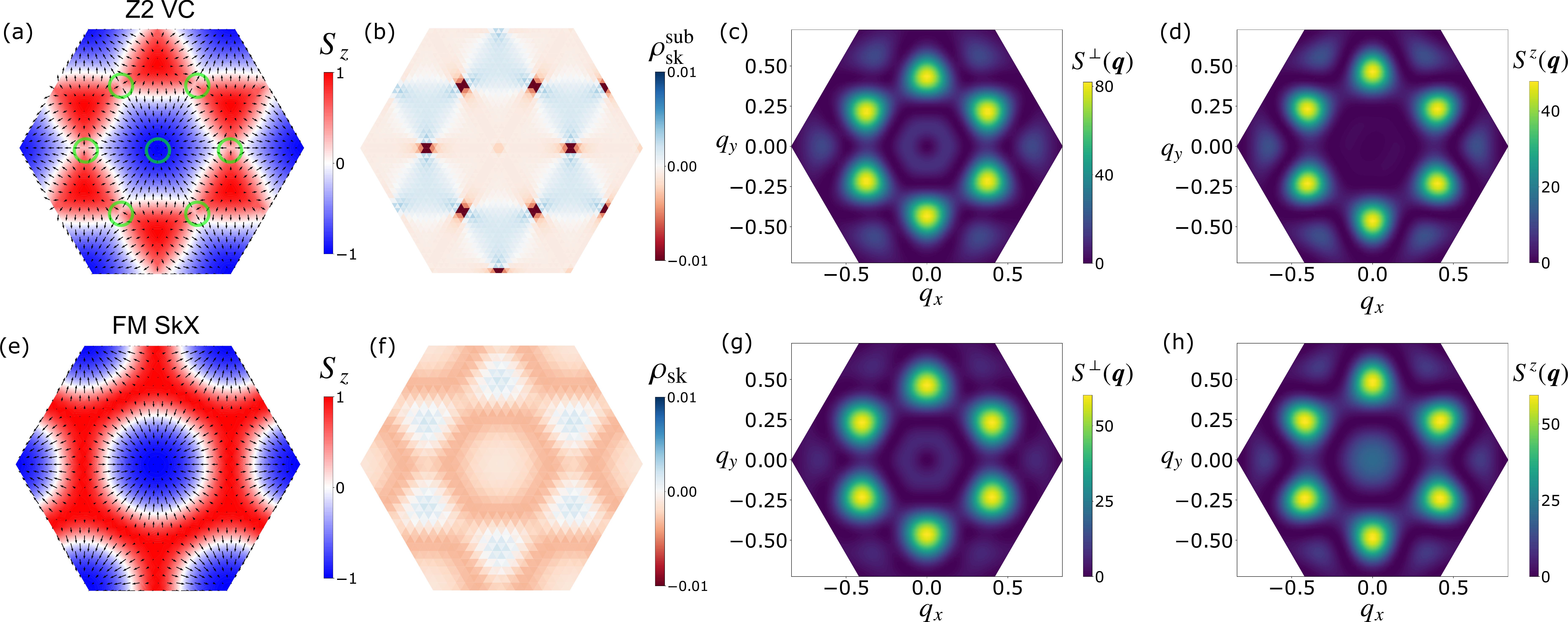}
	\caption{
    Spin textures, skyrmion number density, and static spin structure factors of (a-d) sublattice spins of the $\mathbb{Z}_2$ vortex crystal, only showing one of the four sublattice spins, and (e-h) the ferromagnetic skyrmion crystal.
    In (a) and (e), the color represents the out-of-plane spin component. Positions of $\mathbb{Z}_2$ vortices are indicated by green circles in (a).
    The in-plane static structure factor is plotted in (c) and (g), while the out-of-plane component is plotted in (d) and (h). 
    }
	\label{fig_comparison}
\end{figure}

In this section, we analyze the spin texture of a single sublattice extracted from the  $\mathbb{Z}_2$ VC. 
As discussed in the main text, this sublattice spin texture resembles that of a ferromagnetic skyrmion crystal.
To highlight this similarity, we compare the two spin textures in Fig.~\ref{fig_comparison}(a) and (e).
We note that the spin configuration of the $\mathbb{Z}_2$ vortex crystal was obtained by Monte Carlo annealing with a smaller number of spins than those shown in the manuscript for clarity ($34\times 34$ spins inside a magnetic unit cell).
The ferromagnetic skyrmion crystal is obtained by employing the spin-lattice model described in Ref.~\cite{diazChiralMagnonicEdge2020}.
Both spin textures are relaxed at zero magnetic fields by solving the LLG equation.
Compared to the skyrmion crystals, the skyrmion-like hexagonal domains in the $\mathbb{Z}_2$ VC are more densely packed, with their corners forming singular defects.
The sublattice spin texture of the $\mathbb{Z}_2$  VC can be qualitatively understood as a limiting case of a skyrmion crystal, where individual skyrmions have expanded until they come into contact at their boundaries, with singular points forming at the contact regions.
By applying the magnetic field parallel to the magnetization of skyrmion cores in the skyrmion crystal, we could partially reproduce this scenario. 
However, this configuration is not stable in the ferromagnetic skyrmion crystal due to the significant energy cost from ferromagnetic exchange interactions.

Figure~\ref{fig_comparison}(b) shows the skyrmion number density of sublattice spins in the $\mathbb{Z}_2$ VC, denoted as $\rho_\textrm{sk}^\textrm{sub}$~(see Sec.~\ref{sec: topo_charge} for definition).
Interestingly, the skyrmion number density is strongly enhanced at singular defects with approximately $-1/3$ per defect. 
The triangular and hexagonal domains have almost equal and oppositely signed total skyrmion numbers of $\pm 1/2$. 
As a result, the net skyrmion number of sublattice spins is $N_\textrm{sk}^\textrm{sub}=-1$ per magnetic unit cell of the $\mathbb{Z}_2$ VC due to contributions from singular defects. 
The total skyrmion number ($N_\textrm{sk}$) of the skyrmion crystal is equal to that of sublattice spins in the $\mathbb{Z}_2$ VC.
However, the spatial distribution of the skyrmion number density differs significantly between the two systems: in the skyrmion crystal, the skyrmion number density is spread out widely across the skyrmion cores, as shown in Fig.~\ref{fig_comparison}(f).

We also introduce the static structure factor as
\begin{align*}
	S_p^{\mu\nu}(\vec{q})
	=
	\left|\sum_{\vec{r}_p,\vec{r}'_p} e^{-\textrm{i}\vec{q}\cdot(\vec{r}_p-\vec{r}'_p)}S^{\mu}_{\vec{r}_p}S^{\nu}_{\vec{r}'_p}\right|,
\end{align*}
where $\mu,\nu=x,y,z$ and the summation is taken over the hexagonal periodic cell, as shown in Fig.~\ref{fig_comparison}(a) and (e). Here, $\vec{r}_p$ represents the lattice sites belonging to the $p$-th sublattice in the $\mathbb{Z}_2$ vortex crystal and all lattice sites in the ferromagnetic skyrmion crystal. 
Figure~\ref{fig_comparison} shows the in-plane and out-of-plane static structure factors of (c,d) the $\mathbb{Z}_2$ VC and (g,h) the ferromagnetic skyrmion crystal, which are defined as $S^\perp(\vec{q})=S_1^{xx}(\vec{q})+S_1^{yy}(\vec{q})$ and $S^z(\vec{q})=S_1^{zz}(\vec{q})$, respectively.
Both the $\mathbb{Z}_2$ VC and the ferromagnetic skyrmion crystal exhibit the triple-$Q$ structure in the in-plane and out-of-plane components. 
The main difference is that there is no peak at $\vec{q}=0$ in $S^z(\vec{q})$ of the $\mathbb{Z}_2$ VC while the skyrmion crystal exhibits a broad peak at $\vec{q}=0$.
This absence of a $\vec{q}=0$ peak in the $\mathbb{Z}_2$ VC is consistent with its vanishing net magnetization and its stabilization at zero magnetic field.

\subsection{Calculation of $\mathbb{Z}_2$ vorticity}

\begin{figure}[t]
	\centering
	\includegraphics[width=0.3\textwidth]{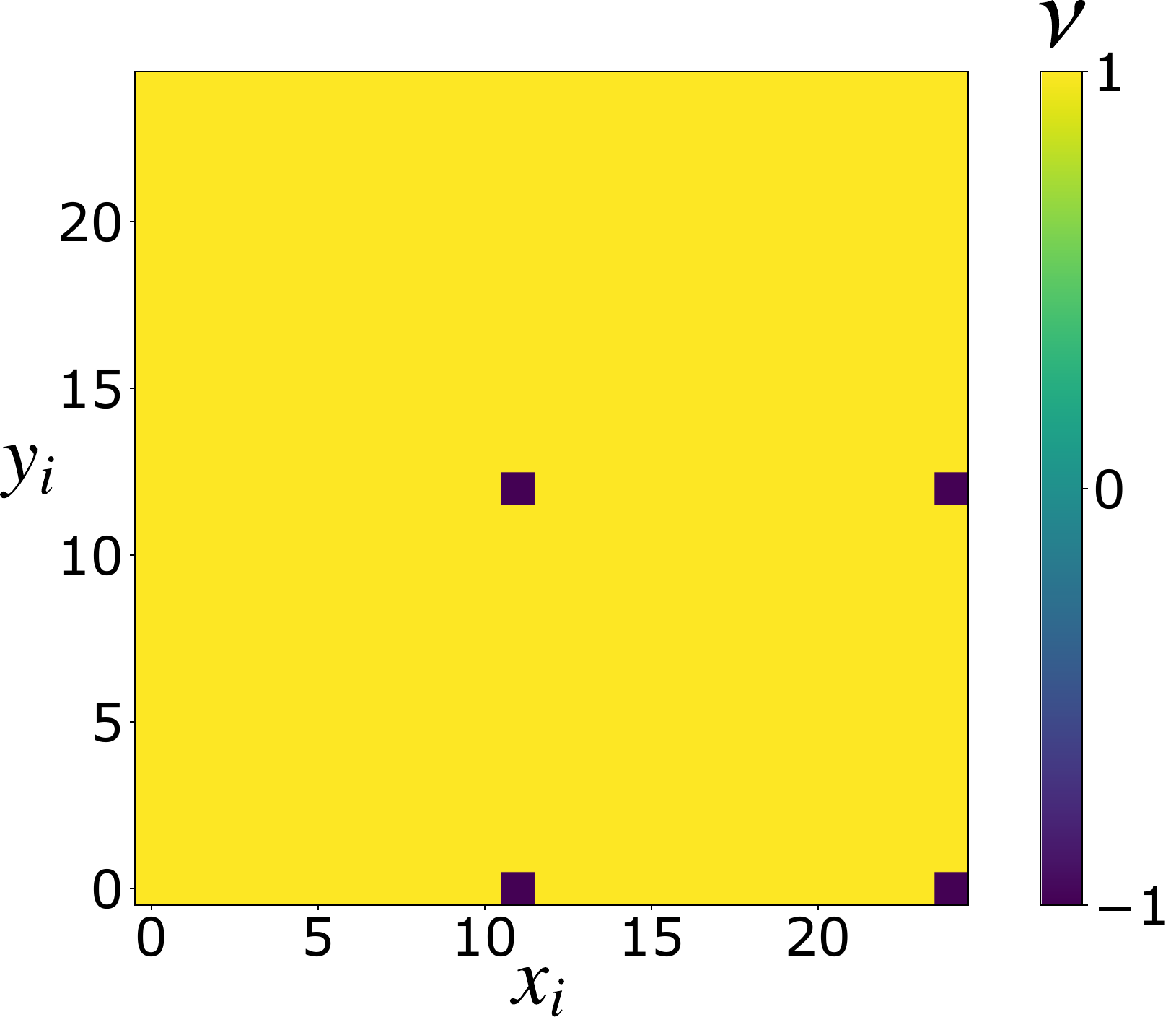}
	\caption{
    The $\mathbb{Z}_2$ vorticity distribution is shown for the $\mathbb{Z}_2$ VC phase at $D_1/J_1=0.5$ and $b/J_1=0.0$. The spin configuration is shown in Fig.~1 of the main text. The horizontal and vertical axes represent the index of lattice points for $25\times 25$ sublattice spins. 
    }
	\label{fig_Z2plot}
\end{figure}

The $\mathbb{Z}_2$ vorticity is computed from the rotation of the $\mathrm{SO}(3)$ order parameter~\cite{Kawamura1984}.
As discussed in the main text, we assume that the $\mathbb{Z}_2$ VCs consist of rigid tetrahedra and use two orthogonal vectors $\hat{\vec{a}}_{i}=\hat{\vec{S}}_{i1}$ and $\hat{\vec{b}}_{i}=\frac{\hat{\vec{S}}_{i1}\times (\hat{\vec{S}}_{i2}-\hat{\vec{S}}_{i3})}{|\,\hat{\vec{S}}_{i1}\times (\hat{\vec{S}}_{i2}-\hat{\vec{S}}_{i3})|}$ to define their rotations~\cite{Barts2021}. 
Firstly, we compute the rotation matrix $R_{ij}^a$ satisfying $\hat{\vec{a}}'_{i}=R_{ij}^a\hat{\vec{a}}_{i}=\hat{\vec{a}}_{j}$, where the rotation axis and the angle of rotation are given by $\vec{\hat{n}}_{ij}^a=\frac{\hat{\vec{a}}_{i}\times\hat{\vec{a}}_{j}}{|\hat{\vec{a}}_{i}\times\hat{\vec{a}}_{j}|}$ and $\omega_{ij}^a=\arccos(\hat{\vec{a}}_{i}\cdot\hat{\vec{a}}_{j})$, respectively.
With $\hat{\vec{b}}'_{i}=R_{ij}^a\hat{\vec{b}}_{i}$, we also compute the rotation matrix $R_{ij}^b$ satisfying $R_{ij}^b\hat{\vec{b}}'_{i}=\hat{\vec{b}}_{j}$.
Since both $\hat{\vec{b}}'_i$ and $\hat{\vec{b}}_j$ are orthogonal to $\hat{\vec{a}}_j$, the axis of rotation $\vec{\hat{n}}_{ij}^b$ is parallel to $\hat{\vec{a}}_{j}$.
Finally, the $\mathrm{SO}(3)$ rotation matrix is given by $R_{ij}=R_{ij}^bR_{ij}^a$, which is parametrized with the axis of rotation $\vec{\hat{n}}_{ij}$ and the angle $\omega_{ij}$. 
The $\mathbb{Z}_2$ vorticity at $i$th site is computed for the smallest closed loop:
\begin{equation}
    \nu_{i}=\frac{1}{2}\textrm{Tr }\left[U_{(x_i-1,y_i-1),(x_i,y_i-1)}U_{(x_i,y_i-1),(x_i,y_i)}U_{(x_i,y_i),(x_i-1,y_i)}U_{(x_i-1,y_i),(x_i-1,y_i-1)}\right],
\end{equation}
with $(x_i,y_i)$ denoting the index of $i$th site.
Figure~\ref{fig_Z2plot} shows the $\mathbb{Z}_2$ vorticity distribution in the $\mathbb{Z}_2$ VC phase. 
Comparing with the spin textures in Fig.~1 of the main text, we identify the nontrivial $\mathbb{Z}_2$ vorticity at singular defects in each sublattice.

\subsection{Monte Carlo simulations and magnetic phase diagram}
\label{sec: phase_diagram}

\begin{figure}[t]
	\centering
	\includegraphics[width=0.75\textwidth]{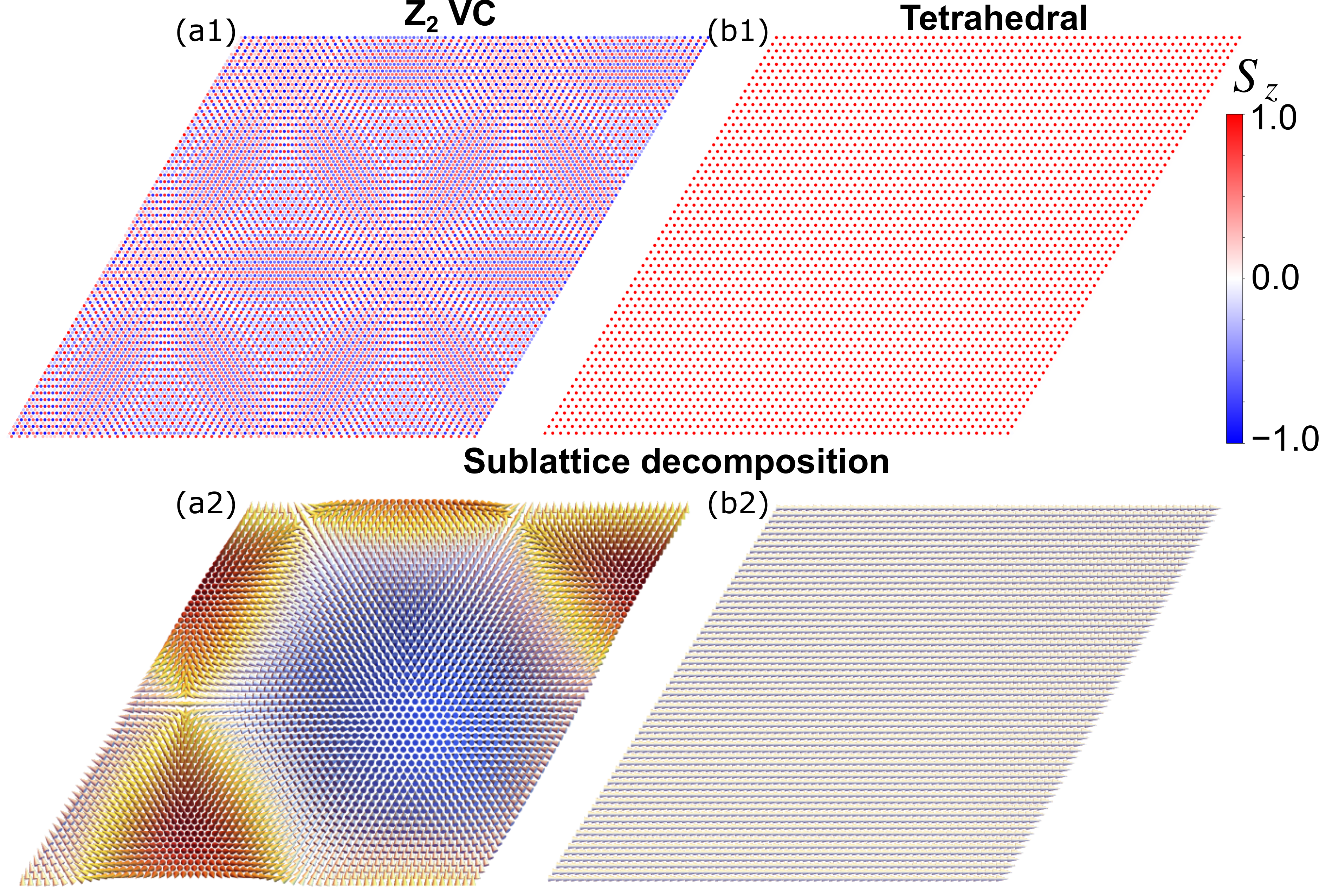}
	\caption{
    (a,b) Spin configurations of the (a)~$\mathbb{Z}_2$ VC and (b)~uniform tetrahedral phase. The out-of-plane spin components are plotted in (a1,b1) while the sublattice spin textures are plotted in (a2,b2). 
    We set $D_1/J_1=0.2$ in both panels with $b/J_1=0.0$ in (a) and $b/J_1=3.0$ in (b).
    }
	\label{fig_MC2}
\end{figure}

\begin{figure}[t]
	\centering
	\includegraphics[width=0.96\textwidth]{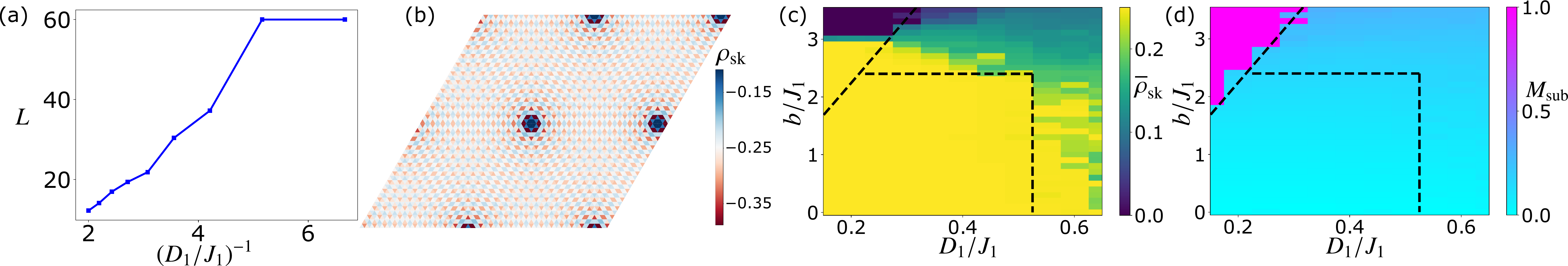}
	\caption{
    (a)~The periodicity ($L=120/\sqrt{\nu}$) of the $\mathbb{Z}_2$ VC is plotted as a function of $(D_1/J_1)^{-1}$, where $\nu$ denotes the total $\mathbb{Z}_2$ vorticity.
    (b)~The skyrmion number density is plotted for a magnetic unit cell of the $\mathbb{Z}_2$ VC.
    (c)~The average skyrmion number density per triangular plaquette and (d) the average magnetization within the sublattice as a function of the relative strength of interfacial DM interactions and external magnetic fields. We use the maximum value out of the four sublattices in (d). Black dashed lines coincide with those in Fig.~3 of the main text, indicating the $\mathbb{Z}_2$ VC phase with a half skyrmion number per vortex.
    }
	\label{fig_topocharge}
\end{figure}

We have performed Monte Carlo simulations of $120\times 120$ spins under periodic boundary conditions at different values of DM interactions and external magnetic fields. 
The other parameters are set as $B_1/J_1=1$ and $\lambda=0.5$. The result was used to compute the skyrmion number per $\mathbb{Z}_2$ vortex in Fig.~3 of the main text. In the following, we summarize additional results from this calculation.

As discussed in the main text, the $\mathbb{Z}_2$ VC is stable over a wide range of parameters. 
We find the phase transition from the $\mathbb{Z}_2$ VC to the uniform tetrahedral phase as the magnetic field is increased.
Figure~\ref{fig_MC2} shows the out-of-plane spin components of the $\mathbb{Z}_2$ VC and the uniform tetrahedral phase, obtained at $D_1/J_1=0.2$ with $b/J_1=0.0$ and $b/J_1=3.0$, respectively.
When a large magnetic field is applied, a ferromagnetic configuration is favored within each sublattice. As a result, the uniform tetrahedral phase becomes stable above the critical magnetic fields, as shown in Fig.~\ref{fig_MC2}(b). At $b/J_1=3.0$, the uniform tetrahedral phase consists of one spin pointing up and the other three spins forming a nearly coplanar 120$^\circ$ order as a compromise between the field-polarized state and the tetrahedral state.

Figure~\ref{fig_topocharge}(a) shows the periodicity of the $\mathbb{Z}_2$ VC against $(D_1/J_1)^{-1}$.
Here, we assume that the spacing between $\mathbb{Z}_2$ vortices is spatially uniform and define the periodicity as $L=120/\sqrt{\nu}$, where $\nu$ denotes the total $\mathbb{Z}_2$ vorticity.
The value of $L$ is averaged over magnetic fields up to $b/J_1<2.0$ in the $\mathbb{Z}_2$ VC phase.
Similarly to the ferromagnetic skyrmion lattice~\cite{mochizukiSpinWaveModesTheir2012}, we find that the periodicity is approximately proportional to $(D_1/J_1)^{-1}$ for $D_1/J_1>0.2$.
However, there is a discontinuous jump in $L$ at $D_1/J_1=0.2$ due to the finite size effect. This is consistent with Fig.~3 of the main text, where the finite size effect is observed for $D_1/J_1\le 0.2$.
The periodicity of the ferromagnetic skyrmion crystal is also proportional to $(D_1/J_1)^{-1}$ as discussed in ~\cite{bogdanovThermodynamicallyStableMagnetic1994}. In that case, the continuum model was employed to derive an analytical expression of the periodicity. Similarly, it is necessary to develop the effective field theory of the four-sublattice tetrahedral order for detailed theoretical analysis, which we leave for future investigations.

Figure~\ref{fig_topocharge}(b) shows the skyrmion number density within a magnetic unit cell of the $\mathbb{Z}_2$ VC.
We find that the average skyrmion number density $\overline{\rho}_\textrm{sk}$ per plaquette remains approximately $\pm1/4$ in the $\mathbb{Z}_2$ VC.
Since the uniform tetrahedral phase exhibits a skyrmion number density of $\pm 1/4$ per triangular plaquette, we could estimate the skyrmion number of a $\mathbb{Z}_2$ vortex by computing the difference in the total skyrmion number between the $\mathbb{Z}_2$ VC and the uniform tetrahedral phase ($\Delta N_\textrm{sk}=N_\textrm{sk}-N_\textrm{sk}^\textrm{tetra}$). 
Crucially, the difference in the total skyrmion number is given by $|\Delta N_\textrm{sk}|=\frac{1}{2}\nu$.
Thus, a $\mathbb{Z}_2$ vortex effectively inserts a half magnetic quantum flux into electronic systems.
In Fig.~\ref{fig_topocharge}(b), we observe large deviations from $\rho_\textrm{sk}=-1/4$ per plaquette at cores of $\mathbb{Z}_2$ vortices, indicating the insertion of a fractional magnetic flux quantum.
It also implies local deformations of the tetrahedral structure. 
The tetrahedral representation in Fig.~1(c) of the main text corroborates this observation.

Figure~\ref{fig_topocharge}(c) shows the average skyrmion number density~$\overline{\rho}_\textrm{sk}$ against interfacial DM interactions and external magnetic fields.
The value of $\overline{\rho}_\textrm{sk}$ remains nearly constant at small $D_1/J_1$ and $b/J_1$, indicating the preservation of tetrahedral order.
We note that the uniform tetrahedral phase has $\overline{\rho}_\textrm{sk}=\pm1/4$ for $D_1/J_1< 0.3$ and $b/J_1<3$ because the three nearly coplanar spins have a small negative $S_z$ component. 
When all spins have positive values in $S_z$ for $b/J_1>3$, the skyrmion number vanishes. Thus, the topological phase transition occurs at higher magnetic fields than the magnetic phase transition from the $\mathbb{Z}_2$ VC to the uniform tetrahedral phase, as indicated by a black dashed line. For $D_1/J_1\ge 0.3$, the topological phase transition occurs at lower fields. The value of $\overline{\rho}_\textrm{sk}$ decreases continuously above the critical magnetic fields in contrast to a sharp drop at $D_1/J_1< 0.3$.

Figure~\ref{fig_topocharge}(d) shows the average magnetization within the sublattice, defined as $M_{\textrm{sub},p}=\sum_{\vec{r}_p}S_{\vec{r}_p}^z/N_p$ with $N_p$ denoting the total number of spins in $p$-th sublattice.  
We plot the maximum value of $M_{\textrm{sub},p}$ across the four sublattices in Fig.~\ref{fig_MC2}(d).
This clearly delineates the phase boundary between the $\mathbb{Z}_2$ VC and the uniform tetrahedral phases, as the uniform tetrahedral phase supports ferromagnetic configurations in each sublattice with fully polarized spins in one sublattice. 
The critical magnetic field of the phase transition is proportional to $D_1/J_1$ as indicated by a black dashed line. 
Given the instability of the uniform tetrahedral phase at $b=0$ and $D_1>0$, we extrapolate the phase boundary between the $\mathbb{Z}_2$ VC and the uniform tetrahedral phases down to $b=D_1=0$, as depicted in Fig.~3 of the main text.

\subsection{Stability of $\mathbb{Z}_2$ vortex crystals compared to the helical phase}

\begin{figure}[t]
	\centering
	\includegraphics[width=0.95\textwidth]{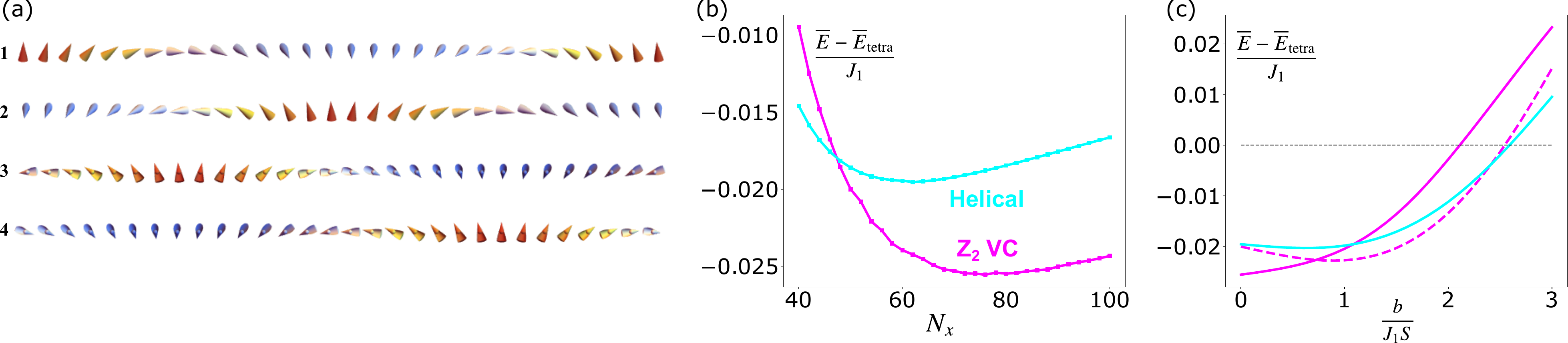}
	\caption{
    (a) Spin configuration of each sublattice in the helical phase, obtained at $B_1/J_1=1$, $\lambda=0.5$, $D_1/J_1=0.5$, and $b/J_1=0$. The numbers on the left denote the sublattice indices.
    (b,c) The average energy per lattice site, $\overline{E}$, is plotted relative to the uniform tetrahedral phase for the $\mathbb{Z}_2$ VC~(magenta) and the helical phase~(cyan).
    In (b), the energy is computed as a function of the system size at $b/J_1=0$. The $\mathbb{Z}_2$ VC is simulated with $N_x\times N_x$ spins while the helical phase is simulated with $N_x\times 4$ spins.
    In (c), the energy is computed as a function of the out-of-plane magnetic field while the system size is fixed. The solid (dashed) magenta line represents the result of $\mathbb{Z}_2$ VC with $N_x=76$ ($N_x=50$) while the size of helical phase is fixed at $N_x=62$.
    }
	\label{fig_helical}
\end{figure}

As discussed in the main text, the $\mathbb{Z}_2$ VC is stabilized by DM interactions even without external fields.
In contrast, a finite magnetic field is required to stabilize the skyrmion crystal in ferromagnetic skyrmion materials because the helical phase is the lowest energy configuration without magnetic fields~\cite{muhlbauerSkyrmionLatticeChiral2009}.
To confirm the stability of the $\mathbb{Z}_2$ VC over the helical phase in our model, we compare the energy of the $\mathbb{Z}_2$ VC and the helical phase under varying out-of-plane magnetic fields.
In the following, we fix $B_1/J_1=1$, $\lambda=0.5$, and $D_1/J_1=0.5$.

First, the optimal size of periodic cells is investigated for the $\mathbb{Z}_2$ VC and the helical phase in the absence of external magnetic fields.
We employ the Monte Carlo annealing method to stabilize each phase with $N_x\times N_x$ spins and $N_x\times 4$ spins, respectively.
Figure~\ref{fig_helical}(a) shows the four-sublattice helical spin configuration with $N_x=60$.
The maximum size of periodic cells is obtained at $N_x=50$ for the $\mathbb{Z}_2$ VC and $N_x=60$ for the helical phase. 
The Monte Carlo simulations are trapped by the local minima at larger values of $N_x$, forming multiple periodic cells within a simulated sample. 
To simulate a larger periodic cell, we interpolate spin textures of the $\mathbb{Z}_2$ VC and the helical phase with a larger $N_x$ and relax them in Monte Carlo annealing and LLG simulations up to $N_x=100$.
As shown in Fig.~\ref{fig_helical}(b), the energy of the $\mathbb{Z}_2$ VC decreases with the system size up to $N_x=76$. The energy slowly increases for $N_x>76$.
The size dependence of the helical phase is less significant and reaches the minimum at $N_x=62$. 
Crucially, the energy of the $\mathbb{Z}_2$ VC is well below the helical phase, confirming the stability of the $\mathbb{Z}_2$ VC against the helical phase without magnetic fields.

We also investigate the classical energies of the $\mathbb{Z}_2$ VC, the uniform tetrahedral phase, and the helical phase under different magnetic field strengths by relaxing the spin configurations in the LLG simulations. 
Figure~\ref{fig_helical}(c) shows the average energy per site of the $\mathbb{Z}_2$ VC and the helical phase compared to the uniform tetrahedral phase. 
While the length of the helical phase is fixed at $N_x=62$~(cyan), we investigate the $\mathbb{Z}_2$ VC with $N_x=76$~(solid magenta) and $N_x=50$~(dashed magenta).
With increasing external fields, the $\mathbb{Z}_2$ VC with a shorter periodicity becomes more stable. 
The phase transition between the $\mathbb{Z}_2$ VC and the uniform tetrahedral phase occurs around $b=2.5$, substantially lower than the critical field in Fig.~3 of the main text. 
This is because the system size is not optimized at finite magnetic fields in this simulation. 
Also, the energy difference between the $\mathbb{Z}_2$ VC and the helical phase becomes smaller near the critical field. 
Interestingly, there is a narrow range of magnetic fields where the helical phase becomes the lowest energy configuration. This is consistent with Fig.~3 of the main text, where the helical phase is obtained at the phase boundary between the $\mathbb{Z}_2$ VC and the uniform tetrahedral phase.
However, further study is necessary to clarify the stability of the helical phase.

\begin{figure}[t]
	\centering
	\includegraphics[width=0.7\textwidth]{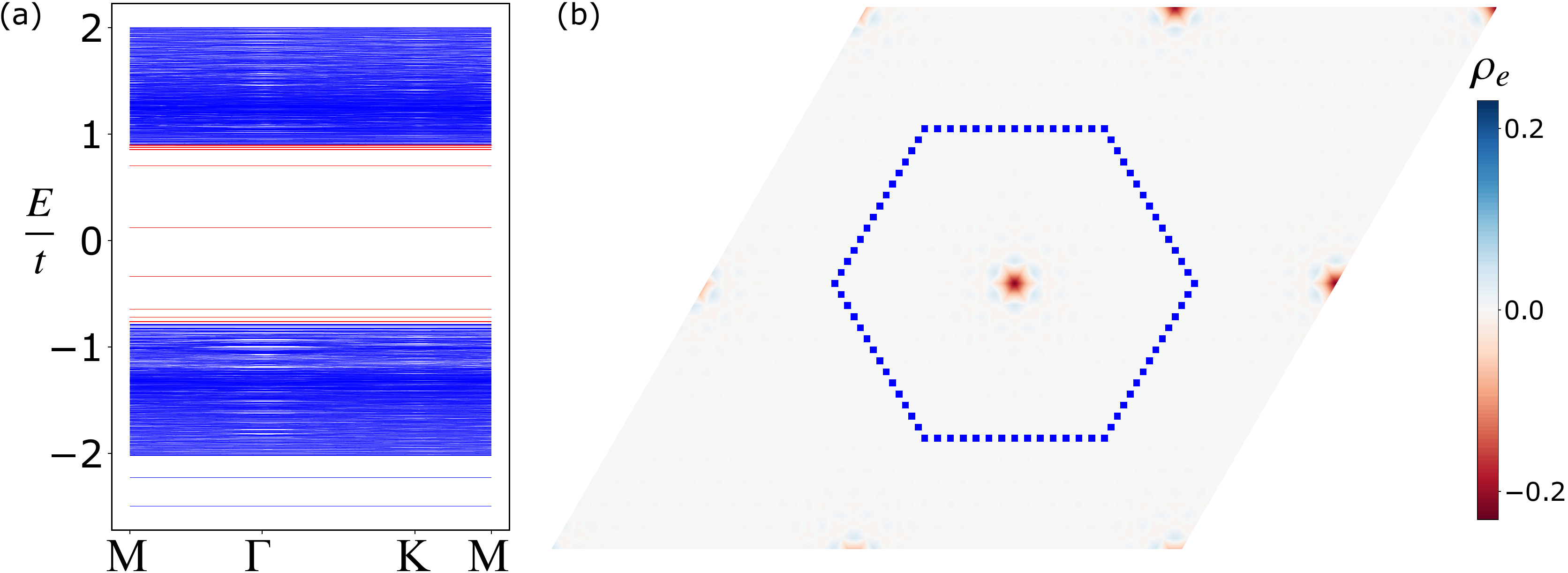}
	\caption{
    (a) Band structure of the Kondo lattice model employing the spin configurations of the $\mathbb{Z}_2$ VC. Bound states of $\mathbb{Z}_2$ vortices near $E=0$ are highlighted in red.
    (b) The charge density of the $\mathbb{Z}_2$ VC phase at the half-filling condition, where the uniform background charge is subtracted. A fractional electric charge bound to $\mathbb{Z}_2$ vortices is computed within a hexagon of length $L$ up to $L=14$, which is indicated by a blue dashed line.
    }
	\label{fig_charge}
\end{figure}

\subsection{Kondo lattice model}
To investigate the fractional electric charge bound to a $\mathbb{Z}_2$ vortex, we consider the Kondo lattice model:
\begin{equation}
     H_\textrm{el}
     =
     -\sum_{\braket{\vec{r},\vec{r}'}}\sum_{s\in\{\uparrow,\downarrow\}}t\, c^\dagger_{\vec{r},s}c^{\,}_{\vec{r}',s}-J\sum_{\vec{r}}\sum_{s,s'\in\{\uparrow,\downarrow\}}c^\dagger_{\vec{r},s}\,\vec{S}_{\vec{r}}\cdot\vec{\sigma}\,c^{\,}_{\vec{r},s'},
\end{equation}
with the Pauli matrices $\vec{\sigma}$, the hopping amplitude $t$, and the Kondo coupling $J$. 
The creation and annihilation operators of conduction electrons with spin $s$ at site $\vec{r}$ are denoted as $c^\dagger_{\vec{r},s}$ and $c_{\vec{r},s}$, respectively.
We employ the spin texture of the magnetic unit cell of the $\mathbb{Z}_2$ VC in Fig.~1 of the main text, which is represented as $\vec{S}_{\vec{r}}=(\sin \theta_{\vec{r}}\cos\phi_{\vec{r}},\sin \theta_{\vec{r}}\sin\phi_{\vec{r}},\cos\theta_{\vec{r}})$.
In the strong coupling limit ($J\gg t$), spins of conduction electrons are polarized by the local spin moment. 
We diagonalize the Kondo coupling term by applying the unitary matrix $U_{\vec{r}}=\vec{n}_{\vec{r}}\cdot \vec{\sigma}$ with $\vec{n}_{\vec{r}}=\left(\sin \frac{\theta_{\vec{r}}}{2}\cos\phi_{\vec{r}},\sin \frac{\theta_{\vec{r}}}{2}\sin\phi_{\vec{r}},\cos\frac{\theta_{\vec{r}}}{2}\right)$~\cite{tataraMicroscopicApproachCurrentdriven2008a}.
The new basis is written as
\begin{equation}
\begin{pmatrix}
 d_{\vec{r},p} \\
 d_{\vec{r},a}
\end{pmatrix}
=
\vec{n}_{\vec{r}}\cdot\vec{\sigma}
\begin{pmatrix}
 c_{\vec{r},\uparrow} \\
 c_{\vec{r},\downarrow}
\end{pmatrix}
=
\begin{pmatrix}
 \cos\frac{\theta_{\vec{r}}}{2} & \sin\frac{\theta_{\vec{r}}}{2}e^{-\textrm{i}\phi_{\vec{r}}}\\
\sin\frac{\theta_{\vec{r}}}{2}e^{\textrm{i}\phi_{\vec{r}}}&-\cos\frac{\theta_{\vec{r}}}{2} 
\end{pmatrix}
\begin{pmatrix}
 c_{\vec{r},\uparrow} \\
 c_{\vec{r},\downarrow}
\end{pmatrix},
\end{equation}
where $d_{\vec{r},p}/d_{\vec{r},a}$ is the annihilation operator of electrons with spins parallel/antiparallel to the local spin moment. 
After the transformation, we have
\begin{equation}
    -J\sum_{\vec{r}}\sum_{s,s'\in\{\uparrow,\downarrow\}}c^\dagger_{\vec{r},s}\,\vec{S}_{\vec{r}}\cdot\vec{\sigma}\,c^{\,}_{\vec{r},s'}
    =-J\sum_{\vec{r}}\left(d^\dagger_{\vec{r},p},d^{\,}_{\vec{r},p}-d^\dagger_{\vec{r},a}d^{\,}_{\vec{r},a}\right).
\end{equation}
When the Kondo coupling is much larger than the hopping amplitude, the low-energy model is obtained by projecting hopping matrices to the new basis $d^\dagger_{\vec{r},p}d_{\vec{r}',p}$. It is given by
\begin{equation}
    H_\textrm{el}=-\sum_{\braket{\vec{r},\vec{r}'}}t\,\left(\cos\frac{\theta_{\vec{r}}}{2}\cos\frac{\theta_{\vec{r}'}}{2}+\sin\frac{\theta_{\vec{r}}}{2}\sin\frac{\theta_{\vec{r}'}}{2}e^{-\textrm{i}(\phi_{\vec{r}}-\phi_{\vec{r}'})}\right)d^\dagger_{\vec{r},p}d^{\,}_{\vec{r}',p}.
    \label{eq:LowE}
\end{equation}
In the main text, the index for parallel spins is dropped.

In the triangular Kondo lattice model, the tetrahedral order becomes the ground state near $1/4$ and $3/4$ fillings~\cite{Martin2008, Akagi2012}, which corresponds to the half-filling condition in the low-energy model.
Performing the Fourier transform, we obtain the band structure of Eq.~\eqref{eq:LowE} as shown in Fig.~\ref{fig_charge}(a).
The four-sublattice structure guarantees the quadruple degeneracy of bands. 
We identify several flat bands near $E=0$, which are bound states of $\mathbb{Z}_2$ vortices (highlighted in red).
Denoting the $n$th energy eigenstate at $i$th sublattice as $\phi_{\vec{k},n,i}$, the charge density is computed as
\begin{equation}
\rho_{e,i}=\int_\textrm{BZ} \frac{d\vec{k}}{\Omega_\textrm{BZ}}\sum_{n\in\textrm{occ}}\left(\phi_{\vec{k},n,i}^*\phi_{\vec{k},n,i}-\frac{1}{2}\right),
\end{equation}
with $\Omega_\textrm{BZ}$ denoting the area of the 1st Brillouin zone. Here, the uniform background charge is subtracted. 
We set $E_F=0$ to define the occupied bands. 
The charge density is plotted in Fig.~\ref{fig_charge}(b), showing fractional electric charges bound to $\mathbb{Z}_2$ vortices. In the main text, we plot the total charge within a hexagon of length $L$ centered at a $\mathbb{Z}_2$ vortex. With 50$\times$50 sublattice sites within the magnetic unit cell of the $\mathbb{Z}_2$ VC, we compute the electric bound charge up to $L=14$ to avoid interferences with other $\mathbb{Z}_2$ vortices, which is indicated in Fig.~\ref{fig_charge}(b).

\subsection{Magnetic excitations}

\begin{figure}[t]
	\centering
	\includegraphics[width=0.6\textwidth]{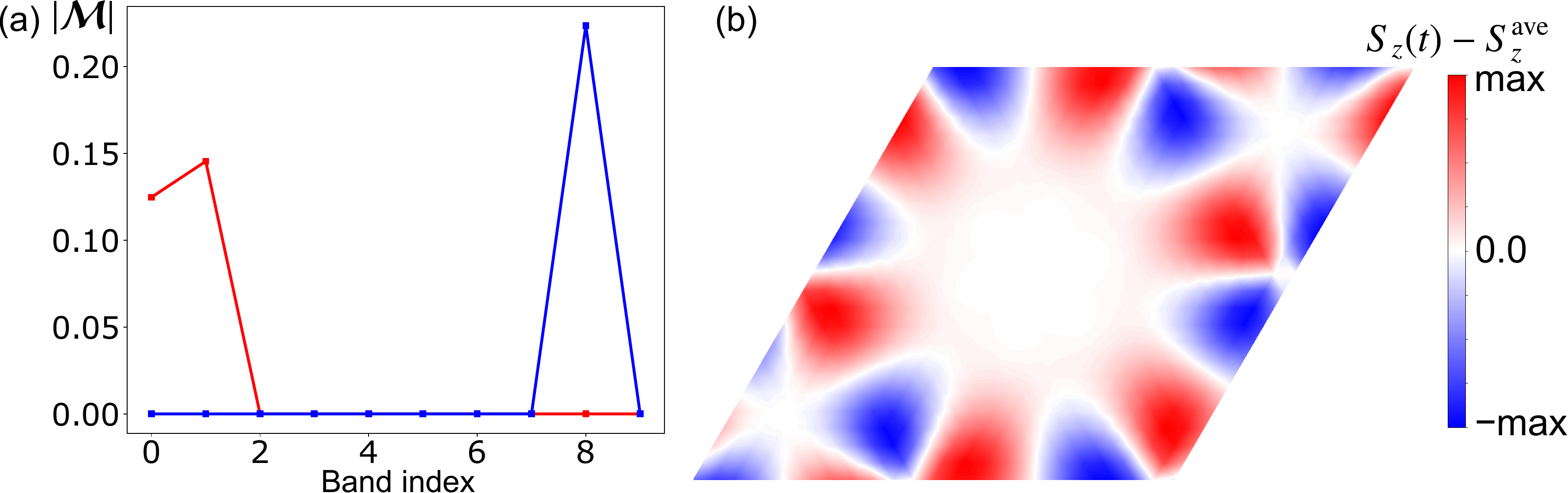}
	\caption{
    (a) The amplitude of dynamical magnetization is plotted for the ten lowest-energy magnon eigenstates at $\Gamma$ point, computed for the $\mathbb{Z}_2$ VC of Fig.~1 in the main text.
    The in-plane component and the out-of-plane component are plotted in blue and red, respectively. The parameters are set as $B_1/J_1=1$, $\lambda=0.5$, $D_1/J_1=0.5$, and $b/J_1=0$.
    (b) Snapshot of the time evolution of spins induced by the 9th magnon eigenmode. The color indicates the time-dependent part of spins after subtracting the time-averaged configurations. 
    }
	\label{fig_dyM}
\end{figure}

In this section, we consider magnetic excitations of the $\mathbb{Z}_2$ VC, focusing on magnetically active magnon modes.
When the $n$th magnon mode is excited, oscillations in spins are given as~\cite{fishmanNormalModesSpin2019}
\begin{eqnarray}
\Delta \vec{S}^{(n)}_{\vec{r}}(\vec{k},t)=2\sqrt{N}\,\textrm{Re}\big[e^{-\textrm{i}\omega_n t}\bra{0}\hat{\vec{S}}_{\vec{r}}\ket{n,\vec{k}} \big],
\end{eqnarray} 
where $\omega_n$ is the frequency of the $n$th magnon mode and $\ket{n,\vec{k}}=a_{\vec{k},n}^{\dagger}\ket{0}$ is the wave function of $n$th magnon mode. 
With the Holstein-Primakoff expansion~\cite{holsteinFieldDependenceIntrinsic1940}, the spin operator is expanded up to quadratic terms as 
\begin{eqnarray}
\hat{\vec{S}}_{\vec{r}}
&=&
\frac{\sqrt{2S}}{2}(a_{\vec{r}}+a_{\vec{r}}^\dagger)\hat{e}^{1}_{\vec{r}}+\frac{\sqrt{2S}}{2\textrm{i}}(a_{\vec{r}}-a_{\vec{r}}^\dagger)\hat{e}^{2}_{\vec{r}}+(S-a_{\vec{r}}^\dagger a_{\vec{r}})\hat{e}^{3}_{\vec{r}}\nonumber\\
&=&
\sum_{\vec{k}'}
\frac{e^{\textrm{i}\vec{k}'\cdot (\vec{R}+\vec{r}_j)}}{\sqrt{N_\textrm{u}}}\left[ \frac{\sqrt{2S}}{2}(a_{\vec{k}',j}+a_{\vec{k}',j}^\dagger)\hat{e}^{1}_{j}+\frac{\sqrt{2S}}{2\textrm{i}}(a_{\vec{k}',j}-a_{\vec{k}',j}^\dagger)\hat{e}^{2}_{j}+(S-a_{\vec{k}',j}^\dagger a_{\vec{k}',j})\hat{e}^{3}_{j}\right].
\end{eqnarray}
Here, we introduce the Bogoliubov transformation for diagonalizing the spin-wave Hamiltonian:
\begin{align}
    \begin{pmatrix}
        a_{\vec{k},j} \\ a^\dagger_{-\vec{k},j}
    \end{pmatrix}
    =
    \sum_{n=1}^{N_\text{b}}
    T^{j,n}_{\vec{k}}
    \begin{pmatrix}
        a_{\vec{k},n} \\ a^\dagger_{-\vec{k},n}
    \end{pmatrix},
    \quad
    T^{j,n}_{\vec{k}}
    =
    \begin{pmatrix}
        u^{j,n}_{\vec{k}} & \left(v^{j,n}_{-\vec{k}} \right)^*\\
        v^{j,n}_{\vec{k}} & \left(u^{j,n}_{-\vec{k}}\right)^*
    \end{pmatrix},
    \label{eq:Bogo}
\end{align}
where $n$ is the band index. 
After substitution, we obtain
\begin{align}
    \bra{0}\hat{\vec{S}}_{\vec{r}_j}\ket{n,\vec{k}}
    &
    =
     \sum_{\vec{k}'}
     \frac{e^{\textrm{i}\vec{k}'\cdot (\vec{R}+\vec{r}_j)}}{\sqrt{N_\textrm{u}}}\bra{0}\left[ \frac{\sqrt{2S}}{2}(a_{\vec{k}',j}+a_{\vec{k}',j}^\dagger)\hat{e}^{1}_{j}+\frac{\sqrt{2S}}{2\textrm{i}}(a_{\vec{k}',j}-a_{\vec{k}',j}^\dagger)\hat{e}^{2}_{j}+(S-a_{\vec{k}',j}^\dagger a_{\vec{k}',j})\hat{e}^{3}_{j}\right]
     a_{\vec{k},n}^{\dagger}\ket{0}
     \nonumber\\
     &
     =
     \frac{e^{\textrm{i}\vec{k}\cdot (\vec{R}+\vec{r}_j)}}{\sqrt{N_\textrm{u}}}
     \frac{\sqrt{2S}}{2}(\hat{e}^1_j-\textrm{i}\hat{e}^2_j)u_{\vec{k}}^{j,n}+\frac{e^{-\textrm{i}\vec{k}\cdot (\vec{R}+\vec{r}_j)}}{\sqrt{N}_\textrm{u}}\frac{\sqrt{2S}}{2}(\hat{e}^1_j+\textrm{i}\hat{e}^2_j) v_{\vec{k}}^{j,n}.
\end{align}
Thus, the time evolution of spins due to the $n$th magnon mode is given by
\begin{align}
    \Delta \vec{S}^{(n)}_{\vec{r}_j}(\vec{k},t)
    &
    =
    \sqrt{2SN_b}\,\textrm{Re}\left[
    e^{-\textrm{i}\omega_n t+\textrm{i}\vec{k}\cdot (\vec{R}+\vec{r}_j)}(\hat{e}^1_j-\textrm{i}\hat{e}^2_j)u_{\vec{k}}^{j,n}
    +
    e^{-\textrm{i}\omega_n t-\textrm{i}\vec{k}\cdot (\vec{R}+\vec{r}_j)}(\hat{e}^1_j+\textrm{i}\hat{e}^2_j) v_{\vec{k}}^{j,n}\right].
\end{align}
We also define the dynamical magnetization of $n$-th magnon eigenstate as~\cite{hirosawaMagnetoelectricCavityMagnonics2022}
\begin{align}
\boldsymbol{\mathcal{M}}^{(n)}_{\vec{k}}
&= \sum_{ j=1 }^{N_\text{b}}
\frac{(\hat{e}^1_j-\textrm{i}\hat{e}^2_j) u^{j,n}_{\vec{k}} + (\hat{e}^1_j+\textrm{i}\hat{e}^2_j) v^{j,n}_{\vec{k}} }{\sqrt{N_\text{b}}} 
.
\label{eq:MagnonM}
\end{align}
The dynamical magnetization is proportional to the amplitude of oscillations of the total spins, characterizing magnetically active excitations. 

Figure~\ref{fig_dyM}(a) shows the dynamical magnetization of the ten lowest-energy magnon eigenstates in the $\mathbb{Z}_2$ VC at $\Gamma$ point. The 1st and 2nd eigenstates exhibit finite in-plane dynamical magnetization, while the 9th eigenstate has out-of-plane dynamical magnetization.
These magnetically active modes are highlighted in the band spectrum of Fig.4(a) in the main text.
Supplementary videos show the time evolution of spins of the 1st, 2nd, and 9th magnon eigenstates, with (1a-3a) for the full spin configurations and (1b-3b) for one of the sublattices. 
We find that skyrmion-like topological defects undergo gyrotropic motion in the 1st and 2nd modes in opposite directions. 
They are similar to the CCW and CW modes in ferromagnetic skyrmion crystals~\cite{mochizukiSpinWaveModesTheir2012}.
However, the time evolution of the 9th eigenmode appears different from the breathing mode in skyrmion crystals. 
While the breathing mode is associated with oscillations in the skyrmion radius, we observe oscillations in the radius with twisted motion in the $\mathbb{Z}_2$ VC.
Figure~\ref{fig_dyM}(b) shows the snapshot of time-evolved spins by the 9th eigenmode, where the static part is subtracted. 
It shows deformations with the six-fold rotational symmetry in skyrmion-like defects, indicating the hybridization between the breathing mode and a sixth-order polygon deformation mode~\cite{schutteMagnonskyrmionScatteringChiral2014}.
This is explained by the breaking of continuous rotational symmetry in skyrmion-like defects of the $\mathbb{Z}_2$ VC, where they form hexagonal domain walls in contrast to circular domain walls of skyrmions.

\subsection{Edge mode calculation}

\begin{figure}[t]
	\centering
	\includegraphics[width=0.95\textwidth]{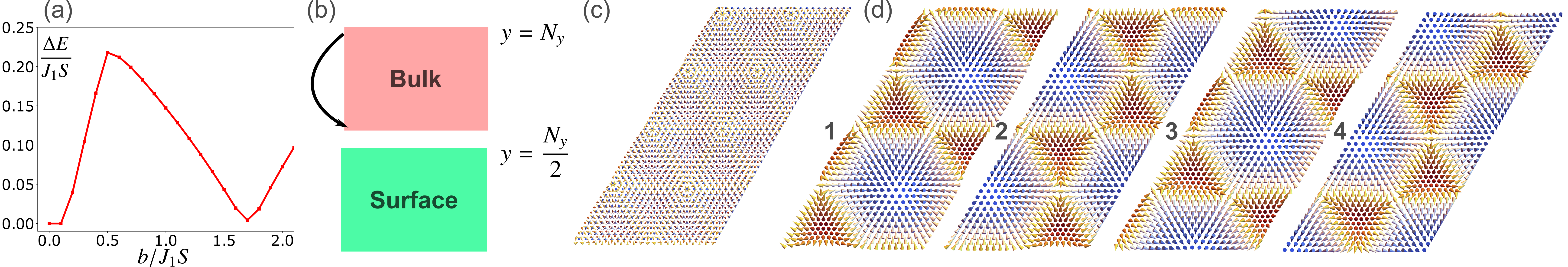}
	\caption{
    (a) The energy gap between the 4th and 5th magnon eigenstates as a function of the external field, computed for the $\mathbb{Z}_2$ VC of Fig.~1 in the main text. 
    (b) Schematic representation of the semi-infinite boundary condition. The lower half of the spins is treated as surface spins, while the upper half is treated as bulk spins.
    (c) Spin configurations of the $\mathbb{Z}_2$ VC under the semi-infinite boundary condition along the $y$-axis~($46\times92$ spins). The periodic boundary condition is imposed along the $x$-axis.
    The parameters are set to $D_1/J_1=0.5$ and $b/J_1=0.5$.
    (d) The sublattice decomposition of the spin textures in (c), with numbers denoting the sublattice index. 
    }
	\label{fig_semiMC}
\end{figure}

The out-of-plane magnetic field induces the topological band gap in the lowest four magnon bands, as discussed in the main text.
Figure~\ref{fig_semiMC}(a) shows the energy gap between the 4th and 5th magnon eigenstates against external magnetic fields.
The topological phase transition occurs around $0.1< b/J_1<0.2$, resulting in the nontrivial Chern number of the lowest bands ($C=-1$). The Chern number remains nontrivial up to $b/J_1\approx 1.7$, where the gap closes again.

The bulk-boundary correspondence guarantees the existence of a chiral magnonic edge state across the band gap.
Here, we study the chiral edge mode in a semi-infinite system to avoid the finite size effect.
The magnetic textures of a semi-infinite system are obtained by Monte Carlo annealing with a special boundary condition, which was originally used for investigating the surface magnetization~\cite{sunMonteCarloCalculations1996}.
Taking the semi-infinite boundary in $y$-axis with $N_y$ spins, we divide the system into two blocks. 
As schematically illustrated in Fig.~\ref{fig_semiMC}(a), spins at $y\le N_y/2$ are treated as surface spins while those at $y>N_y/2$ are bulk spins. 
The semi-infinite boundary condition is introduced by assuming the periodicity within the bulk block, where spins at the upper boundary of the bulk are coupled with those at the lower boundary of the bulk.
However, spins at the lower boundary of the bulk are coupled only to the nearest and next nearest neighbors.
Thus, asymmetric spin interactions are introduced in the semi-infinite boundary condition. 
Figure~\ref{fig_semiMC}(b) and (c) show the obtained spin configurations of the $\mathbb{Z}_2$ VC under the semi-infinite boundary condition at $D_1/J_1=0.5$ and $b/J_1=0.5$. Although the system size is only twice as large as the magnetic unit cell, the obtained spin configurations are almost identical to the bulk spin textures. 

We employ the renormalization method to compute the local density of states~(LDOS) of a semi-infinite system~\cite{HENK199369, Mook2014edge}. 
The semi-infinite system is decomposed into blocks of principle layers, where the $0$th layer corresponds to the surface block and $n$th layer is the bulk block ($n>0$). 
Assuming the periodic boundary along the $x$-axis, the LDOS of magnons within the surface block is given by
\begin{align}
    N_0(\omega,q_x)=-\frac{1}{\pi} \textrm{Im}\textrm{Tr}\Big[G_{00}(\omega+\textrm{i}\kappa, q_x)\Big],
\end{align}
with the subscript indicating the layer index.
We also introduce the damping rate $\kappa$.
The Green's function $G_{nm}$ satisfies
\begin{align}
    \delta_{nm}&=\sum_{j=0}^\infty (z\,\Sigma\delta_{nj}-H_{nj})G_{jn}, 
\end{align}
with $z=\hbar(\omega+\textrm{i}\kappa)$ and $H_{nm}$ denotes the Hamiltonian matrix element between $n$th and $m$th principle layers. Here, we multiply $z$ by $\Sigma$ to account for the paraunitary nature of bosonic wave functions. 
We derive the edge Green's function $G_{00}$ by renormalizing the Hamiltonian matrix to a block diagonal form~($|H_{nm}|\rightarrow 0$).
Figure~4 of the main text is obtained with $\hbar\kappa/J_1=0.02$.